\documentclass[12pt,preprint]{aastex}
\usepackage{epsf}

\def\linebreak{\hfil\break}

\def\
{\hfil\linebreak}

\def\orch{{\it Orchestra}}
\def\nbody{{$n$-body}}
\def\nbodies{{$n$-bodies}}
\def\deg{\ifmmode {^\circ}\else {$^\circ$}\fi}
\def\degree{\ifmmode {^\circ}\else {$^\circ$}\fi}
\def\mum{\ifmmode {\rm \mu {\rm m}}\else $\rm \mu {\rm m}$\fi}
\def\arcsec{\ifmmode ^{\prime \prime}\else $^{\prime \prime}$\fi}
\def\secpoint{\mbox{$''\mskip-7.6mu.\,$}}

\def\inch{\ifmmode ^{\prime \prime}\else $^{\prime \prime}$\fi}
\def\arcmin{\ifmmode ^{\prime}\else $^{\prime}$\fi}
\def\qprime{\ifmmode q^{\prime}\else $q^{\prime}$\fi}

\def\degree{\ifmmode {^\circ}\else {$^\circ$}\fi}
\def\arcsec{\ifmmode ^{\prime \prime}\else $^{\prime \prime}$\fi}
\def\secpoint{\mbox{$''\mskip-7.6mu.\,$}}

\def\inch{\ifmmode ^{\prime \prime}\else $^{\prime \prime}$\fi}
\def\arcmin{\ifmmode ^{\prime}\else $^{\prime}$\fi}

\def\mjup{\ifmmode { M_J}\else $ M_J$\fi}
\def\rjup{\ifmmode { R_J}\else $ R_J$\fi}
\def\mearth{\ifmmode { M_{\oplus}}\else $ M_{\oplus}$\fi}
\def\rearth{\ifmmode { R_{\oplus}}\else $ R_{\oplus}$\fi}
\def\ldust{\ifmmode { L_d}\else $ L_d$\fi}
\def\ldstar{\ifmmode { L_d / L_{\star}}\else $ L_d / L_{\star}$\fi}
\def\lstar{\ifmmode { L_{\star}}\else $ L_{\star}$\fi}
\def\lsun{\ifmmode { L_{\odot}}\else $ L_{\odot}$\fi}
\def\mlz{\ifmmode m_{L,0}\else $ m_{l,0}$\fi}
\def\ncr{\ifmmode n_{c,rel}\else $ n_{c,rel}$\fi}
\def\bl{\ifmmode b_l\else $ b_l$\fi}
\def\mcl{\ifmmode M_d\else $ M_d$\fi}
\def\xcl{\ifmmode x_d \else $ x_d$\fi}
\def\ap{\ifmmode a_p\else $ a_p$\fi}
\def\rp{\ifmmode R_p\else $ R_p$\fi}
\def\mp{\ifmmode M_p\else $ M_p$\fi}
\def\mpl{\ifmmode M_p\else $ M_p$\fi}
\def\mstar{\ifmmode M_{\star}\else $ M_{\star}$\fi}
\def\msun{\ifmmode M_{\odot}\else $ M_{\odot}$\fi}
\def\tstar{\ifmmode T_{\star}\else $ T_{\star}$\fi}
\def\rstar{\ifmmode R_{\star}\else $ R_{\star}$\fi}
\def\rsun{\ifmmode R_{\odot}\else $ R_{\odot}$\fi}
\def\mjup{\ifmmode M_{J}\else $ M_{J}$\fi}
\def\rjup{\ifmmode R_{J}\else $ R_{J}$\fi}
\def\mjupyr{\ifmmode { M_J~yr^{-1}}\else $ M_J~yr^{-1}$\fi}
\def\msunyr{\ifmmode { M_{\odot}~yr^{-1}}\else $ M_{\odot}~yr^{-1}$\fi}
\def\gyr{\ifmmode {\rm g~yr^{-1}}\else $\rm g~yr^{-1}$\fi}
\def\ergg{\ifmmode {\rm erg~g^{-1}}\else $\rm erg~g^{-1}$\fi}
\def\kms{\ifmmode {\rm km~s^{-1}}\else $\rm km~s^{-1}$\fi}
\def\ms{\ifmmode {\rm m~s^{-1}}\else $\rm m~s^{-1}$\fi}
\def\rhill{\ifmmode R_H\else $R_H$\fi}
\def\rfast{\ifmmode R_{fast}\else $R_{fast}$\fi}
\def\rgap{\ifmmode R_{gap}\else $R_{gap}$\fi}
\def\vhill{\ifmmode v_H\else $v_H$\fi}
\def\qb{\ifmmode Q_b\else $Q_b$\fi}
\def\qg{\ifmmode Q_g\else $Q_g$\fi}
\def\qc{\ifmmode Q_c\else $Q_c$\fi}
\def\qdstar{\ifmmode Q_D^\star\else $Q_D^\star$\fi}
\def\mpro{\ifmmode m_{pro}\else $m_{pro}$\fi}
\def\mesc{\ifmmode m_{esc}\else $m_{esc}$\fi}
\def\rmin{\ifmmode r_{min}\else $r_{min}$\fi}
\def\rmax{\ifmmode r_{max}\else $r_{max}$\fi}
\def\mmax{\ifmmode m_{max}\else $m_{max}$\fi}
\def\rmind{\ifmmode r_{min,d}\else $r_{min,d}$\fi}
\def\rmaxd{\ifmmode r_{max,d}\else $r_{max,d}$\fi}
\def\mmaxd{\ifmmode m_{max,d}\else $m_{max,d}$\fi}
\def\qz{\ifmmode q_{0}\else $q_{0}$\fi}
\def\qi{\ifmmode q_{i}\else $q_{i}$\fi}
\def\ql{\ifmmode q_{l}\else $q_{l}$\fi}
\def\qs{\ifmmode q_{s}\else $q_{s}$\fi}
\def\r0{\ifmmode r_{0}\else $r_{0}$\fi}
\def\m0{\ifmmode m_{0}\else $m_{0}$\fi}
\def\M0{\ifmmode M_{0}\else $M_{0}$\fi}
\def\xm{\ifmmode x_{m}\else $x_{m}$\fi}
\def\gyr{\ifmmode {\rm g~yr^{-1}}\else ${\rm g~yr^{-1}}$\fi}
\def\cms{\ifmmode {\rm cm~s^{-1}}\else ${\rm cm~s^{-1}}$\fi}
\def\gcms{\ifmmode {\rm g~cm^{-2}}\else $\rm g~cm^{-2}$\fi}
\def\gcmc{\ifmmode {\rm g~cm^{-3}}\else $\rm g~cm^{-3}$\fi}

\def\2470{[24]--[70]}

\newbox\grsign \setbox\grsign=\hbox{$>$} \newdimen\grdimen \grdimen=\ht\grsign
\newbox\simlessbox \newbox\simgreatbox
\setbox\simgreatbox=\hbox{\raise.5ex\hbox{$>$}\llap
     {\lower.5ex\hbox{$\sim$}}}\ht1=\grdimen\dp1=0pt
\setbox\simlessbox=\hbox{\raise.5ex\hbox{$<$}\llap
     {\lower.5ex\hbox{$\sim$}}}\ht2=\grdimen\dp2=0pt

\begin{document}

\title{Collisional Cascade Caclulations for Irregular Satellite Swarms in Fomalhaut b}
\vskip 7ex
\author{Scott J. Kenyon}
\affil{Smithsonian Astrophysical Observatory,
60 Garden Street, Cambridge, MA 02138} 
\email{e-mail: skenyon@cfa.harvard.edu}

\author{Benjamin C. Bromley}
\affil{Department of Physics \& Astronomy, University of Utah, 
201 JFB, Salt Lake City, UT 84112} 
\email{e-mail: bromley@physics.utah.edu}
%
%

\begin{abstract}

We describe an extensive suite of numerical calculations for the collisional
evolution of irregular satellite swarms around 1--300~\mearth\ planets 
orbiting at 120~AU in the Fomalhaut system. For 10--100~\mearth\ planets, swarms 
with initial masses of roughly 1\% of the planet mass have cross-sectional areas 
comparable to the observed cross-sectional area of Fomalhaut b.  Among 
30--300~\mearth\ planets, our calculations yield optically thick swarms of 
satellites for ages of 1--10~Myr. Observations with HST and ground-based AO 
instruments can constrain the frequency of these systems around stars in the 
$\beta$~Pic moving group and possibly other nearby associations of young stars.

\end{abstract}


\section{INTRODUCTION}

Fomalhaut b is a planet candidate on an eccentric orbit at a distance of $\sim$ 120~AU 
from the A-type star Fomalhaut 
\citep[e.g.,][]{kalas2008,currie2012,galich2013,kalas2013,beust2014}.  The optical 
colors and lack of detections beyond 1~\mum\ suggest emission from a cloud of dust 
instead of a planetary photosphere 
\citep[e.g.,][]{maren2009,janson2012b,currie2012,kalas2013,janson2015}. 
The observed level of optical emission requires grains with a total cross-sectional 
area of roughly $10^{23}$~cm$^2$ \citep[e.g.,][]{kalas2008,currie2012,galich2013,kalas2013}.

Two types of collision models can produce a clump of dust emission at large distances 
from an A-type star. In the simplest picture, two objects with radii of roughly 
100~km collide at high velocity and generate an expanding cloud of small particles
\citep[e.g.,][]{kalas2008,galich2013,kcb2014,lawler2015}.  Clouds expanding at 
the escape velocity of a pair of 100~km objects are unresolved on 50--100~yr 
time scales \citep[e.g.,][]{galich2013,kcb2014,lawler2015}. Smaller ejected 
particles with larger optical depth generally expand more rapidly 
\citep[e.g.,][]{gault1963,housen2003}.  If these particles contain a reasonable 
amount of mass, HST or JWST images should resolve Fomalhaut b within the next 
decade \citep[e.g.,][]{tamayo2013,kcb2014}. When expanding clouds have an 
internal velocity dispersion, differential motion shears the cloud into a ring
\citep{kb2005,kcb2014}. Over the next decade, HST or JWST observations can also
test this prediction.

Alternative models posit a disk-shaped or a semi-spherical swarm of 
irregular
satellites 
orbiting a super-Earth mass planet \citep{kalas2008,kw2011a,kcb2014}. In this 
approach, a collisional cascade among particles with radii $r \lesssim$ 
500--1000~km maintains a large population of dust grains over the 200--400~Myr 
age of Fomalhaut. Analytic results for the long-term evolution favor particles 
with a power-law size distribution in swarms with masses of roughly 
0.1~\mearth\ around planets with masses $\mpl \approx$ 10~\mearth.  However, 
the radius of the largest object in the cascade (\rmax) depends on the slope 
$q$ of the power-law size distribution, where smaller $q$ requires larger 
\rmax. Large \rmax\ requires very massive satellite swarms.

Observational tests of this model are possible but more complicated 
\citep[e.g.,][]{kcb2014}. Radiation pressure from the central star 
prevents small particles with $r \lesssim$ 100~\mum\ from remaining 
bound to the planet. Ejection of these particles produces a distinct 
trail along the planet's orbit. Simple estimates suggest the density 
of small particles ejected from massive planets is detectable
\citep{kcb2014}. Testing this aspect of the model requires more 
detailed analyses of the dust ejected from the satellite swarm.

To explore this picture in more detail, we examine a suite of numerical 
simulations for spherical swarms of satellites orbiting super-Earth mass 
planets.  Aside from deriving the evolution of dust clouds as a function of 
the mass of the planet and the surrounding satellite system, we consider how 
the amount of dust orbiting the planet depends on the initial radius of the 
largest satellite, the bulk strength of satellites, and the recipe for 
distributing the debris from a collision into lower mass objects. For a 
standard model, swarms with initial masses \mcl\ = 0.01~\mpl\ and 
\rmax\ $\approx$ 200--400~km orbiting planets with masses \mpl\ = 
10--100~\mearth\ match the observed cross-sectional area of Fomalhaut b.  
Calculations with weaker satellites allow lower mass swarms to match the data.

For planets with \mpl\ $\approx$ 30--300~\mearth, $a \approx$ 100~AU, and ages 
of 1--10~Myr, satellite swarms have large optical depth $\tau \approx$ 0.1--1. 
Relative to the central star, predicted contrast ratios of $10^{-6} - 10^{-7}$ 
are at least a factor of 100 larger than observed in Fomalhaut b. Observations 
with ground-based AO systems \citep[e.g.,][]{biller2013} or HST 
\citep[e.g.,][]{schneider2014} can place limits on the frequency of optically 
thick satellite swarms around planets orbiting nearby young stars.

To connect our results with previous analytic work, we begin our discussion 
with the derivation of a simple analytic model (\S\ref{sec: anmod}). After 
summarizing the numerical approach (\S\ref{sec: num}), we describe the outcomes 
of simulations as a function of various input parameters (\S\ref{sec: calcs}).
The paper concludes with a brief discussion (\S\ref{sec: disc}) and a summary
of the major results (\S\ref{sec: conc}).

\section{ANALYTIC MODEL}
\label{sec: anmod}

To interpret observations of debris disks c. 2000, \citet{wyatt2002} and \citet{dom2003} 
developed an analytic model for the long-term evolution of a swarm of large solid objects 
in a circumstellar disk. \citet{kw2011a} later extended this approach to spherical swarms
of satellites orbiting a massive planet \citep[see also][]{kw2011b}. In a swarm of satellites,
objects with radius $r$, mass $m$, and mass density $\rho$ orbit within a spherical shell 
with width $\delta a$ centered at a distance $a$ from a planet with mass \mp\ and radius \rp.
The planet orbits with semimajor axis $a_p$ from a central star with mass \mstar\ and luminosity 
\lstar.  Destructive collisions between satellites produce a collisional cascade which slowly 
grinds solids into smaller and smaller objects. Defining an upper mass limit $m_{max}$ for 
solids participating in the cascade, the analytic model yields a simple formula for 
$N_{max}(t)$, the number of these large objects as a function of time. If radiation pressure 
sets $m_{min}$, a lower mass limit for solids with stable orbits around the planet, then the 
cascade produces a power-law size distribution between $m_{min}$ and $m_{max}$ 
\citep[e.g.,][]{dohn1969,will1994,obrien2003,koba2010a}. 
Setting the slope of this size distribution yields another simple formula for the time evolution 
of the surface area of the dust cloud $A_d(t)$. Adopting optical properties for solids in the
cloud yields the dust luminosity $L_d(t)$.

\subsection{Time evolution}
\label{sec: anmod-time}

To derive expressions for $N_{max}(t)$ and $A_d(t)$, \citet{wyatt2002}, \citet{dom2003}, and
\citet{kw2011a} adopt the particle-in-a-box model, where kinetic theory sets the collision 
rate $\dot{N}$.  Defining $V$ as the volume of the spherical shell, $\sigma$ as the geometric 
cross-section, and $v$ as the relative particle velocity, each particle has a collision time 
$t_c \approx N_0^{-1} (V / 2 \sigma v) $ where $N_0$ is the initial number of 
particles\footnote{ Formally, collisions destroy two identical particles on the time scale 
$2 t_c$; $t_c$ is then the time scale to destroy a single particle.}.  
If all collisions 
between particles are destructive, the number of large particles declines at a rate 
$\dot{N}_{max} \approx -N_{max}^2 / N_0 t_c$.  Solving for $N_{max}(t)$:
\begin{equation}
N_{max}(t) = { N_{max,0} \over {1 + t / t_c} } ~ ,
\label{eq: nmax} 
\end{equation}
where $N_{max,0}$ is the number of large objects at $t$ = 0.

With $N_{max}(t)$ known, the total cross-sectional area and dust luminosity follow. 
For any power-law size distribution with $N(r) \propto r^{-q}$, the total mass $M_d$ 
and cross-sectional area $A_d$ of the swarm are simple functions of the minimum size, 
the maximum size, and the slope $q$ \citep[e.g.,][]{wyatt2002,dom2003}.
The stellar energy intercepted by the solids is $L_d = A_d / 4 \pi a_p^2$. If $m_{min}$,
$m_{max}$, and $q$ never change, the cross-sectional area $A_{d,0}$ and the initial dust 
luminosity $L_{d,0}$ are simple functions of $N_{max}$ and the parameters of the size
distribution. Thus,
\begin{equation}
A_d(t) = { A_{d,0} \over {1 + t / t_c} } ~ .
\label{eq: adt} 
\end{equation}
and
\begin{equation}
L_d(t) = { L_{d,0} \over {1 + t / t_c} } ~ .
\label{eq: ldt} 
\end{equation}
At early times ($t \ll t_c$), the cross-sectional area, dust luminosity and total mass in 
the disk are roughly constant. At late times ($t \gg t_c$), the area, luminosity, and mass
decline as $t^{-1}$ \citep{wyatt2002,dom2003}.

\subsection{Destructive Collisions}
\label{sec: anmod-coll}

The simple relations in eqs. (1)--(3) hinge on maintaining a power-law size distribution 
with an invariant slope for particles with $m_{min} \lesssim m \lesssim m_{max}$. This 
outcome requires destructive collisions among equal mass objects.  Collision outcomes 
depend on the ratio $\qc/\qdstar$, where $\qdstar$ is the collision energy per unit mass 
needed to eject half the mass of a pair of colliding planetesimals to infinity and 
$\qc$ is the center of mass collision energy per unit mass 
\citep[see also][]{weth1993,will1994,tanaka1996b,stcol1997a,kl1999a,obrien2003,koba2010a}.
For impact velocity $v$, $\qc = \mu v^2 / 2 (m_1 + m_2)$, where $\mu = m_1 m_2 / (m_1 + m_2)$
is the reduced mass for a pair of colliding planetesimals with masses $m_1$ and $m_2$. 
For equal mass objects, $\qc = v^2 / 8$.

Following standard practice,
\begin{equation}
\qdstar = Q_b r_c^{\beta_b} + Q_g \rho_p r_c^{\beta_g}
\label{eq: qd}
\end{equation}
where $Q_b r_c^{\beta_b}$ is the bulk component of the binding energy,
$Q_g \rho_g r_c^{\beta_g}$ is the gravity component of the binding energy,
and $r_c$ is the radius of a merged pair of planetesimals 
\citep[e.g.,][]{benz1999,lein2008,lein2009}.  For icy objects, we adopt 
parameters -- $Q_b \approx 10^5$~erg~g$^{-1}$~cm$^{-\beta_b}$,
$\beta_b \approx -0.40$, $Q_g \approx$ 0.11~erg~g$^{-2}$~cm$^{3-\beta_g}$,
and $\beta_g \approx$ 1.3 -- which are broadly consistent with analytic 
estimates, laboratory experiments, and numerical simulations 
\citep[see also][]{davis1985,hols1994,love1996,housen1999,ryan1999,
arakawa2002,giblin2004,burchell2005}.  

Setting $\qc \approx \qdstar$ establishes constraints on the particles 
destroyed by an adopted collision velocity.  Among large particles with 
$r \gtrsim$ 1~km, the collision energy must overcome the gravitational
component of the binding energy: $\rmax = (v^2 / 8 Q_g \rho)^{0.77}$. 
When two particles with $r \lesssim$ 1~cm collide, the impact kinetic 
energy must exceed the strength component of the binding energy:
$\rmin = (8 Q_b / v^2)^{0.4}$. In between these two limits, \qdstar\ is
always smaller than \qc. Thus, collisions with velocity $v$ destroy 
particles with $\rmin \le r \le \rmax$. Because the debris from these 
collisions produces an equilibrium size distribution between \rmin\ and
\rmax, continued destructive collisions maintain this size distribution.

To relate these constraints to the properties of the planet and the 
central star, it is convenient to use the Hill radius,
\begin{equation}
\rhill = \left ( { \mp \over 3 \mstar } \right )^{1/3} a_p ~ ,
\label{eq: rhill}
\end{equation}
which establishes a volume where the gravity from the planet overcomes 
the gravity from the central star.  For particles orbiting with random 
inclination in a spherical shell surrounding the planet, the collision 
velocity is a simple function of the orbital velocity $v_K$, 
$v = f_v v_K$ \citep{kw2011a}.  For a shell with $a = \eta_1 R_H$, 
$v = 3^{1/6} (G / \eta_1)^{1/2} f_v \mp^{1/3} \mstar^{1/6} a_p^{-1/2}$. 

For Fomalhaut b, we adopt a set of fiducial parameters to evaluate 
the collision velocity and other aspects of the collisional cascade
\citep[see also][]{kw2011a}. A 10~\mearth\ planet orbits a 1.9~\msun\ star 
at $a_p \approx$ 120~AU. Satellites with $\rho$ = 1~\gcmc\ and $r \approx$ 
100~km lie in a spherical shell with $\eta_1$ = 0.2 and $\eta_2$ = 0.5. 
Orbits with much larger $\eta_1$ ($\approx$ 0.3--1) are unstable
\citep[e.g.,][]{ham1992,ham1997,toth1999,shen2008,martin2011}; 
more compact configurations evolve too quickly. Clouds with (i) typical 
mass \xcl\ = \mcl/\mpl\ = 0.01 and (ii) collision velocities slightly 
larger than Keplerian, $f_v$ = 1.25 \citep{kw2011a}, then have orbital 
velocity 
\begin{equation}
\label{eq: vorb}
v \approx 0.32
\left ( { f_v \over 1.25 } \right )
\left ( { \eta_1 \over 0.2 } \right )^{-1/2}
\left ( { \mp \over 10~\mearth\ } \right )^{1/3}
\left ( { \mstar \over 1.9~\msun\ } \right )^{1/6}
\left ( { \ap \over {\rm 120~AU} } \right )^{-1/2} ~ \kms ~ .
\end{equation} 

Substituting this velocity into our expressions for \rmin\ and \rmax:
\begin{eqnarray*}
~~~~~~~~~~~ \rmax & \approx & 100 
\left ( {f_v \over {1.25} } \right )^{1.54}
\left ( {\eta_1 \over 0.2 } \right )^{-0.77}
\left ( {\rho \over {\rm 1~\gcmc} } \right )^{-0.77}
\left ( {Q_g \over {\rm 0.11~{\rm erg~g^{-2}~cm^{1.7}} } } \right )^{-0.77} \\
 & & ~~~~~ \left ( { \mp \over 10~\mearth } \right )^{0.51}
\left ( { \mstar \over 1.9~\msun } \right )^{0.255} 
\left ( { a_p \over {\rm 120~AU} } \right )^{-0.77} ~ {\rm km} ~ . ~~~~~~~~~~~~~~~~~~~~~~~~ (7) \\
\label{eq: r-max}
\end{eqnarray*}
and 
\addtocounter{equation}{1}
\begin{eqnarray*}
~~~~~~~~~~~~~~~~~ \rmin & \approx & 1.8 \times 10^{-4}
\left ( { f_v \over 1.25 } \right )^{-5}
\left ( { \eta_1 \over 0.2 } \right )^{5/2}
\left ( { Q_b \over {\rm 10^5~erg~g^{-1}~cm^{0.4}} } \right )^{5/2} \\
& & ~~~~~~~~~~~~~~ \left ( { \mp \over 10~\mearth } \right )^{-5/3}
\left ( { \mstar \over 1.9~\msun } \right )^{-5/6}
\left ( { a_p \over {\rm 120~AU} } \right )^{5/2} ~ \mum . ~~~~~~~~~~~ (8)
\label{eq: r-min}
\end{eqnarray*}
Objects with intermediate sizes -- 
$\rmin < r < \rmax$ -- have smaller \qdstar\ than particles with $r = \rmin$ or 
$r = \rmax$. Radiation pressure typically ejects particles with $r \lesssim r_b$,
where $r_b \gg \rmin$. Thus, all particles with $r \lesssim \rmax$ are either
destroyed or ejected. 

\addtocounter{equation}{1}

For simplicity, many analytic models adopt a \qdstar\ which is independent of radius.
Eqs. \ref{eq: r-max}--\ref{eq: r-min} justify this assumption: any \qdstar\ which
ensures the destruction of objects with some maximum radius guarantees that collisions
will also destroy all smaller objects.

\subsection{Size Distribution}
\label{sec: anmod-size}

With \rmax\ known as a function of \qg, deriving $A_{d,0}$ requires values 
for \rmin\ and $q$. In numerical simulations of collisional cascades, the
slope of the equilibrium power-law size distribution is $q \approx$ 3.5--3.7
\citep[e.g.,][]{dohn1969,will1994,obrien2003,koba2010a}. In models where \qdstar\ is
constant with particle radius, $q \approx$ 3.5. When the bulk strength component of
\qdstar\ declines with radius, $q \approx$ 3.7. We adopt $q$ = 3.5.

In an optically thin swarm of satellites, radiation pressure sets \rmin. For dust
grains orbiting the central star, radiation pressure removes particles smaller than
the `blowout' radius, $r_b \approx (3 \lstar Q_{pr} / 8 \pi c G \rho \mstar)$, where
$Q_{pr}$ is the radiation pressure coefficient which accounts for absorption and
scattering \citep[e.g.,][]{burns1979}. Fomalhaut has \mstar\ = 1.9 \msun\ and 
\lstar $\approx$ 20~\lsun; thus, $r_b \approx$ 7~\mum\ for icy grains with 
$Q_{pr}$ = 1 and $\rho$ = 1~\gcmc. 

When particles orbit a planet, ejection depends on the orbital velocity of a particle 
around the planet relative to the orbital velocity of the planet around the star 
\citep[e.g.,][]{burns1979}. 
Defining $\beta = F_r / F_g$ as the ratio of the radiative force to the gravitational
force, radiation ejects particles orbiting a planet when $\beta \gtrsim \beta_0 (v / v_p)$, 
where $v$ is the orbital velocity of a particle around the planet, $v_p$ is the orbital 
velocity of the planet around the star, and $\beta_0 \approx$ 1/3 to 1. Thus,
$r_b \lesssim (3 \lstar Q_{pr} / 8 \pi c G \rho \mstar) (v_p / \beta_0 v) $.
In physical units,
\begin{eqnarray*}
~~~~~~~~~~~~~~~~~~~~~~ r_b & \lesssim & 100 
\left ( { 2/3 \over \beta_0 } \right )
\left ( { \eta_1 \over 0.2 } \right )^{1/2}
\left ( { Q_{pr} \over 1 } \right )
\left ( { \rho \over 1~\gcmc\ } \right )^{-1} \\
& & ~~~~~~~ \left ( { \mp \over 10 ~ \mearth } \right )^{-1/3}
\left ( { \mstar \over 1.9 ~ \msun } \right )^{-2/3}
\left ( { \lstar \over 20 ~ \lsun } \right ) ~ \mum ~ . ~~~~~~~~~~~~~~~~~~~~~ (9)
\label{eq: rblow}
\end{eqnarray*}
More massive planets hold onto smaller particles. For particles orbiting at
a fixed fraction of the Hill radius, $r_b$ is independent of \ap.

\addtocounter{equation}{1}

Setting \rmin\ = $r_b$ and integrating over a power-law size distribution 
with $q$ = 3.5 yields the initial surface area.
For convenience, we separate the linear dependence of $A_{d,0}$ on the 
cloud mass \mcl\ into a linear dependence on $\xcl \mp$:
\begin{equation}
A_{d,0} = 1.4 \times 10^{24} 
\left ( { \xcl \over 0.01 } \right )
\left ( { \mp \over 10~\mearth\ } \right )
\left ( { \rmax \over {\rm 100~km} } \right )^{-1/2}
\left ( { \rmin \over {\rm 100~\mum} } \right )^{-1/2} ~ {\rm cm^2} ~ .
\label{eq: ad0}
\end{equation}
With these parameters, the initial surface area is roughly 10~times the
observed surface area of a dust cloud in Fomalhaut b.

\subsection{Collision Time}
\label{sec: anmod-tcoll}

Deriving the long-term evolution of $A_d$ requires a numerical estimate for the
collision time. The simplest approaches adopt the lifetime of the largest particle 
against collisions with identical particles. More elaborate treatments include the 
impact of collisions with much smaller particles \citep[e.g.,][]{wyatt2007a,wyatt2007b,
koba2010a,kw2011a,kw2011b}.  Because the lifetime depends on a variety of 
relatively unknown parameters, we consider the time scale for destructive 
collisions among identical particles.

To estimate the collision time, we again consider a spherical shell with 
semimajor axis $a = \eta_1 \rhill$ and thickness $\delta a = \eta_2 a $. 
The volume of this shell is 
$V = 4 \pi a^2 \delta a = 4 \pi \eta_1^3 \eta_2 a_p^3 \mp / (3 \mstar)$. 
If the cloud consists of a monodisperse set of particles with mass $m$, $N$ 
= $\mcl / m$.  To express the mass of the cloud relative to the mass of the 
planet, we define $\xcl = \mcl / \mp $ and use $m = 4 \pi \rho \rmax^3 / 3$. 
Then, $N = 3 \xcl \mp / 4 \pi \rho \rmax^3$.  For mono-disperse particles, the 
collisional cross-section is $\sigma = 4 \pi \rmax^2$. Thus, 
$N \sigma = 3 \xcl \mp / 2 \rho \rmax$. 
Combining all of these relations and defining $t_c = V / (2 N_0 \sigma v)$,
\begin{equation}
t_c = \left ( { 2 \pi ~ \eta_1^{7/2} ~ \eta_2 ~ \rho ~ \rmax ~ a_p^{7/2} \over 3^{13/6} ~ G^{1/2} ~ f_v ~ \xcl ~ \mstar^{7/6} \mp^{1/3}} \right ) ~ .
\label{eq: tc}
\end{equation}

When the cascade contains substantial mass in particles with sizes smaller 
than \rmax, the collision time is different from the $t_c$ in eq.~\ref{eq: tc}
\citep[e.g.,][]{wyatt2007a, wyatt2007b,koba2010a,kw2011a,kw2011b}. 
For the fragmentation parameters used in eq. \ref{eq: qd} and the collision
velocity from eq.~\ref{eq: vorb}, collisions between one object with 
$r \approx \rmax$ and another object with $r \gtrsim$ 0.1~\rmax\ destroy 
both objects. Cratering collisions with a much smaller object, $r \ll \rmax$, 
eject roughly 2.7 times the mass of the smaller object.  Thus, every collision 
removes mass from the largest object.  Accounting for cratering collisions and
a broader range of catastrophic collisions 
(i) allows larger objects to participate in the cascade, increasing \rmax, and 
(ii) increases the rate large objects lose mass, shortening $t_c$.  

Deriving the impact of these additional collisions requires integrating the 
collision rate over the size distribution. However, the range of sizes included 
in the integration depends on the collision rate. Although it is possible to 
construct an iterative solution, most investigators simply set \rmax\ as a free 
parameter and derive the collision time for an \rmin\ set by the blowout radius
$r_b$. The revised collision time is then $\sim$ 0.01--4 $t_c$ 
\citep{wyatt2007a,koba2010a,kw2011a}. 
However, this factor depends on $v$, \qdstar, \rmin, and the details of the 
size distribution.  For simplicity, we add a multiplicative parameter $\alpha$ 
($\le$ 4) to our expression for the collision time. In \S\ref{sec: calcs-an}, 
comparisons between this analytic model and our numerical results allow us 
to infer $\alpha$ for irregular satellite systems in Fomalhaut b.

Converting the parameters in eq.~\ref{eq: tc} to physical units, the collision
time is:
\begin{eqnarray*}
~~~~~~~~~~t_c & = & 4.3 \times 10^{6} 
\left ( { \alpha \over 1 } \right )
\left ( { \eta_1 \over 0.2 } \right )^{7/2}
\left ( { \eta_2 \over 0.5 } \right )
\left ( { f_v \over 1.25 } \right )^{-1}
\left ( { \xcl \over 0.01 } \right )^{-1}
\left ( { \mp \over 10~\mearth } \right )^{-1/3}
\left ( { \mstar \over 1.9~\msun } \right )^{-7/6} \\
& & ~~~~~~~~~~~~~~~~~~~~~~~~~~ \left ( { \rho \over 1~\gcmc\ } \right )
\left ( { r \over {\rm 100~km} } \right )
\left ( { a_p \over {\rm 120~AU} } \right )^{7/2} ~ {\rm yr} ~ . ~~~~~~~~ (12)
\label{eq: tcn}
\end{eqnarray*}
More massive clouds, planets, and central stars shorten the collision time. 
Clouds consisting of larger, denser satellites orbiting planets with larger
semimajor axes lengthen the collision time. Collisional cascades remove $\sim$ 
90\% of the initial mass in $\sim$ 10 collision times. Thus, the lifetime of
the cascade is a significant fraction of the lifetime of an A-type star like
Fomalhaut.

\addtocounter{equation}{1}

\subsection{Example}
\label{sec: anmod-exam}

\begin{figure} 
\includegraphics[width=6.5in]{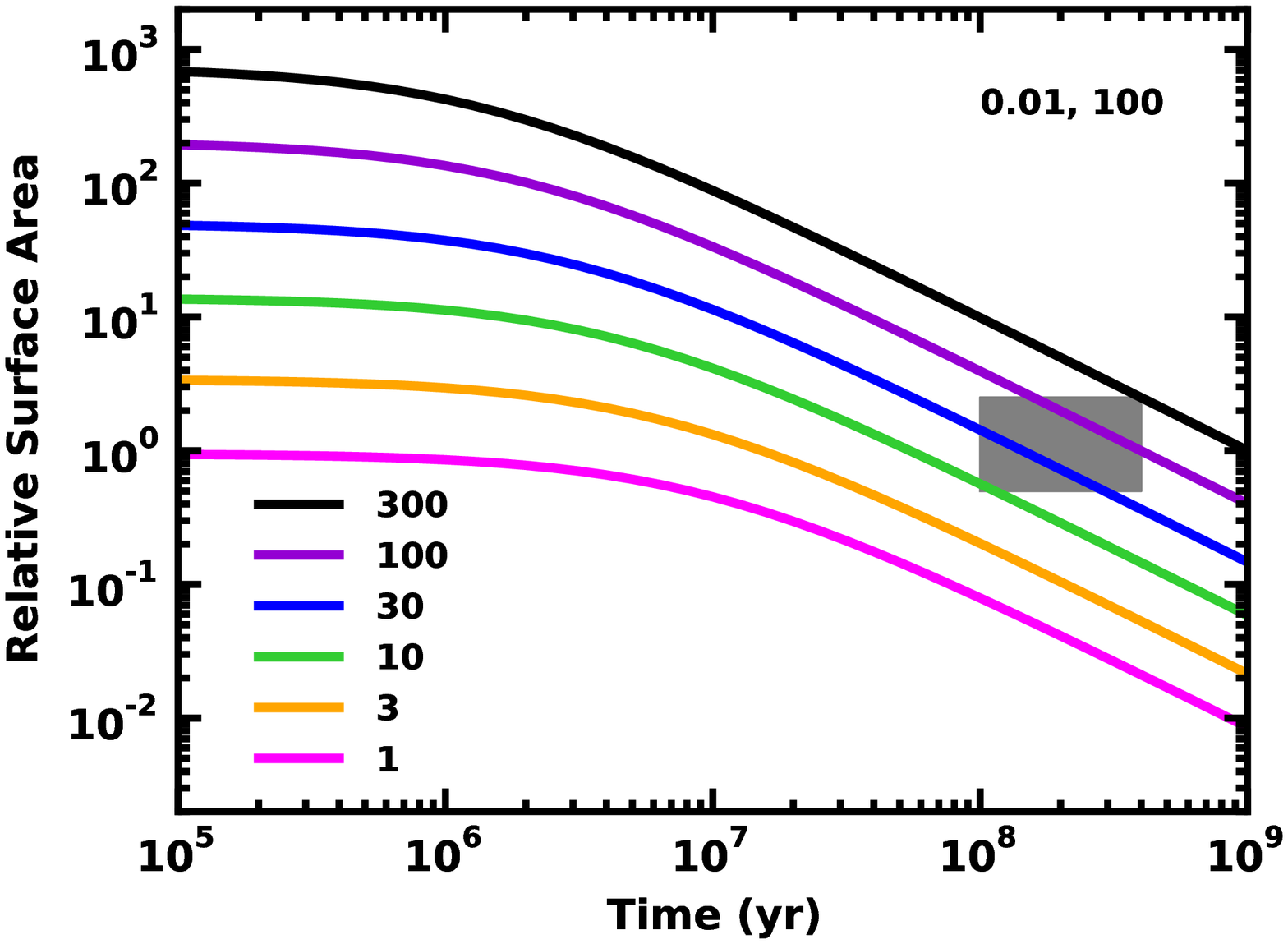}
\vskip 3ex
\caption{
Time evolution of the relative surface area for the analytic model with
$\alpha$ = 1 described in the text. The legend in the lower left indicates 
\mpl\ the mass of the planet in \mearth\ for each solid curve.  
The legend 
in the upper right indicates \xcl\ the mass of circumplanetary material 
relative to the mass of the planet and \rmax\ the radius (in km) of the 
largest object in the swarm. Eq.~\ref{eq: r-min} sets values for \rmin\ the 
radius (in \mum) of the smallest particle in the swarm. The grey shaded box 
indicates the 
locus of allowed points for Fomalhaut b, using the surface area derived 
from the measured brightness, the age of Fomalhaut, and 1$\sigma$ errors.
For the adopted combination of \xcl\ and \rmax, clouds orbiting planets 
with \mpl\ $\approx$ 30--100~\mearth\ match the observations.  Models with 
\mpl\ = 10~\mearth\ and 300~\mearth\ barely miss the shaded box; those with 
\mpl\ = 1~\mearth\ have too little surface area at all times.
\label{fig: area1}
}
\end{figure}

To illustrate the analytic model, we consider a simple example. With standard 
parameters, \mstar\ = 1.9~\msun, \lstar\ = 20~\lsun, \ap\ = 120~AU, $\eta_1$ = 0.2, 
$\eta_2$ = 0.5, $\rho$ = 1~\gcmc, \qb\ = $10^{-5}$~erg~g$^{-1}$~cm$^{0.4}$, 
$f_v$ = 1.25, and \rmax\ = 100~km, eqs. (\ref{eq: adt}) and (\ref{eq: ad0})
yield the time evolution of the surface area.  To compare with observations of 
Fomalhaut b, we adopt nominal values and uncertainties for the surface area, 
$A_b \approx 1_{-0.5}^{+1} \times 10^{23}$~cm$^2$ and age, 
$t_b \approx 200_{-100}^{+200}$~Myr 
\citep[e.g.,][and references therein]{mama2012,kcb2014}.  The relative surface 
area of a model satellite swarm is then $A_d(t) / A_b$.  For satellite swarms 
with \xcl\ = 0.01 and a range of masses for the central planet, 
Figure~\ref{fig: area1} compares the time evolution of the relative surface 
area evolves with observations of Fomalhaut b.

For the adopted parameters, satellite swarms around super-Earths with
\mp\ = 30--100~\mearth\ match the data. Models with \mp\ = 10~\mearth\ and
\mp\ = 300~\mearth\ almost match the data. Satellite evolution around lower
mass or more massive planets do not match the data.

Adopting other parameters leads to similar conclusions.  Factor of two (ten) 
changes in \rmax\ (\xcl) modify the relative surface area by $\sim$ 25\% at
ages of 100--400~Myr. A modest, 20\% increase in $\eta_1$ increases $t_c$
by a factor of two at the expense of a 5\% reduction in $A_{d,0}$. This
change yields a better match to the data for models with \mp\ = 10~\mearth,
at the cost of worse matches for models with \mp\ = 100~\mearth. 

\subsection{Advantages and Limitations}
\label{sec: anmod-ad}

The analytic model has several clear advantages. It is conceptually simple,
easy to modify, and straightforward to calculate. The observable quantities 
$A_d$ and $L_d$ have obvious relationships to the physical parameters.
Generating ensembles of debris clouds for plausible variations in the
physical parameters allows robust comparisons with large sets of data.
Comparisons between data and models yield important insights into the
evolution of circumstellar debris disks and swarms of circumplanetary
satellite systems \citep[e.g.,][and references therein]
{wyatt2008,kw2010,kw2011a,kw2011b,matthews2014}.

Despite its broad success, the model does not address several interesting 
issues in the time evolution of irregular satellites. In current theory, 
planets capture material from a circumstellar disk to supply the 
circumplanetary swarm \citep[e.g.,][]{koch2011,quill2012,hgaspar2013,nesv2014}.  
If some captured objects have $r > \rmax$, these objects may accrete material 
from the swarm and reduce the lifetime of smaller particles considerably. 
Dynamical interactions among growing large objects could lead to significant 
scattering of particles within the swarm and interactions with larger 
satellites closer to the planet.

The analytic model also does not allow time variations in \rmin, \rmax, 
and the slope of the power law size distribution. In a real collisional
cascade, small particles gradually chip away at the larger objects. 
Significant reductions in \rmax\ shorten the lifetime of the cascade. 
Among smaller particles with $r \approx$ \rmin, the typical collision time
of $10^2 - 10^4$ yr (for $A_d \approx 10^{25} - 10^{23}$ cm$^2$) is not
much longer than the typical orbital period of $\sim$ 100~yr. Radiation
pressure typically takes many orbital periods to eject small particles
\citep[e.g.,][and references therein]{poppe2011}. Thus, \rmin\ might be
significantly smaller than $r_b$ at early times, enabling a larger initial
surface area which declines more rapidly with time. In between \rmin\ and
\rmax, it takes many collision times to establish a power law size 
distribution with $q$ = 3.5. If the initial size distribution is far from
equilibrium, the early evolution of $A_d$ might differ significantly from
predictions of the analytic model.

Addressing these issues requires numerical simulations. For a satellite 
swarm where the orbital elements are fixed in time, it is straightforward 
to conduct a suite of coagulation calculations to learn how time variations
in \rmin, \rmax, $q$, and other physical parameters impact the long term 
evolution of the satellite swarm. We describe our numerical approach in
\S\ref{sec: num} and then discuss the results of the simulations in 
\S\ref{sec: calcs}. 

\section{NUMERICAL MODEL}
\label{sec: num}

To perform numerical calculations of the collisional evolution of an irregular satellite 
system, we use \orch, an ensemble of computer codes for the formation and evolution of 
planetary systems.  \orch\ includes a multiannulus 
coagulation code which derives the time evolution of a swarm of solid objects orbiting 
a central mass \citep{kb2004a,kb2008,kb2012}. Although this code was originally designed 
to follow solids within a circumstellar disk, it is straightforward to modify the algorithms 
to track solids orbiting within a spherical shell. \orch\ also includes an $n$-body 
code which follows the trajectories and dynamical interactions of large objects
\citep{bk2006,bk2011a,bk2013}. In these calculations, we disable dynamical interactions 
between coagulation particles and the $n$-bodies mediated by tracer particles.

\subsection{Numerical Grid}
\label{sec: num-grid}

We conduct coagulation calculations of particles orbiting with semimajor axis $a$ 
inside a spherical shell of width $\delta a$ around a planet with mass \mp. Within
this shell, there are $M$ mass batches with characteristic mass $m_k$ and radius 
$r_k$ \citep{weth1993,kl1998}. Batches are logarithmically spaced in mass, with 
mass ratio $\delta \equiv m_{k+1} / m_{k}$.  Each mass batch contains $N_k$ 
particles with total mass $M_k$ and average mass $\bar{m}_k = M_k / N_k$. Particle 
numbers $N_k < 10^{15}$ are always integers.  Throughout the calculation, the average 
mass is used to calculate the average physical radius $\bar{r}_k$, collision 
cross-section, collision energy, and other necessary physical variables.  As mass 
is added and removed from each batch, the average mass changes \citep{weth1993}.

Numerical calculations with $\delta \gtrsim$ 1 lag the result of an ideal 
calculation with infinite mass resolution (see the Appendix). Simulations with 
$\delta$ = 1.05--1.19 yield somewhat better solutions to the evolution of 
10--100~km objects than calculations with $\delta$ = 1.41--2.00. However,
the evolution of the cross-sectional area of a swarm of solids is fairly
independent of $\delta$. To track the evolution of the size distribution 
reasonably well, we consider a suite of calculations with $\delta$ = 1.19 
($ = 2^{1/4}$).

In these calculations, we follow particles with sizes ranging from a minimum size 
\rmin\ to the maximum size \rmax. The algorithm for assigning material to the mass 
bins extends the maximum size as needed to accommodate the largest particles.  When 
collisions produce objects with radii $r < \rmin$, this material is lost to the grid.

When the average mass in a bin exceeds a pre-set promotion mass \mpro, the code 
creates a set of $n$-bodies with masses equal to the average mass in the bin. In 
these calculations, \mpro\ = $10^{24}$~g (\rmax\ $\approx$ 600~km). Promoted 
objects are assigned a random semi-major axis, $a_{pro}$, in the range 
$(a - \delta a, a + \delta a)$, a random inclination sin~$i$, a random orbital 
phase, and orbital eccentricity $e$ = 0. For this first exploration of the evolution,
we include the gravity of the central planet but ignore gravitational forces of 
nearby stars.

\subsection{Initial Conditions}
\label{sec: num-init}

All calculations begin with a swarm of planetesimals with initial maximum size 
\r0\ and mass density $\rho$ = 1~\gcmc. These particles have initial number density
$n_0$ and total mass $M_0$. For the simulations in this paper, we consider two
different initial size distributions for the planetesimals. To follow the analytic
model as closely as possible, one set of calculations begins with a power law size
distribution, $n(r) \propto r^{-q}$ and $q$ = 3.5. To study whether our calculations
produce this equilibrium size distribution, we begin a second set of calculations 
with a mono-disperse set of planetesimals.

\subsection{Evolution}
\label{sec: num-evol}

The mass distribution of the planetesimals evolves in time due to inelastic collisions.  
All planetesimals have the same collision velocity, which is fixed at the start of
each calculation.  As summarized 
in \citet{kb2004a,kb2008}, we solve a coupled set of coagulation equations which
treats the outcomes of mutual collisions between particles in every mass bin.
We adopt the particle-in-a-box algorithm, where the physical collision rate is 
$n \sigma v f_g$, $n$ is the number density of objects, $\sigma$ is the geometric 
cross-section, $v$ is the relative velocity from eq.~\ref{eq: vorb}, and $f_g$ is 
the gravitational focusing factor \citep{weth1993,kl1998}. The collision algorithm 
treats collisions in the dispersion regime -- where relative velocities are 
large -- and in the shear regime -- where relative velocities are small 
\citep{kl1998,kb2014}. 

For swarms of satellites, all collisions are in the dispersion regime, where we 
adopt a variant of the piecewise analytic approximation of 
\citet[][see also Kenyon \& Luu 1998; Kenyon \& Bromley 2012]{spaute1991}. When
collisions involve particles with $r \lesssim$ 300~km, $f_g \lesssim$ 3--4. As
satellites reach sizes of 1000--2000~km, $f_g \lesssim$ 50. Compared to simulations
where $f_g \gtrsim 10^3 - 10^4$ \citep[e.g.,][]{kb2008}, gravitational focusing has
a modest impact on the evolution. 

For each pair of colliding planetesimals and the collision energies $Q_c$ and 
\qdstar\ defined in \S\ref{sec: anmod}, the mass of the merged planetesimal is
\begin{equation}
m = m_1 + m_2 - \mesc ~ ,
\label{eq: msum}
\end{equation}
where the mass of debris ejected in a collision is
\begin{equation}
\mesc = 0.5 ~ (m_1 + m_2) \left ( \frac{Q_c}{Q_D^*} \right)^{b_d} ~ .
\label{eq: mej}
\end{equation}
The exponent $b_d$ is a constant of order unity 
\citep[e.g.,][]{davis1985,weth1993,kl1999a,benz1999,obrien2003,koba2010a,lein2012}. 
Here, we consider $b_d$ = 1.

To place the debris in the grid of mass bins, we set the mass of the 
largest collision fragment as 
\begin{equation}
\mmaxd = m_{L,0} ~ \left ( \frac{Q_c}{Q_D^*} \right)^{-b_L} ~ \mesc ~ .
\label{eq: mmaxd1}
\end{equation}
To explore the sensitivity of the evolution to this algorithm, we set
$m_{L,0} $ = 0.2 and $b_L = $ 0 or 1 \citep{kb2008,koba2010a,weid2010}. 
Lower mass objects have a differential size distribution $n(r) \propto r^{-q}$. 
After placing a single object with mass \mmaxd\ in an appropriate bin, 
we place material in successively smaller mass bins until (i) the mass 
is exhausted or (ii) mass is placed in the smallest mass bin. Any material 
left over is removed from the grid \citep[see also][]{kb2015}.  

\section{CALCULATIONS}
\label{sec: calcs}

To examine the long-term evolution of satellite swarms, we consider 
a baseline model where the central star has mass 1.90 \msun\ and 
luminosity 20 \lsun. Planets with 
\mp\ = 1, 3, 10, 30, 100, or 300~\mearth\ orbit with a semimajor axis 
\ap\ = 120~AU. The spherical cloud has \xcl\ = 0.01, $\eta_1$ = 0.2, and 
$\eta_2$ = 0.5. Particles within the shell have $\rho$ = 1~\gcmc, 
\rmin\ = 100~\mum\, and \rmax\ = 50, 100, 200, or 400~km. The initial size 
distribution is a power law with $n(r) \propto r^{-q}$ and $q$ = 3.5. 
To establish collision outcomes, we adopt the fragmentation parameters 
-- \qb, \qg, $\beta_b$, and $\beta_g$ -- summarized in 
\S\ref{sec: anmod-coll}. For the largest object in the debris we set $b_L$ = 0.

To check the sensitivity of the calculations to these parameters, we vary
one parameter and hold others fixed. In turn, we derive results for
\xcl\ = 0.001 or 0.0001;
\rmin\ = 10~\mum\ or 1~mm;
(\qb,\qg) = $(5 \times 10^4, 0.055)$ or $(2.5 \times 10^4, 0.0275)$; and
$b_L$ = 1. Although we perform these calculations for all combinations 
of \mp\ and \rmax\ in the baseline model, we focus our discussion primarily
on calculations with \mp\ = 10~\mearth.

In describing the results of the simulations, we consider the long-term 
evolution of the size of the largest object (\S\ref{sec: calcs-rmax}), 
the size distribution (\S\ref{sec: calcs-sd}), and the cross-sectional 
area (\S\ref{sec: calcs-area}).  
In many of the simulations, the masses of
2--6 objects reach the promotion mass. After promotion into the 
\nbody\ portion of \orch, continued growth leads to strong dynamical
interactions among a few \nbodies.  
Several examples of the long-term 
actions of the \nbodies\ illustrate their likely impact on the rest of the 
swarm and satellites orbiting closer to the planet
(\S\ref{sec: calcs-nb}).
In \S\ref{sec: calcs-an}, comparisons with predictions of the analytic model 
allow us to clarify how the behavior of the swarm differs in the two approaches. 

\subsection{Evolution of the Size of the Largest Object}
\label{sec: calcs-rmax}

\subsubsection{Baseline model}
\label{sec: calcs-rmax-base}

\begin{figure} 
\includegraphics[width=6.5in]{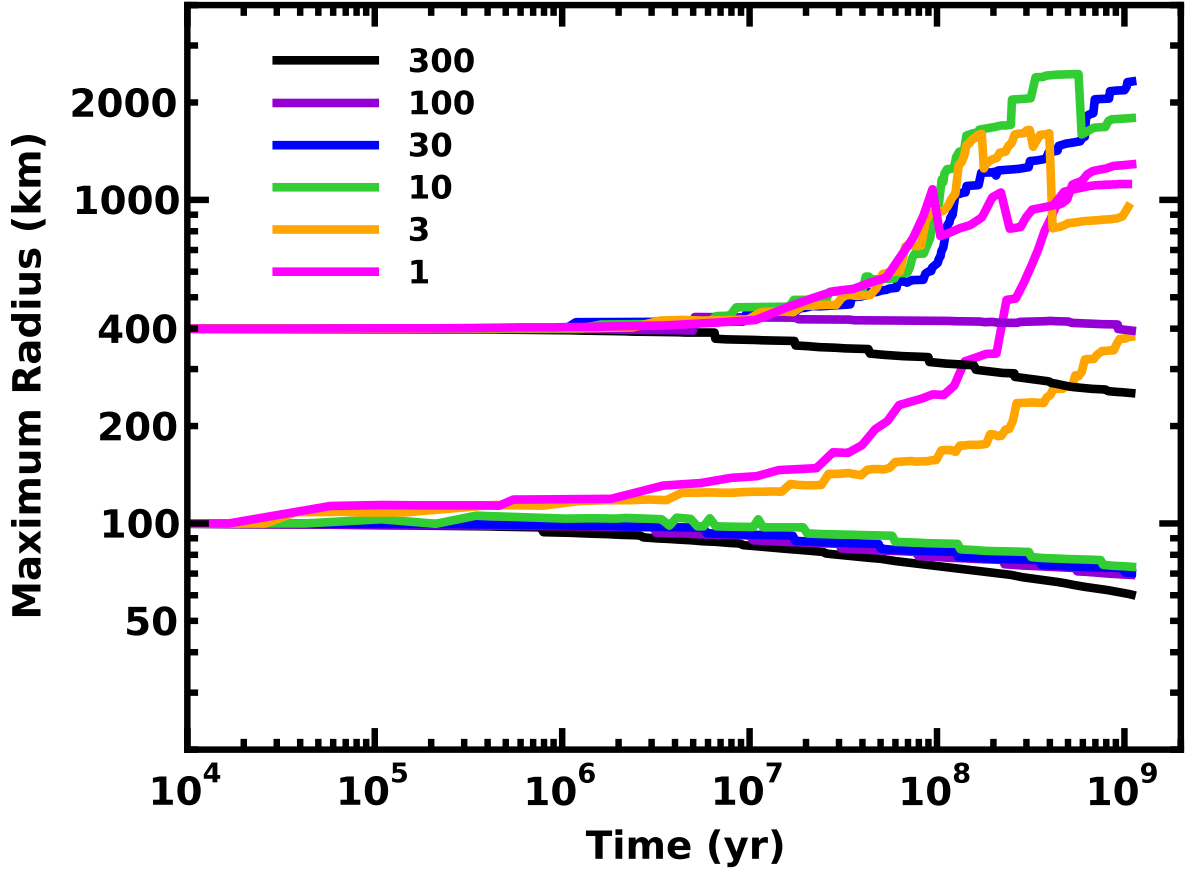}
\vskip 3ex
\caption{
Time evolution of the radius of the largest object derived from coagulation
calculations of circumplanetary clouds of particles with the nominal 
fragmentation parameters, \xcl\ = 0.01,
\rmin\ = 100~\mum, \mlz\ = 0.2, and \bl\ = 0.0.  The legend in the upper 
left indicates \mpl\ for each solid curve. For calculations with massive 
planets, the size of the largest object smoothly declines with time. When 
the mass of the planet is smaller, the largest objects sweep up small 
particles and grow into much larger objects. In calculations where \rmax\ is 
400~km (50~km), the largest objects are more (less) likely to grow with time.
\label{fig: rmax1}
}
\end{figure}

\begin{figure} 
\includegraphics[width=6.5in]{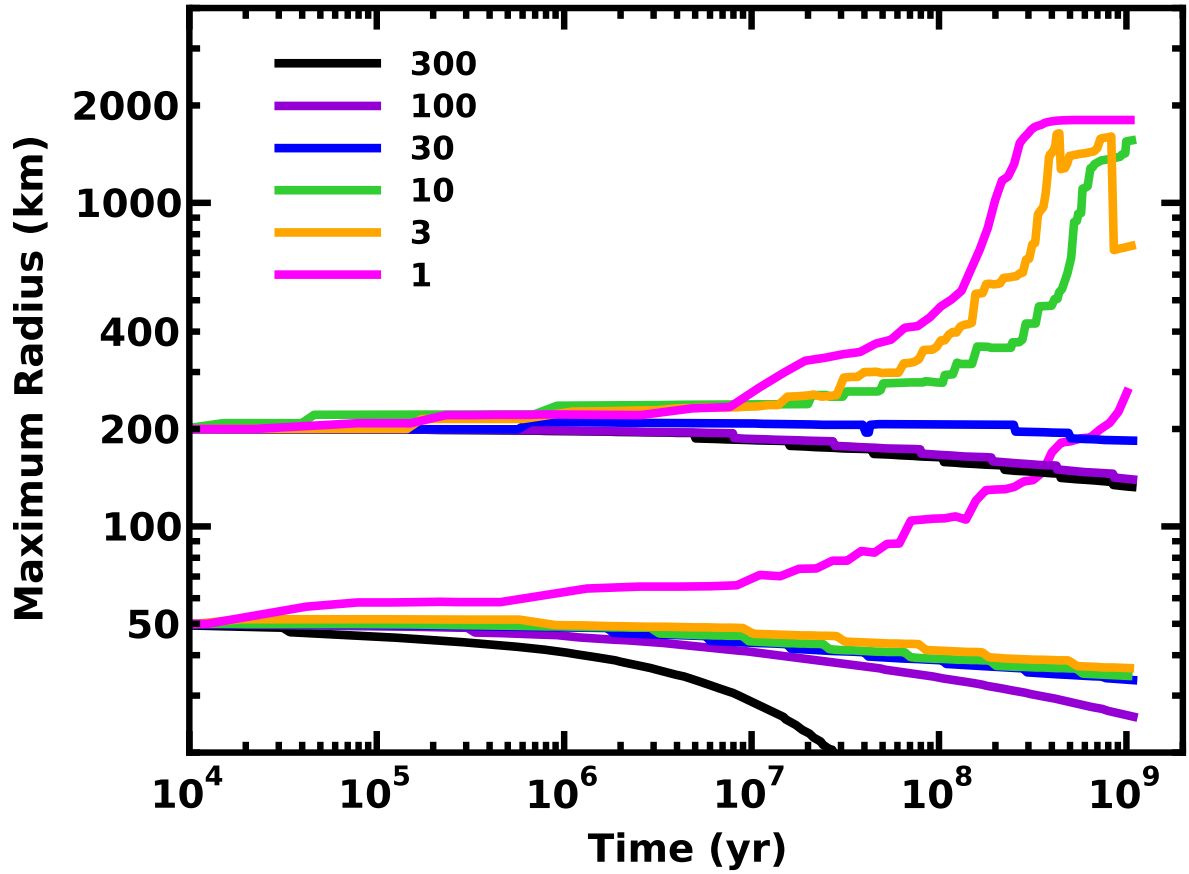}
\vskip 3ex
\caption{
As in Fig.~\ref{fig: rmax1} for calculations starting with \rmax\ = 
50~km and 200~km. When \mpl\ is large, the collisional cascade 
gradually destroys the largest objects. Around lower mass planets,
the largest objects grow with time.
\label{fig: rmax2}
}
\end{figure}

In the baseline model, the radius of the largest object \rmax\ depends
on the mass of the central planet and the initial 
\rmax\ (Figs.~\ref{fig: rmax1}--\ref{fig: rmax2}). 
When \mpl\ = 30--300~\mearth, the orbital velocity at 0.2\rhill\ is 
sufficient to destroy 100~km objects. Occasional catastrophic collisions 
among pairs of 100~km objects destroy them completely; the population of 
these objects gradually diminishes with time. More frequent cratering 
collisions with much smaller objects chip away at the mass in every 
large object. The average radius of these objects gradually declines 
from 100~km at $t$ = 0 to 60--70~km at $t$ = 1~Gyr. 

When \mpl\ = 10~\mearth, collisions almost shatter pairs of 100~km objects.
Barring collisions with smaller objects, these objects grow slowly. 
However, cratering collisions with other satellites reduce their mass
faster than collisions with the largest objects increase their mass. 
Thus, \rmax\ gradually declines with time.

When \mpl\ = 1--3~\mearth, the orbital velocity is not sufficient to
destroy 100~km objects (Fig.~\ref{fig: rmax1}, lower orange and magenta
curves). Cratering collisions are also insufficient to reduce their mass.
Although collisions destroy all smaller objects, these objects grow 
slowly with time. After 1~Gyr, satellites orbiting a 3~\mearth\ planet
reach radii of 400~km; satellites orbiting a 1\mearth\ planet have 
\rmax\ $\gtrsim$ 1000~km. In some cases, these calculations yield several 
$n$-bodies which interact dynamically. We discuss several of these 
outcomes in \S\ref{sec: calcs-nb}.

In calculations with smaller \rmax, it is easier to destroy the largest
planetesimals. Cratering collisions are also more efficient. For
\mpl\ = 3--300~\mearth, \rmax\ declines more rapidly with time
(Fig.~\ref{fig: rmax2}). By the end of the calculation at 1~Gyr, the
largest satellites have
\rmax\ = 2.5~km for \mpl\ = 300~\mearth\ up to
\rmax\ = 35~km for \mpl\ = 3~\mearth.
When \mpl\ = 1~\mearth, the collision energy is still insufficient to 
destroy 50~km objects. Thus, these objects grow slowly to $\sim$ 300~km
over 1~Gyr.

When \rmax\ is initially larger than 100~km, it is easier for large objects
to grow over time (Figs.~\ref{fig: rmax1}--\ref{fig: rmax2}). Satellites
with initial radii of 200~km (400~km) reach radii of 1000--2000~km in 
0.3--1~Gyr around 1--10~\mearth\ (1--30~\mearth) planets. In several 
calculations, pairs of large satellites promoted into the $n$-body code
scatter one another out of the grid. Scattering leaves behind a few much 
smaller objects, reducing \rmax\ within the grid. 

Collisional evolution with larger satellites orbiting more massive planets 
always reduces the size of the largest objects. For 300~\mearth\ planets,
catastrophic and cratering collisions diminish the sizes of the largest 
satellites by 25\% to 40\%. The reduction in size is smaller, $\sim$ 10\% 
to 25\% for 200--400~km satellites orbiting 100~\mearth\ planets.

\subsubsection{Initial cloud mass}
\label{sec: calcs-rmax-cloud}

Changing the initial mass of the cloud has an obvious impact on the long-term
evolution of the largest objects. When the mass of the cloud is smaller, the
collision time is longer. Evolution is correspondingly slower. Thus, the sizes
of the largest objects remain closer to their initial values.

\begin{figure} 
\includegraphics[width=6.5in]{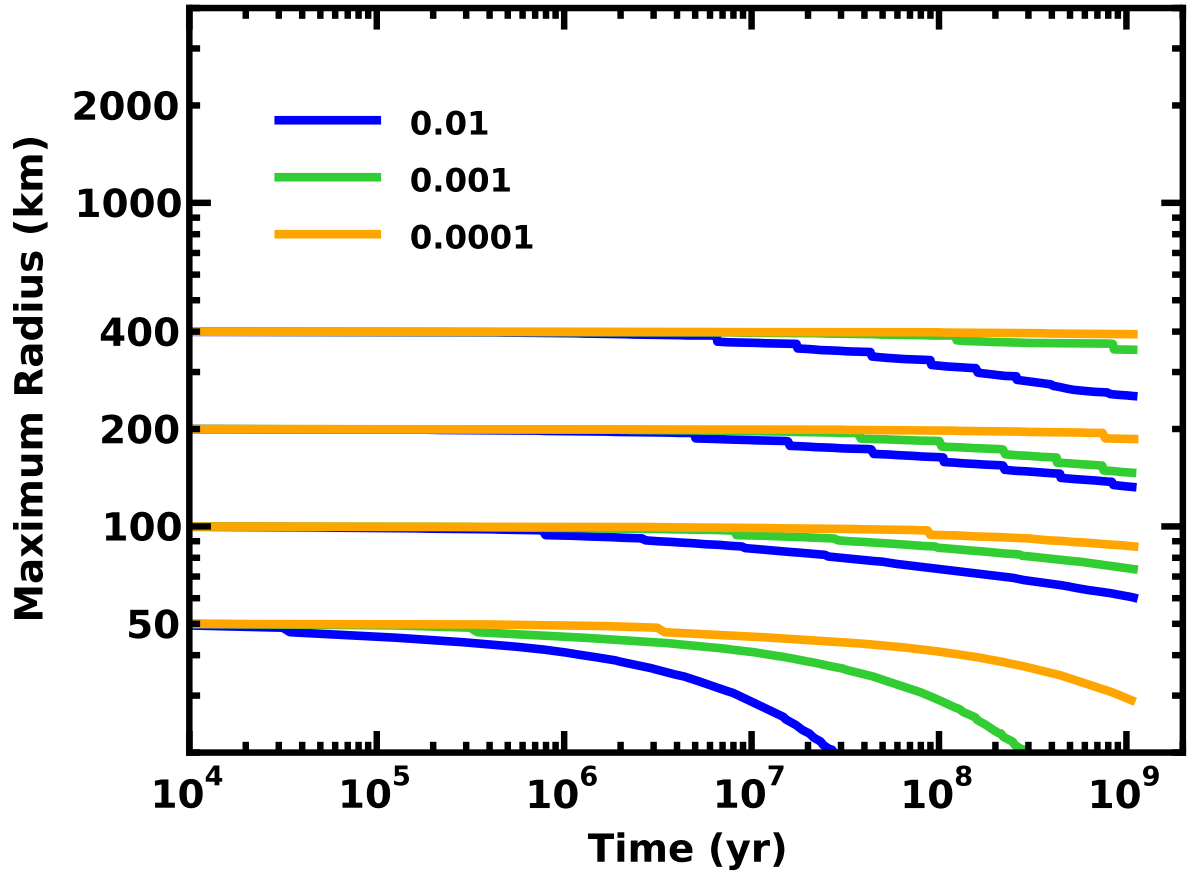}
\vskip 3ex
\caption{
As in Figs.~\ref{fig: rmax1}--\ref{fig: rmax2} for calculations with
\mpl\ = 300~\mearth\ and \xcl\ as indicated in the legend.
When the mass of the cloud is smaller, the radius of the largest object 
changes more slowly.
\label{fig: rmax3}
}
\end{figure}

Fig.~\ref{fig: rmax3} illustrates the impact of initial cloud mass on particle
size for satellites orbiting 300~\mearth\ planets. When the initial \rmax\ is 
small, differences in the evolution are obvious. When \xcl\ = 0.01, it takes 
only 30~Myr for \rmax\ to decline to 20~km. A factor of 10 reduction in the
initial cloud mass increases this time scale to 300~Myr. Another factor of
ten reduction increases the time sale to 3~Gyr. 

As we increase the initial \rmax, the initial cloud mass has a smaller and 
smaller impact on the overall evolution.  Although reducing the cloud mass
increases collision times, large objects already have long collision times.  
Slow evolution simply becomes slower. Larger objects are also more impervious
to collisional destruction; reducing \xcl\ makes them even more impervious. 

\subsubsection{Size of the smallest particles}
\label{sec: calcs-rmax-small}

\begin{figure} 
\includegraphics[width=6.5in]{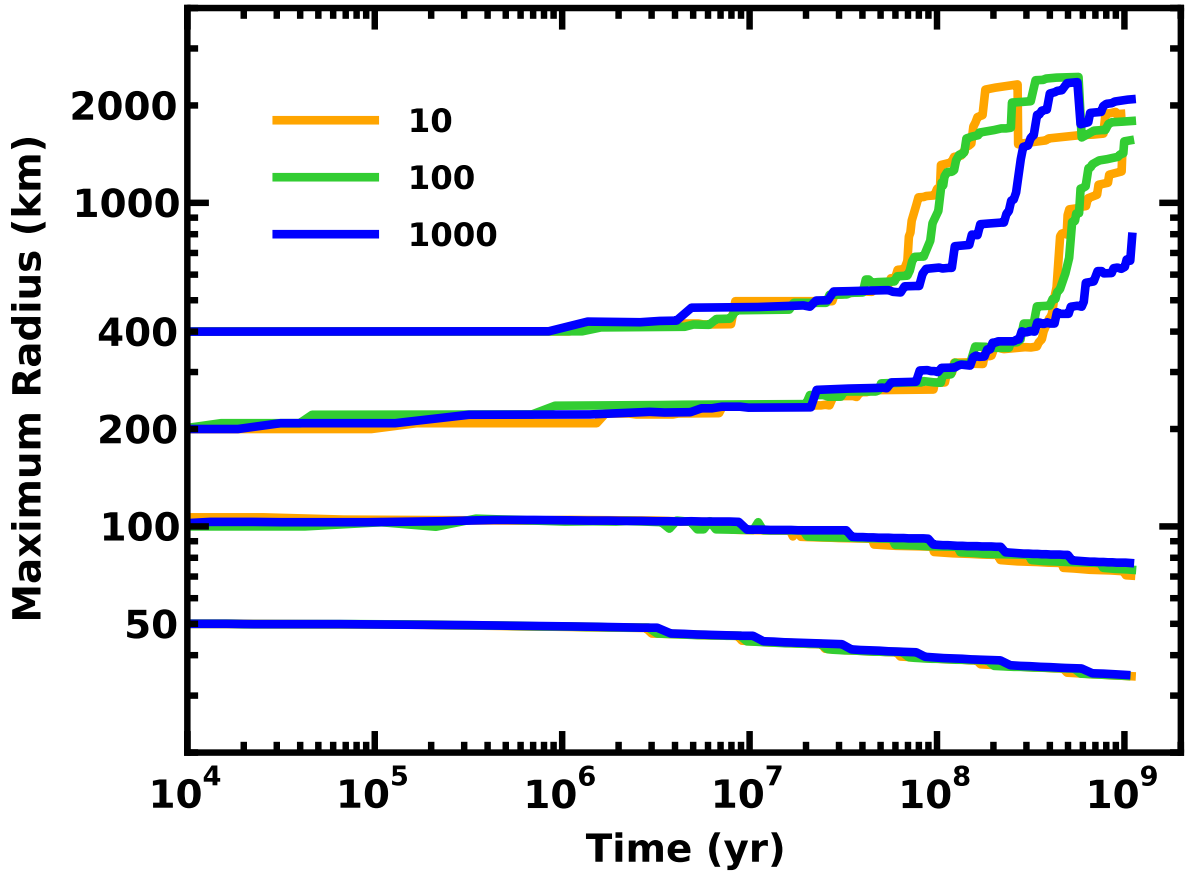}
\vskip 3ex
\caption{
As in Figs.~\ref{fig: rmax1}--\ref{fig: rmax2} for calculations with
\mpl\ = 10~\mearth\ and
\rmin\ = 10~\mum\ (orange curves), \rmin\ = 100~\mum\ (green curves),
and \rmin\ = 1000~\mum\ (1~mm; blue curves). The growth of large
objects is fairly independent of the size of the smallest particles 
in the grid.
\label{fig: rmax4}
}
\end{figure}

Not surprisingly, modifying the initial size of the smallest objects in the 
grid has little impact on the evolution of the largest objects in the grid
(Fig.~\ref{fig: rmax4}). When \mpl\ = 10~\mearth\ and the initial \rmax\ is 
50--100~km, the collisional cascade effectively destroys the largest 
objects in the grid. Collisions with the smallest particles in the grid
remove little mass from the largest objects.  Changing \rmin\ by a factor 
of ten has little impact on \rmax. 

When the initial \rmax\ is larger, the largest objects grow slowly with time.
Large objects accrete little mass from the ensemble of objects with radii
close to \rmin. Changing \rmin\ barely modifies the accretion rate.

At late stages in the growth of large satellites, stochastic variations in
the collision rate among large satellites produces large changes in \rmax.
As large objects grow, the rate of collisions with other large objects 
declines and becomes more random. These random collisions produce large
fluctuations in the time scale for objects to reach sizes of 1000~km
(Fig.~\ref{fig: rmax4}, top curves). When these objects are promoted into
the $n$-body code, random dynamical interactions then yield random drops
(or spikes) in \rmax.

\subsubsection{Mass of the largest object in the debris}
\label{sec: calcs-rmax-mass}

\begin{figure} 
\includegraphics[width=6.5in]{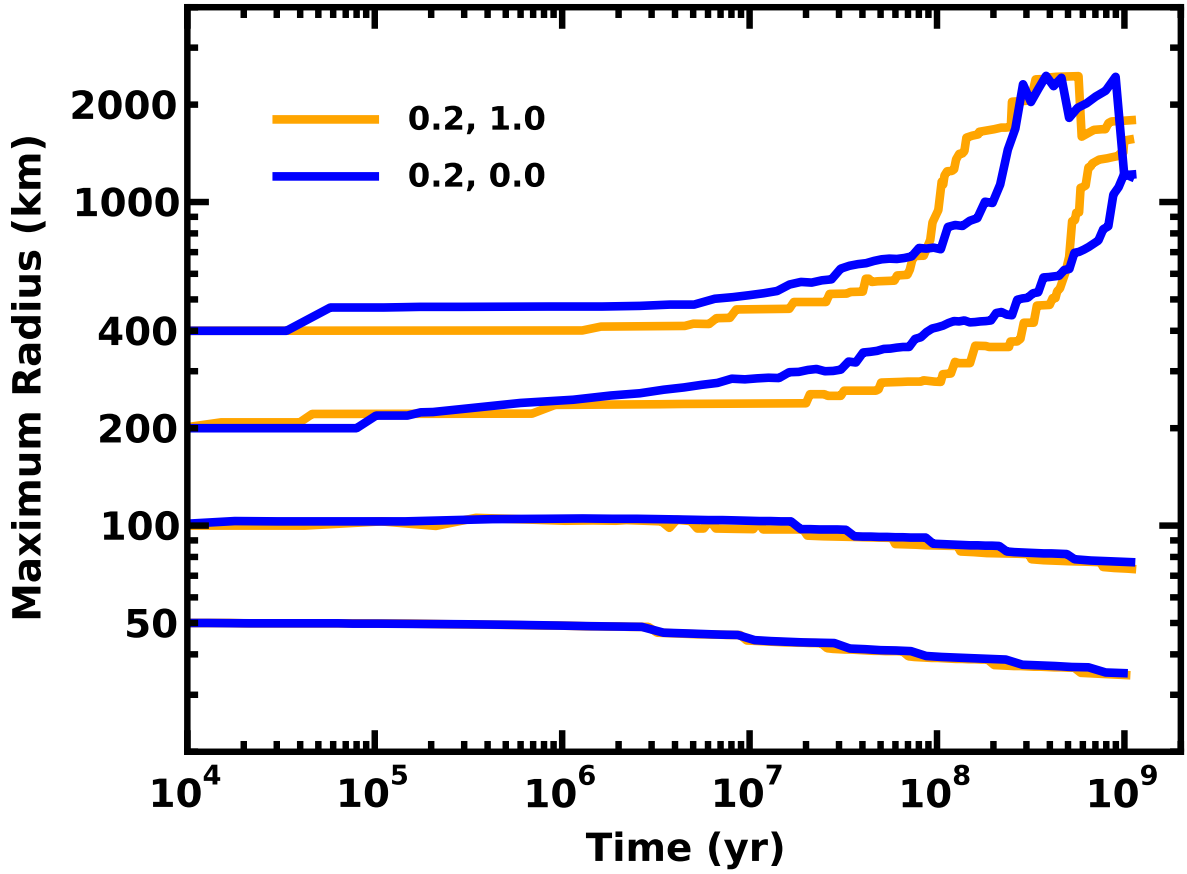}
\vskip 3ex
\caption{
As in Figs.~\ref{fig: rmax1}--\ref{fig: rmax2} for calculations with
\mpl\ = 10~\mearth, $m_{L,0}$ = 0.2, and either \bl\ = 0.0 (blue curves) 
or \bl\ = 1.0 (orange curves).  The growth of large objects is fairly 
independent of the exponent in the relation between the mass of the 
largest object and the collision energy.
\label{fig: rmax5}
}
\end{figure}

How we distribute the debris into smaller mass bins also has modest impact on
the evolution of the largest objects (Fig.~\ref{fig: rmax5}). In our baseline
model with $m_{L,0}$ = 0.2 and $b_L$ = 0, debris tends to fill bins with larger
masses than in calculations with $m_{L,0}$ = 0.2 and $b_L$ = 1. When the
cascade destroys large objects, the mass loss rate from the grid depends on
the rate collisions transport mass to smaller and smaller objects. Thus, we
expect calculations with $b_L$ = 1 to lose mass somewhat more rapidly than those
with $b_L$ = 0. When larger objects grow, cratering collisions are less important;
the exponent $b_L$ then has negligible importance.

Our results confirm these expectations. In the top curves of Fig.~\ref{fig: rmax5},
200~km and 400~km objects grow with time. Superimposed on a gradual rise in
\rmax\ with time, stochastic variations produce a few random increases in \rmax.
Later, random dynamical interactions eject objects promoted into the $n$-body grid. 
The final radii of large objects is fairly independent of $b_L$.

In the lower curves of the figure, 50~km and 100~km satellites get smaller and 
smaller with time. Despite the somewhat larger mass loss of models with $b_L$ = 1,
the final radii are independent of $b_L$.

\subsubsection{Binding energy}
\label{sec: calcs-rmax-qd}

As discussed in \S\ref{sec: anmod}, \qdstar\ -- the binding energy of satellites -- 
sets the size of the largest object destroyed in collisions with fixed impact velocity. 
Smaller \qdstar\ allows collisions to destroy larger objects. 

\begin{figure} 
\includegraphics[width=6.5in]{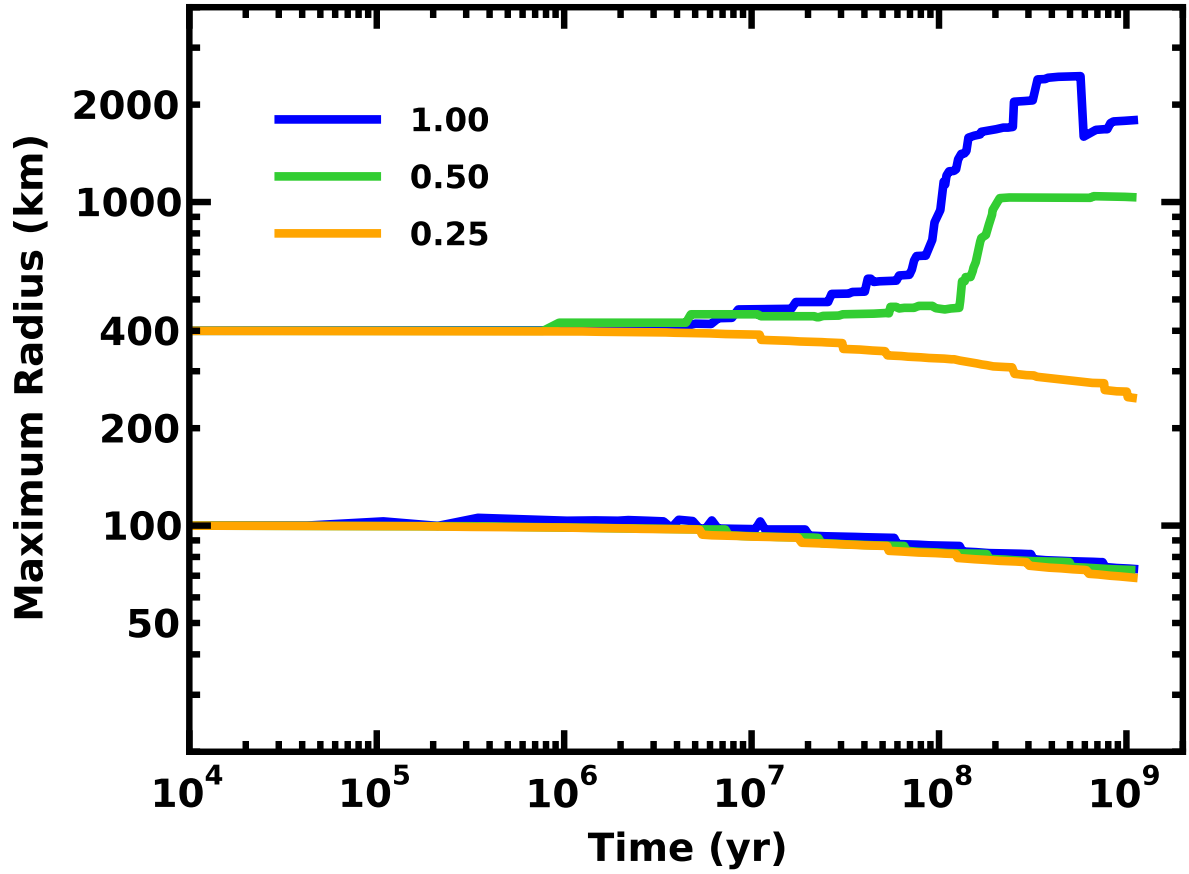}
\vskip 3ex
\caption{
As in Figs.~\ref{fig: rmax1}--\ref{fig: rmax2} for calculations with
\mpl\ = 10~\mearth\ and
different \qdstar. The legend indicates the value of \qdstar\ relative
to the nominal fragmentation parameters listed in the main text.
\label{fig: rmax6}
}
\end{figure}

Fig.~\ref{fig: rmax6} illustrates the impact of smaller \qdstar\ on the evolution
of 100~km and 400~km satellites orbiting 10~\mearth\ planets. When \rmax\ = 100~km,
the \qdstar\ in our baseline calculations is small compared to the collision energy.
Collisions completely shatter these satellites. Although reducing \qdstar\ makes 
them easier to destroy, the amount of debris lost in a collision is fairly similar.
Thus, the evolution of 100~km satellites is independent of \qdstar.

In our baseline models, 400~km objects grow throughout the calculation. At late
times, these objects reach sizes approaching 2000~km. Reducing \qdstar\ has a 
clear impact on the growth of these objects. A factor of two reduction in 
\qdstar\ prevents large objects from growing past 1000~km. Collisions among
smaller objects are more destructive, removing objects from the grid more 
rapidly. With fewer objects to accrete, the growth of the largest objects stalls.

Another factor of two reduction in \qdstar\ completely halts the growth of 400~km
objects. Catastrophic and cratering collisions reduce these sizes of these objects
by 35\% to 40\% in 1~Gyr.

\subsubsection{Initial size distribution}
\label{sec: calcs-rmax-sd}

As a final example in this sequence, we consider the impact of the initial size
distribution. When calculations start with a mono-disperse set of satellites,
cratering collisions do not occur. If collisions between the large objects are
catastrophic, the debris populates smaller size bins. Cratering collisions begin.
Continued cratering and catastrophic collisions populate smaller and smaller size
bins. Once all of the mass bins have debris, catastrophic and cratering collisions
remove mass from all bins in the grid. Compared to our baseline calculations,
these calculations have somewhat more mass in larger size bins. For the largest
objects, collisions with larger small particles have a larger collision energy
than collisions with smaller small particles. Thus, calculations with a mono-disperse
set of satellites evolve somewhat faster than our baseline calculations.

If large object collisions promote growth, there is less debris in smaller mass
bins. Cratering collisions are less effective in filling smaller mass bins. 
Collisions between objects in these smaller bins are less frequent; mass loss
from the grid is smaller. As these calculations proceed, there is much more mass 
in large objects (which are stronger) than in small objects (which are weak).
The largest objects then grow faster.

\begin{figure} 
\includegraphics[width=6.5in]{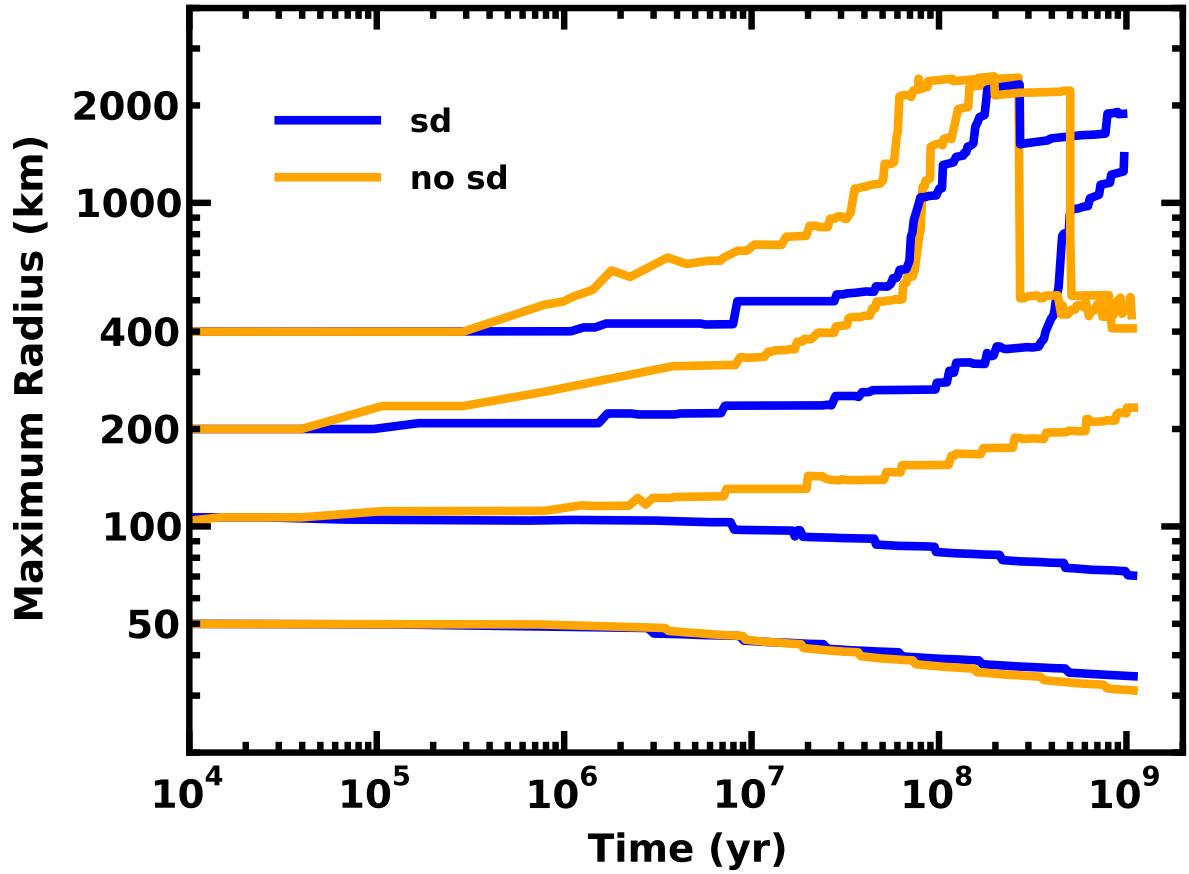}
\vskip 3ex
\caption{
As in Figs.~\ref{fig: rmax1}--\ref{fig: rmax2} for calculations with
\mpl\ = 10~\mearth\ and different initial size distributions.  The 
legend indicates whether the calculation starts with a mono-disperse
set of satellites (no sd) or a power law size distribution. When the
initial population is mono-disperse, satellites grow faster.
\label{fig: rmax7}
}
\end{figure}

Our simulations confirm these expectations (Fig.~\ref{fig: rmax7}). When the 
initial \rmax\ is 50~km, collisions are destructive. The largest objects get 
smaller and smaller with time. At late times, the satellites in the 
mono-disperse calculations are somewhat smaller than those in calculations 
with an initial power-law size distribution. 

When the initial \rmax\ is 100~km, the evolution is very sensitive to the
initial size distribution. In \S\ref{sec: anmod}, the analytic model suggests 
\rmax\ = 100~km
for satellites orbiting a 10~\mearth\ planet. For this initial size, cratering
collisions are critical. When they are present, \rmax\ declines with time.
When they are not present, \rmax\ grows with time. The figure shows this 
dichotomy clearly. With no initial size distribution, 100~km objects grow slowly
with time. With the power law initial size distribution, \rmax\ gets smaller
and smaller with time.

Among larger satellites with \rmax\ = 200~km or 400~km, the evolution also
depends on the initial size distribution. With no cratering collisions among
smaller objects in the grid, the cascade in fairly inactive. The mass in the
grid is nearly constant in time. With more mass in the grid, the largest 
objects grow more rapidly. Promotion into the $n$-body grid occurs earlier;
dynamical interactions are more severe. In these calculations, scattering of
$n$-bodies produced from a mono-disperse set of satellites leaves behind 
a few small objects which have suffered few collisions and are close to their 
original sizes.

\subsection{Evolution of the Size Distribution}
\label{sec: calcs-sd}

In standard collisional cascade models, destructive collisions generate 
a roughly constant mass flow from the largest objects to the smallest 
objects \citep[e.g.,][]{dohn1969,will1994,obrien2003,koba2010a}.  
When \rmin\ $\approx$ 0, catastrophic and cratering collisions maintain 
a power law cumulative size distribution $n(>r) \propto r^{-q_c}$ with 
$q_c \approx$ 2.5--2.7. When $\rmin\ > 0$, cratering collisions between 
objects with $r < \rmin$ and those with $r > \rmin$ do not occur. Fewer
collisions reduces the mass flow rate for $r < \rmin$. Mass then builds 
up in bins with $r \gtrsim$ \rmin. This excess of particles increases 
the rate of cratering collisions among larger particles, creating a
deficit among these particles. Together, the excess and the deficit produce a 
`wave' in the power law size distribution \citep[e.g.,][]{campo1994,obrien2003}.
Over time, continued collisional evolution tends to produce other waves
among particles with larger radii.

Addressing wave production in a numerical simulation requires an artificial
extension of the size distribution to much smaller sizes. For an adopted
size distribution for $r < \rmin$, it is possible to calculate the collision
rate analytically and to correct the mass flow rate for particles with 
$r \approx$ \rmin\ \citep[e.g.,][]{obrien2003}.  For Fomalhaut b, however, 
radiation pressure removes particles with $r < \rmin$ on time scales much 
smaller than the collisional time. While some of these particles might lie
on orbits which occasionally bring them back into the satellite swarm, most
never return. Thus, real satellite swarms likely have wavy size distributions.

In the rest of this section, we examine how `equilibrium' wavy size 
distributions depend on model parameters.  To discuss the time evolution 
of these size distributions, we derive the relative cumulative size 
distribution. At each $r_k$ in the grid, the cumulative size distribution 
$n(>r)$ is the number of objects with radius larger than $r$.  To isolate 
the waviness about a power law, we define the relative cumulative size 
distribution
\begin{equation}
n_{c,rel} = n(>r) / n_0 r^{-q_n} ~ ,
\label{eq: ncum}
\end{equation}
where $n_0$ is a normalization factor. For these calculations, we adopt 
$q_n$ = 2 and normalize the relative cumulative size distribution to 1
at 10~km or at 100~km. 

By normalizing every relative cumulative size distribution at 10~km or
100~km, we suppress the natural evolution of $n_0$ with time. In all 
calculations, $n_0$ follows the standard evolution of the total mass
in satellites: roughly constant at early times and then declining linearly
with time at later times. In this section, we focus on the evolution of
the shape of the size distribution. We return to the long-term evolution
of the surface area in \S\ref{sec: calcs-area}.

\subsubsection{Baseline model}
\label{sec: calcs-sd-base}

In the baseline model, the initial size distribution is a power law from
\rmin\ = 100~\mum\ to \rmax\ = 50, 100, 200, or 400~km. Large satellites 
tend to grow. Catastrophic and cratering collisions slowly reduce the 
sizes of satellites. For the adopted starting conditions, the collision
time is 10--100~Myr. We expect the debris from destructive collisions 
to establish an equilibrium size distribution on this time scale.

Fig.~\ref{fig: sd1} illustrates the evolution of $n_{c,rel}$ for calculations 
with \mpl\ = 10~\mearth\ and \rmax\ = 100~km. This evolution is very different 
from standard predictions for a collisional cascade, $\ncr \propto r^{-q_r}$ 
and $q_r \approx$ 0.5--0.75\footnote{In our convention, we have three power
law slopes, $q$ (differential power law), $q_c$ (cumulative power law) and
$q_r$ (relative cumulative power law). These have a simple relationship:
$q \approx$ $q_c + 1$ $\approx$ $q_r + 3$.} 
\citep[e.g.,][]{dohn1969,will1994,obrien2003,koba2010a}.  
After only 0.1~Myr, \ncr\ develops a characteristic shape, consisting of
(i) a steep rise from \rmax\ to $r \approx$ 50~km,
(ii) a gradual rise from 50~km to 0.1 km,
(iii) a steep drop from 0.1~km to 30~cm, and
(iv) a steep rise from 30~cm to 100~\mum.
The rise from 50~km to 0.1~km has several distinct large-scale 
oscillations; other fluctuations are small relative to the overall
trend. After 1--10~Myr, \ncr\ is independent of time except at large
sizes $r \gtrsim$ 10--30~km.

In this calculation, the lack of particles with $r \lesssim$ 100~\mum\ produces
the pronounced wave in the size distribution from 100~\mum\ to 0.1~km. With no
very small particles ($r < \rmin$), collisions remove objects with $r \approx$ 
\rmin\ at a lower rate, producing the excess of 100--1000~\mum\ satellites. In
turn, these objects remove larger particles at a faster rate, creating the large
deficit at 1~cm to 10~m. At larger sizes, the waves are due to
(i) stochastic collisions among the largest particles, which leads to an
intermittent supply of debris among smaller particles and
(ii) less frequent destructive collisions among 0.1--10~km particles and
1--100~m particles.

The time scale to set up the equilibrium size distribution is many collision 
times for the smallest particles. In this example, the collision time for a
typical small particle is $v A_d / V \approx$ $10^4$~yr. The main features 
of the wavy size distribution develop in just a few collision times,
$\sim$ 0.1~Myr. By roughly 1~Myr, \ncr\ establishes a distinctive pattern
from 100~\mum\ to roughly 10~km, which remains fixed for 1~Gyr. At the 
largest sizes, fluctuations in the collision rate produce a time varying 
feature which slowly grows with time.

Fig.~\ref{fig: sd2} shows how the depth of the wave depends on the mass of
the central planet. In this example, the minimum in \ncr\ at 30~\mum\ grows
shallower and shifts to smaller sizes with decreasing planet mass. These
changes are solely a function of the collision time. 
With $t_c \propto \mpl^{-1/3}$ (eq.~\ref{eq: tc}), a factor of 300 in
\mpl\ corresponds to a factor of 6.7 in the collision time. Thus, satellites 
orbiting a 300~\mearth\ planet experience roughly 7 times as many collisions
over a fixed time interval as those orbiting a 1~\mearth\ planet. With fewer
collisions, the size distribution of satellites orbiting 1~\mearth\ planets 
follows the initial power law more closely. Despite the longer collision 
time, collisions still establish the steep size distribution among the
smallest particles and produce clear waves throughout the size distribution.

For any mass of the central planet, the waviness within \ncr\ is also sensitive
to the cloud mass. Gradually reducing the initial \xcl\ produces the same 
progression in the depth and position of the deep minimum as in 
Fig.~\ref{fig: sd2}. With $t_c \propto \xcl^{-1}$ (eq.~\ref{eq: tc}), factor 
of 7 reductions in \xcl\ for \mpl\ = 300~\mearth\ yield \ncr\ similar to the
magenta curve in Fig.~\ref{fig: sd2}. With larger reductions, \ncr\ more
closely resembles the smooth power law of the initial size distribution.	

Changing the initial \rmax\ has little impact on \ncr\ at 0.1--1~Gyr
(Fig.~\ref{fig: sd3}). For \rmax\ = 50--400~km, the normalized size 
distributions at 100~Myr are essentially identical from 100~\mum\ to 
10--20~km. At larger sizes, the deviations from a power law depend 
on \rmax.

\subsubsection{Size of the smallest particles}
\label{sec: calcs-sd-small}

For calculations with fixed \mpl, \xcl, and \rmax, changing \rmin\ has
the same impact on \ncr\ as changing \mpl\ or \xcl\ (Fig.~\ref{fig: sd4}).
For the set of parameters in our calculations, collisions between a
large particle with radius $r_l$ and a small particle with radius 
$r_s \ll r_l$ remove 2--3 times the mass of the small particle from 
the large particle.  When \rmin\ increases, collisions remove less
mass from all remaining particles. Among 0.1~km and larger particles,
the smaller mass loss has a fairly small impact on \ncr. However,
less mass loss produces a steeper size distribution for particles 
with $r \approx$ \rmin\ and a larger deficit in particles at 
10--100~cm. Larger \rmin\ also shifts the minimum in \ncr\ to larger 
radii. 

When \rmin\ decreases, collisions remove somewhat more mass from all 
particles. The size distribution for the smaller particles becomes
less steep and the pronounced minimum at 10--100~cm grows smaller.
In our example, calculations with \rmin\ $\approx$ 10~\mum\ nearly
eliminate the deep minimum in \ncr\ at 10--100~cm. This example 
retains the waves in \ncr\ for $r \gtrsim$ 1~cm and the steep gradient
for $r \lesssim$ 1~cm.

\subsubsection{Mass of the largest object in the debris}
\label{sec: calcs-sd-large}

Distributing the mass in the debris differently also has a clear impact 
on \ncr\ at small sizes. In our calculations, all collisions among
particles with $r \lesssim$ 1--10~km have $\qc / \qdstar \gtrsim$ 1.
Both objects shatter.  When $b_L \approx$ 0, most of the debris is
placed in bins with masses close to the mass of the original particles.
When $b_L \approx$ 1, more debris is distributed among particles with
much lower mass. Spreading debris around more mass bins fills in the
minimum in \ncr.

Fig.~\ref{fig: sd5} illustrates this point. In calculations with 
$b_L$ = 0, there is a clear minimum in \ncr\ at 10--100~cm. When
$b_L$ = 1, the minimum is less deep; the slope of the size 
distribution to smaller radii is shallower. At large radii
($r \gtrsim$ 0.1~km), $b_L$ has little impact on the \ncr: all
relative size distributions are wavy and roughly constant from 
0.1--100~km.

In both examples in the figure, the shape of the size distribution 
is independent of \rmax. All of the size distributions have small-scale 
fluctuations about the general trend, but these are small compared to 
the overall trends as a function of radius.

\subsubsection{Binding energy}
\label{sec: calcs-sd-qd}

Reducing \qdstar\ has a similar impact on \ncr\ as changing $b_L$
(Fig.~\ref{fig: sd6}). Smaller \qdstar\ leads to more ejected mass
per collision. More ejected mass enhances the mass excess at the
smallest sizes, steepening the size distribution. A larger population
of smaller particles enhances the deficit at somewhat larger sizes. 
As with \rmin, changing \qdstar\ changes the depth and the location
of the deficit. Larger \qdstar\ reduces the deficit and shifts it to
smaller sizes. Smaller \qdstar\ adds to the deficit and shifts it to
larger sizes.

Reducing \qdstar\ also adds to the waviness of \ncr\ at larger sizes.
With more mass loss in every collision with much smaller particles, 
large particles distribute more mass among smaller mass bins. 
Truncating the size distribution at non-zero \rmin\ creates a waviness
in this mass loss, which is enhanced as \qdstar\ is reduced.

\subsubsection{Initial size distribution}
\label{sec: calcs-sd-sd}

To judge how the size distribution evolves when the calculations start 
from a mono-disperse set of satellites, we consider the baseline model 
with \rmin\ = 10~\mum\ instead of 100~\mum. The smaller \rmin\ give us a
better view of the evolution of the population of small particles and
yields a good comparison for models with and without an initial size
distribution of particles with radii smaller than \rmax.

Fig.~\ref{fig: sd7} illustrates the evolution of a baseline model with 
\mpl\ = 10~\mearth\ and \rmax\ = 100~km. At the start of this calculation,
collisions between pairs of 100~km objects produce satellites somewhat 
larger than 100~km and substantial debris. After 0.1~Myr, \ncr\ has a 
clear excess of particles with radii of a few km and some smaller particles.
By 1~Myr, the excess has grown considerably; there is a substantial debris
tail down to 1~m. In another 9~Myr, the debris populates the full range in
allowed particles sizes and establishes a characteristic size distribution
which remains fixed for the rest of the calculation. 

The equilibrium size distribution in Fig.~\ref{fig: sd7} has the same features 
as in the baseline model. At small sizes (10~\mum\ to 1~cm), there is a steep
and slightly wavy power law with slope $q \approx$ 4.5 ($q_c \approx$ 3.5; 
$q_r \approx$ 1.5). At larger sizes, the 
power law slope is closer to $q \approx$ 2 with significant waves. Close to
\rmax, the slope again steepens. 

Fig.~\ref{fig: sd8} compares snapshots of the size distribution for calculations
with (`sd') and without (`no sd') an initial power law size distribution among
particles with radii smaller than \rmax. At 10~Myr, calculations with a 
mono-disperse set of large particles have more mass. In these calculations, the 
largest particles lose less mass through collisions with much smaller objects.
Thus, they diminish in size less rapidly with time. For particles which are
large enough to escape destruction, the extra mass in the largest particles
allows them to grow more rapidly.

At 1~Gyr, the \ncr\ for 10~\mum\ to 30--50~km particles is independent of the
initial size distribution. In the `no sd' models, the largest objects reach 
sizes of 200--300~km in 1~Gyr. Satellites in the `sd' models gradually lose
mass and reach sizes of 70~km after 1~Gyr. Despite this difference, smaller 
debris particles produced in the collisions of the largest objects have an 
identical size distribution. Thus, the long-term evolution of the smallest 
objects is independent of the starting point.

\subsection{Evolution of the Cross-sectional Area}
\label{sec: calcs-area}

For Fomalhaut b, observations cannot measure the size of the largest object or
discern the size distribution across any range of sizes. Aside from the size
and the color of the cloud as a whole, the only observable is the total 
brightness. For grains with radii $\gtrsim$ 10~\mum, we relate the brightness to 
the cross-sectional area \citep[e.g.,][]{currie2013,galich2013,kalas2013,kcb2014}.
To compare model results with the observations, we rely on the time variation 
of the cross-sectional area in each calculation relative to the adopted area 
for Fomalhaut b, $A_d \approx 10^{23}$~cm$^2$. A successful calculation matches 
this area at the adopted age of Fomalhaut, $t_F \sim$ 200~Myr.  We assign a 
factor of two uncertainty to $A_d$ and $t_F$. The model `target' is then a 
rectangular box in $(t_F, A_d)$ space (see also Fig.~\ref{fig: area1}).

\subsubsection{Baseline model}
\label{sec: calcs-area-base}

\begin{figure} 
\includegraphics[width=6.5in]{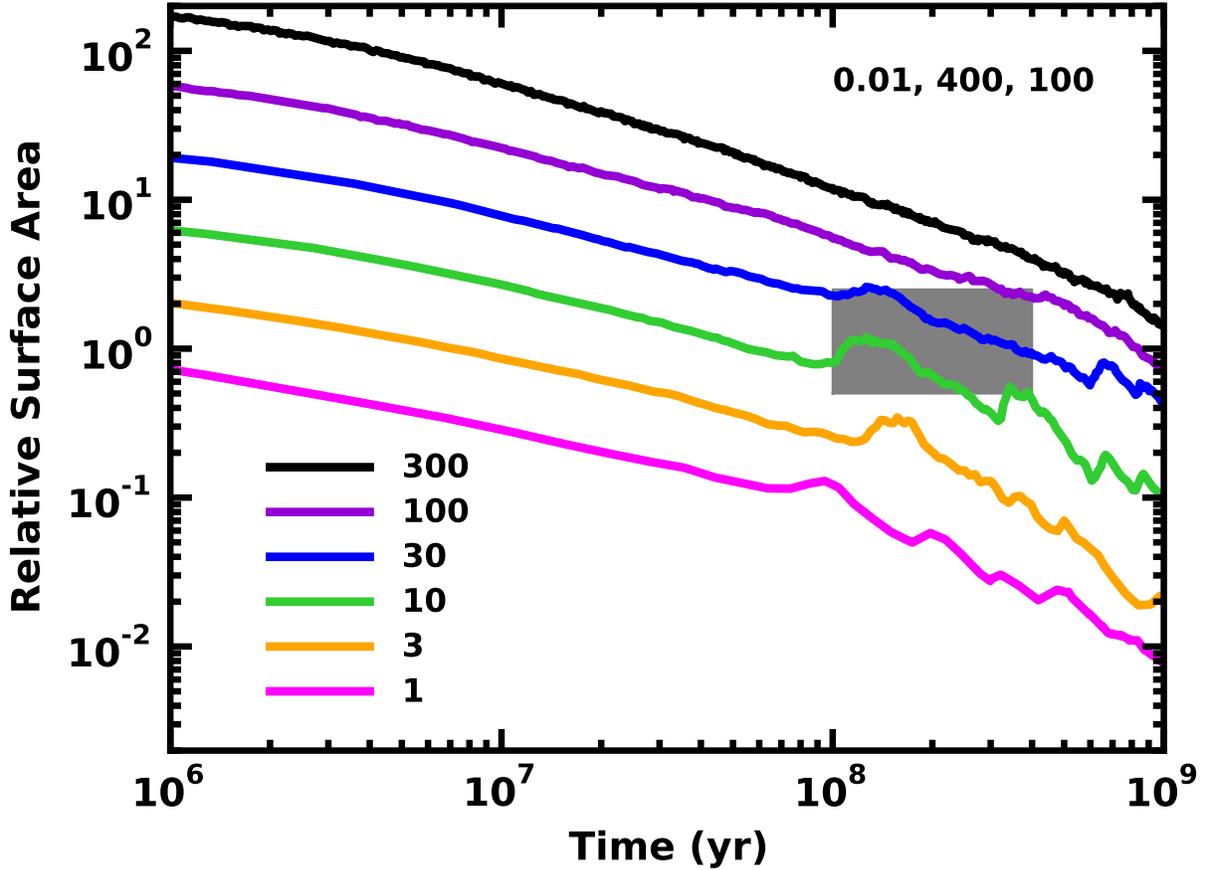}
\vskip 3ex
\caption{
Time evolution of the relative surface area derived from coagulation 
calculations of circumplanetary clouds of particles with the nominal 
fragmentation parameters, 
\mlz\ = 0.2, and \bl\ = 0.0.  The legend in the upper right indicates
the initial \xcl, \rmax\ (in km), and \rmin\ (in \mum). The legend in 
the lower left indicates the mass of the central planet. When the largest
object does not grow (\mpl\ = 100, 300~\mearth), the relative surface 
area declines smoothly with time. At late times in simulations with 
growing satellites, the surface area fluctuates about a gradual decline.
\label{fig: area2}
}
\end{figure}

Fig.~\ref{fig: area2} illustrates the time variation of $A_d$ for baseline models
with initial \rmax\ = 400~km. 
Initially, the surface area -- $A_d \propto \mcl \propto \xcl \mp$ -- depends 
only on the initial cloud mass.  
Collisions then redistribute mass through the grid and gradually change the 
relative surface area. From eqs.~\ref{eq: tc} and \ref{eq: tcn}, the collision 
time is $t_c \propto \mp^{2/3} \mcl^{-1}$.  For fixed \xcl, swarms around more 
massive planets have larger cloud masses and shorter $t_c$.  Thus, calculations 
with \mpl\ = 100--300~\mearth\ evolve more rapidly than those with \mpl\ = 
1--3~\mearth.
After a brief re-adjustment where $A_d$ 
grows substantially, the relative surface area declines from $\sim$ 100 (where 
the swarm may become optically thick, see the discussion in 
\S\ref{sec: disc-theory-rmin} below) at 1~Myr to $\sim$ 1 at 1~Gyr.  Although 
the 100~\mearth\ models graze the target box at 400~Myr, these models generally 
fail to match the observations.

Calculations with \mpl\ = 1--3~\mearth\ also fail. These models begin with 
relative surface areas close to the target and fairly long collisions times 
of 20--30~Myr. However, the largest objects in these calculations grow with
time, removing small particles from the grid. After 100~Myr of evolution, 
swarms of satellites orbiting 1--3~\mearth\ planets have relative surface 
areas at least a factor of two below the observations. 

Swarms of satellites around 10--30~\mearth\ planets match the observations 
throughout the 100--400~Myr target period. In these systems, the initial
cross-sectional area is roughly 10--20~times the area of Fomalhaut b. 
Although the largest objects in these simulations also grow with time, 
debris from the collisions of smaller objects maintains a large surface 
area for over 100~Myr. After 400~Myr (1~Gyr), satellites around the 
10~\mearth\ (30~\mearth) planet have a surface area smaller than Fomalhaut b.
Thus, there is a substantial cushion between the predicted time evolution 
and the evolution required to match the observations.

\begin{figure} 
\includegraphics[width=6.5in]{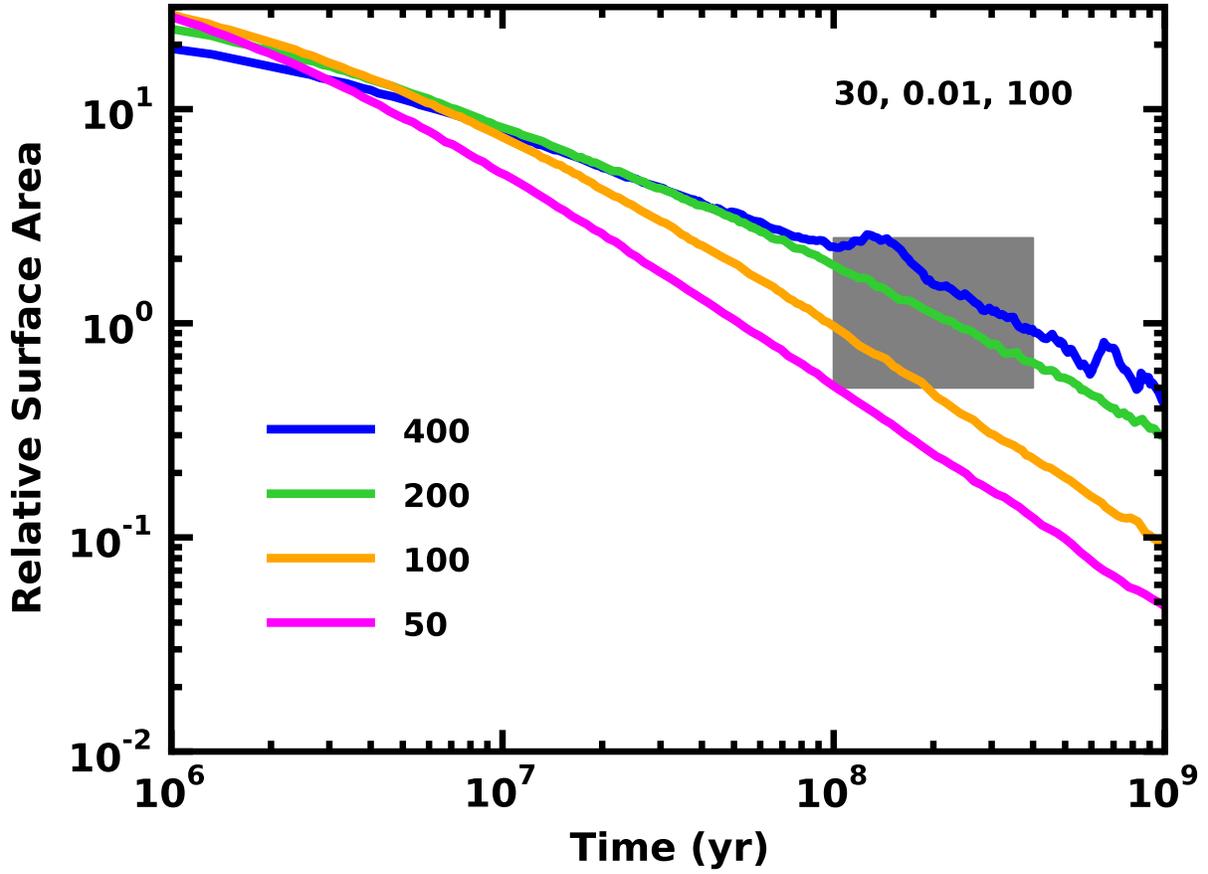}
\vskip 3ex
\caption{
As in Fig.~\ref{fig: area2} for 
the \mpl, \xcl, and \rmin\ indicated in the upper right corner for
the range of \rmax\ (in km) indicated in the lower left corner. The
slow decline of the relative surface area is smoother for smaller
\rmax.
\label{fig: area3}
}
\end{figure}

Calculations starting with smaller \rmax\ yield smaller cross-sectional area 
at late times (Fig.~\ref{fig: area3}). With fixed total mass, swarms with 
smaller initial \rmax\ have larger surface area ($A_d \propto \rmax^{-1/2}$). 
However, the collision time scales linearly with \rmax\ (eq.~\ref{eq: tc}).
Thus, destructive collisions remove material more rapidly from swarms with
smaller \rmax. More rapid mass loss results in smaller cross-sectional area. 
For \mpl\ = 30~\mearth, simulations with initial \rmax\ = 100--200~km still
match the observations at 200~Myr. Satellites with initial \rmax\ = 50~km
fall below the target.

When \rmax\ $\lesssim$ 200~km, calculations with \mpl\ = 10~\mearth\ also
fail to match the observations. Collisional evolution is too rapid for 
these models to match the observed surface area at 100--400~Myr. In 
contrast, models with more massive planets, \mpl\ = 100--300~\mearth, 
and \rmax\ = 100--200~km, pass through the target. However, swarms with
\rmax\ = 50~km succeed only for \mpl\ = 100~\mearth.

\subsubsection{Initial cloud mass}
\label{sec: calcs-area-cloud}

The initial cloud mass has a more dramatic impact on the evolution of the 
surface area than \rmax. For fixed \mpl, the collision time scales inversely 
with \xcl\ and \rmax. Smaller \xcl\ and smaller \rmax\ lengthen the collision 
time. However, the initial surface area scales linearly with \xcl\ and as 
$\rmax^{-1/2}$. Models with smaller \xcl\ therefore start out with much smaller 
area and have more trouble matching observations.

\begin{figure} 
\includegraphics[width=6.5in]{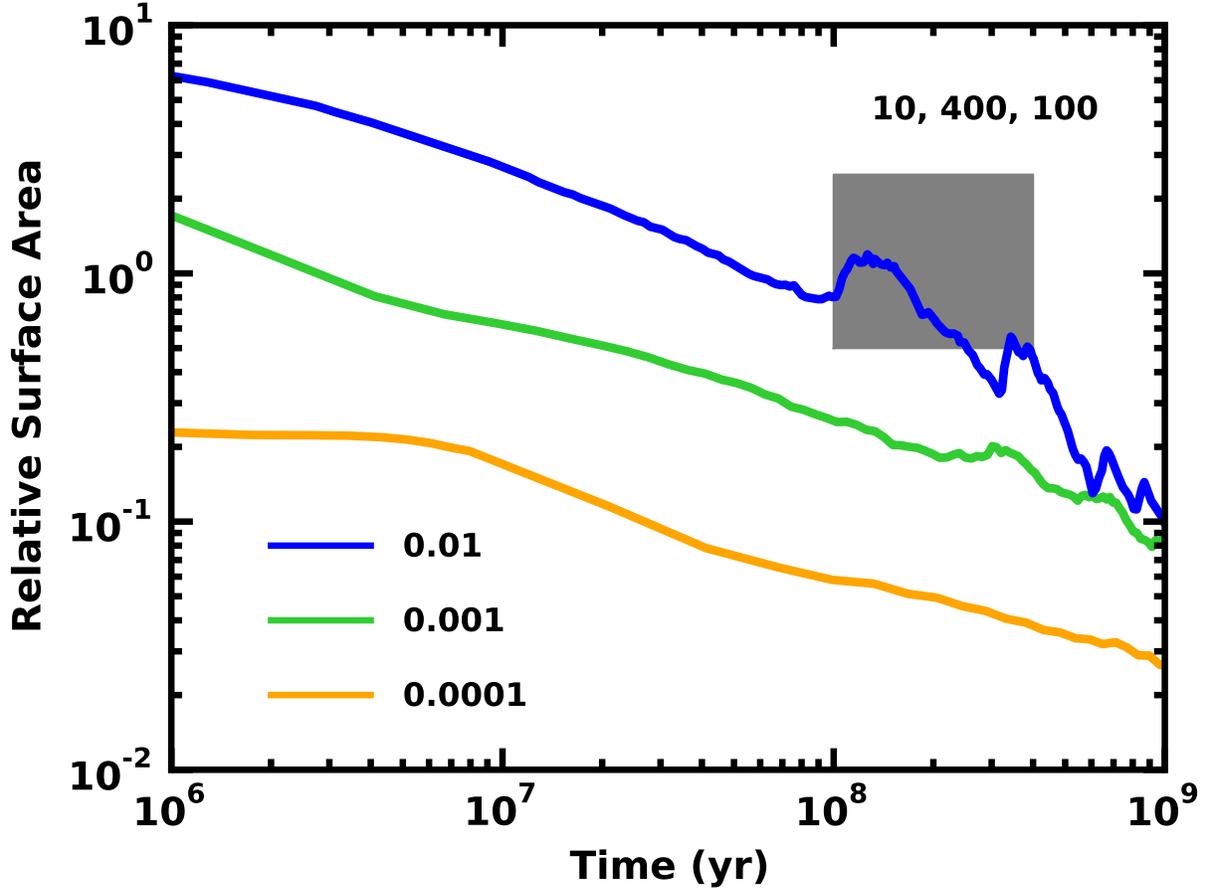}
\vskip 3ex
\caption{
As in Fig.~\ref{fig: area2} for the \mpl, \rmax, and \rmin\ indicated
in the upper right corner for various \xcl\ as indicated in the lower
left corner. Lower mass clouds have smaller relative surface areas.
\label{fig: area4}
}
\end{figure}

Fig.~\ref{fig: area4} illustrates these points for \mpl\ = 10~\mearth\ and
\rmax\ = 400~km. The baseline model with \xcl\ = 0.01 matches the data 
for ages of 100--400~Myr. Factor of ten smaller swarms have an initial 
surface area somewhat larger than Fomalhaut b, but collisions reduce the 
area by almost a factor of ten after 100~Myr. Another factor of ten reduction
in \xcl\ leaves the surface area below observations throughout the evolution
of the swarm.

\begin{figure} 
\includegraphics[width=6.5in]{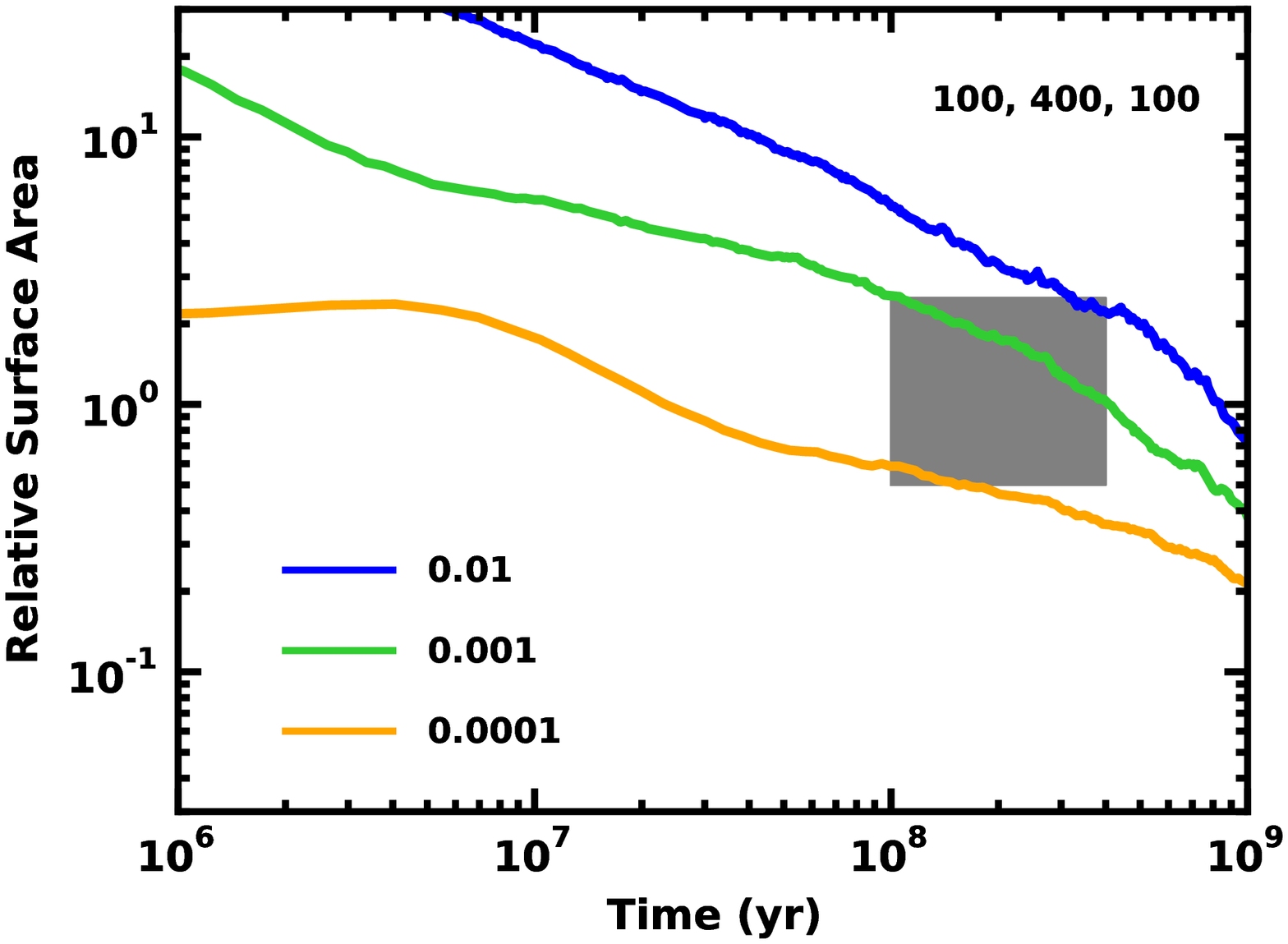}
\vskip 3ex
\caption{
As in Fig.~\ref{fig: area4} for \mpl\ = 100~\mearth.
\label{fig: area5}
}
\end{figure}

Fig.~\ref{fig: area5} shows that low mass swarms around massive planets can
also match observations. When \mpl\ = 100~\mearth, satellites with \xcl\ = 0.01
have a surface area that grazes the upper edge of the target box at 300--400~Myr.
Reducing the initial mass by a factor of 100 yields a surface area that grazes 
the lower edge of the target box at 100--200~Myr. Intermediate cloud masses 
match the data well; models with \xcl\ = 0.001 pass through the upper middle 
of the target box.

\subsubsection{Size of the smallest particles}
\label{sec: calcs-area-small}

In the analytic model, the surface area of a satellite swarm scales with 
$\rmin^{-1/2}$. Changing \rmin\ by an order of magnitude thus modifies the
initial surface area by a factor of $\sim$ 3. Because \rmin\ has little
impact on the cloud mass or the collision time, the early evolution of
satellites with different \rmin\ is nearly identical. Over time, however, 
the smallest particles shape the size distribution at larger sizes 
(Fig.~\ref{fig: sd4}). When \rmin\ is smaller (larger), more (fewer) 
intermediate particles with $r \approx$ 10~cm to 10~m survive the cascade.
If these particles contribute much to the total surface area of the swarm,
then we expect the surface area at late times to scale more steeply with 
the minimum particle size.

\begin{figure} 
\includegraphics[width=6.5in]{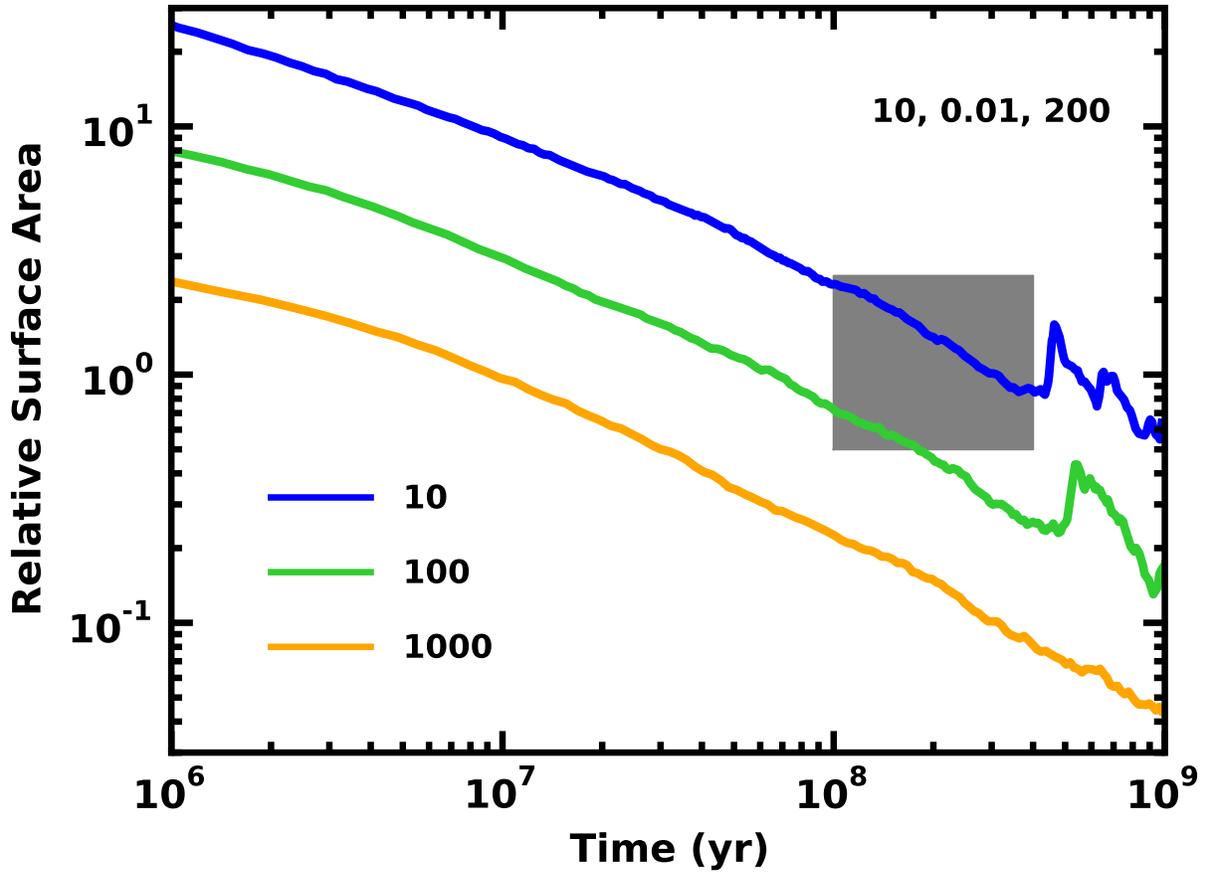}
\vskip 3ex
\caption{
As in Fig.~\ref{fig: area2} for the \mpl, \xcl, and 
\rmax\ summarized in the upper right corner for the 
\rmin\ listed in the lower left corner. Calculations 
with smaller \rmin\ have more small particles and 
larger relative surface area.
\label{fig: area6}
}
\end{figure}

Fig.~\ref{fig: area6} demonstrates that transformations to the size 
distribution have little impact on the evolution of the surface area. 
In the baseline model with \mpl\ = 10~\mearth, \rmax\ = 200~km, and 
\rmin\ = 100~\mum, the evolution of the surface area passes through 
the lower left corner of the target area. In calculations with 
\rmin\ = 1~mm, the surface area is $\sqrt{10}$ smaller and falls well
below the target. Factor of ten smaller \rmin\ yields a factor of 
$\sqrt{10}$ larger surface area which lies in the upper half of the
target. Thus, the surface area scales exactly as $\rmin^{-1/2}$.

Despite significant differences in the size distributions of calculations
with different \rmin, these changes have little impact on the evolution
of the surface area. Particles with radii of 1~cm to 100~m have a limited
fraction of the mass of the larger particles and a negligible surface area 
compared to much smaller particles. Augmenting (or reducing) the population
of these particles by factors of 100--1000 has no observable impact on 
the total surface area of the cloud.

\subsubsection{Mass of the largest object in the debris}
\label{sec: calcs-area-large}

In our collision model, we use a simple algorithm to distribute debris
into the mass bins. For the baseline model with $b_L$ = 0, the 
largest object in the debris has 20\% of the total mass in the debris. 
In comparison models with $b_L = 1$, larger relative collision energies 
place more material in lower mass bins. Material leaves the grid more 
rapidly. Although the distribution of debris has little impact on the 
evolution of the largest objects (Fig.~\ref{fig: rmax5}), the relative 
number of the smallest particles depends on the mass of the largest 
object in the debris (Fig.~\ref{fig: sd5}). In the complete ensemble of 
calculations, $N(\rmin)$ changes by a factor of 3--5; we expect similar
changes in the cross-sectional area.

Fig.~\ref{fig: area6} confirms this conjecture for calculations with
\mpl\ = 10--30~\mearth, \xcl\ = 0.01, \rmax\ = 400~km, and \rmin\ = 
100~\mum. When $b_L$ = 0, the predicted surface area passes through the target
box. For $b_L$ = 1, calculations with \mpl\ = 30~\mearth\ pass through
the lower edge of the target; models with \mpl\ = 10~\mearth\ completely
miss the target.

Larger $b_L$ allows satellite swarms around more massive planets to match
observations at 100--400~Myr. In the baseline model, calculations with 
\mpl\ = 100--300~\mearth\ and \xcl\ = 0.01 have much larger surface area 
than Fomalhaut b. When $b_L = 1$, satellites orbiting planets with \mpl\ =
100~\mearth\ (300~\mearth) pass through the lower (upper) half of the target
box.

\subsubsection{Binding energy}
\label{sec: calcs-area-qd}

In our suite of calculations with different \qdstar, smaller 
\qdstar\ prevents the largest objects from growing 
(Fig.~\ref{fig: rmax6}). The additional debris produced from 
collisions between the largest objects decreases the relative 
numbers of satellites with radii larger than 1--10~cm 
(Fig.~\ref{fig: sd7}). However, smaller \qdstar\ has little impact 
on the relative population of the smallest objects which dominate 
the cross-sectional area. Thus, swarms of satellites with different
binding energies have comparable surface areas.

\begin{figure} 
\includegraphics[width=6.5in]{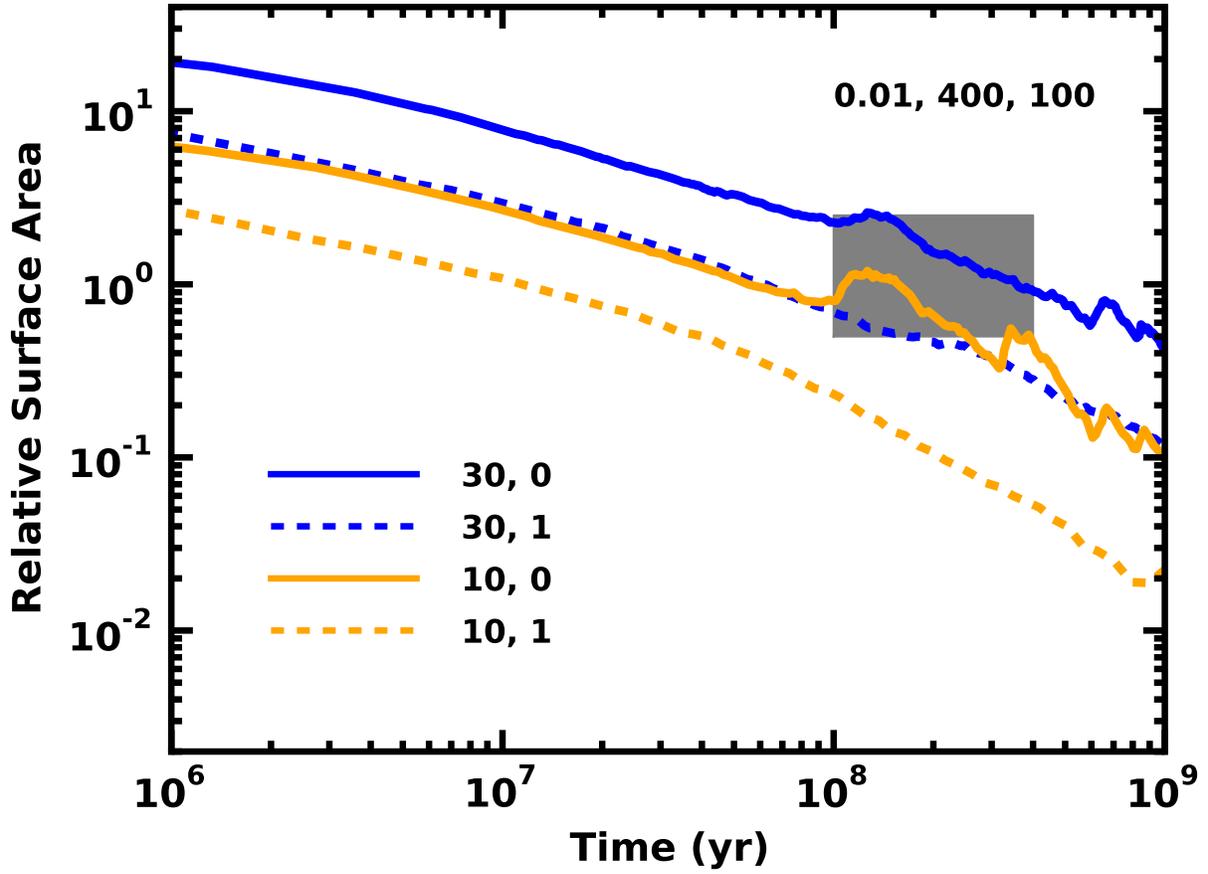}
\vskip 3ex
\caption{
As in Fig.~\ref{fig: area6} for the \xcl, \rmax, and \rmin\ listed
in the upper right corner and various \mpl\ and $b_L$ as listed in
the lower right corner. Calculations with $b_L$ = 1 have smaller
relative surface area than those with $b_L$ = 0.
\label{fig: area7}
}
\end{figure}

Fig.~\ref{fig: area7} shows the modest variations in cross-sectional
area for calculations with different \qdstar. In the baseline model,
400~km satellites reach radii of 2000~km on time scales of 100--400~Myr.
Large fluctuations in the surface area resulting from stochastic variations 
in debris production from the collisions of the largest objects begin at
100~Myr and continue beyond 1~Gyr. Lowering \qdstar\ by a factor of two
limits the growth of these objects to 1000~km. Slower growth delays the
large oscillations in the surface area and minimizes them at later times.
Another factor of two reduction in \qdstar\ completely eliminates the
growth of the largest objects. Less growth enables more debris and
larger surface area at early times. As the calculation proceeds, debris
production is somewhat more modest than calculations with larger \qdstar,
resulting in a slightly lower surface area at later times.

In calculations with $b_L$ = 1, smaller \qdstar\ speeds up the time evolution 
of the surface area. When \qdstar\ is smaller and $b_L$ = 1, collisions place
more debris in smaller mass bins. Placing debris in smaller mass bins allows
the cascade to eject mass from the grid more rapidly. Thus, the surface area
declines more rapidly with time. For the smallest \qdstar\ considered in our
calculations, models with $b_L$ = 1 require very large and probably unrealistic
initial cloud masses to match the target for Fomalhaut b. 

\subsubsection{Initial Size Distribution}
\label{sec: calcs-area-sd}

\begin{figure} 
\includegraphics[width=6.5in]{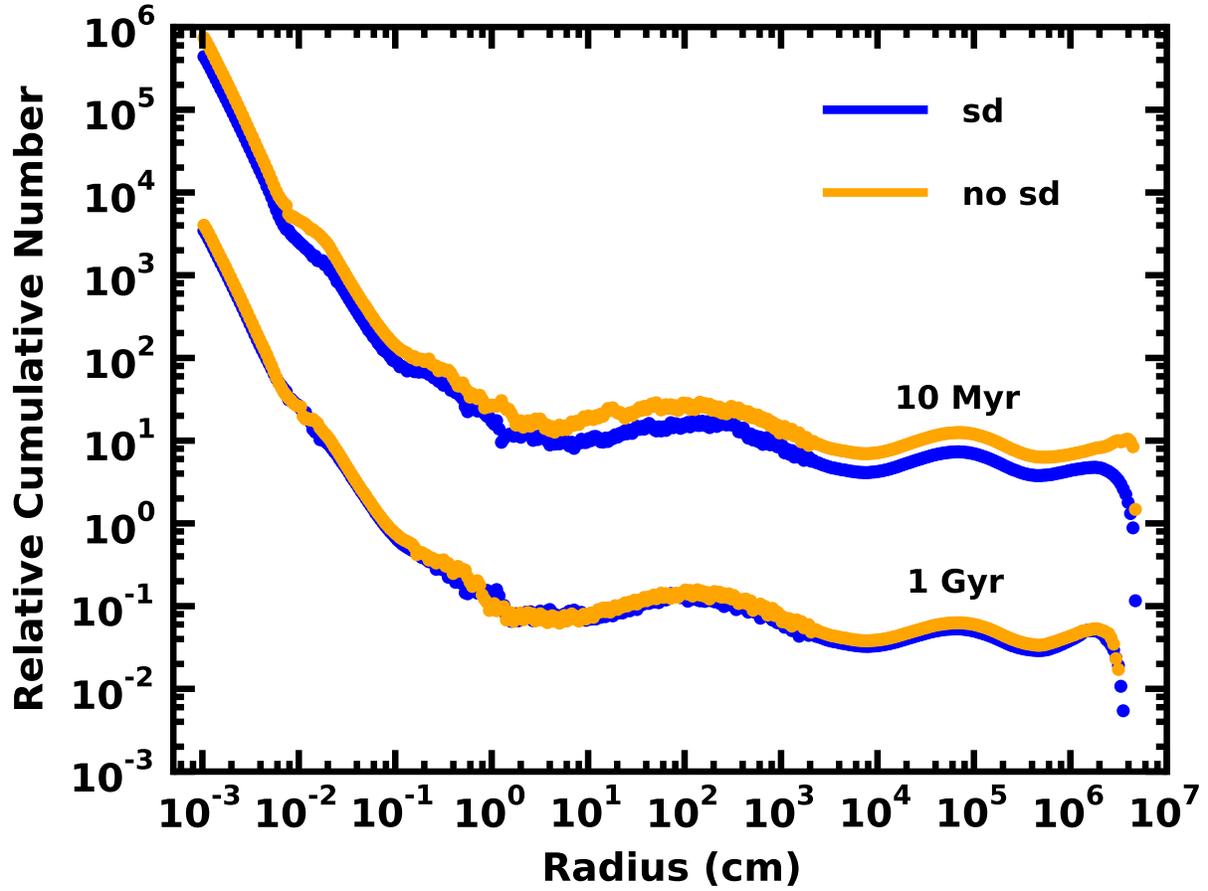}
\vskip 3ex
\caption{
Comparison of model relative size distributions for calculations
starting from an initial power law size distribution of particles 
(`sd') and a mono-disperse set of particles (`no sd') at 10~Myr
(upper set of curves) and at 1~Gyr (lower set of curves).  Aside
from minor deviations at the largest sizes, the two sets of 
calculations yield identical size distributions.
\label{fig: sd8}
}
\end{figure}

When calculations begin with an initial power-law size distribution, the
initial surface area is substantial (Fig.~\ref{fig: area2}). With a
mono-disperse set of large particles, the initial surface area is 
negligible. For \rmax\ = 100~km and \rmin\ = 100~\mum, it takes several
collision times to populate the low mass end of the size distribution
(Fig.~\ref{fig: sd8}) and increase the surface area.

\begin{figure} 
\includegraphics[width=6.5in]{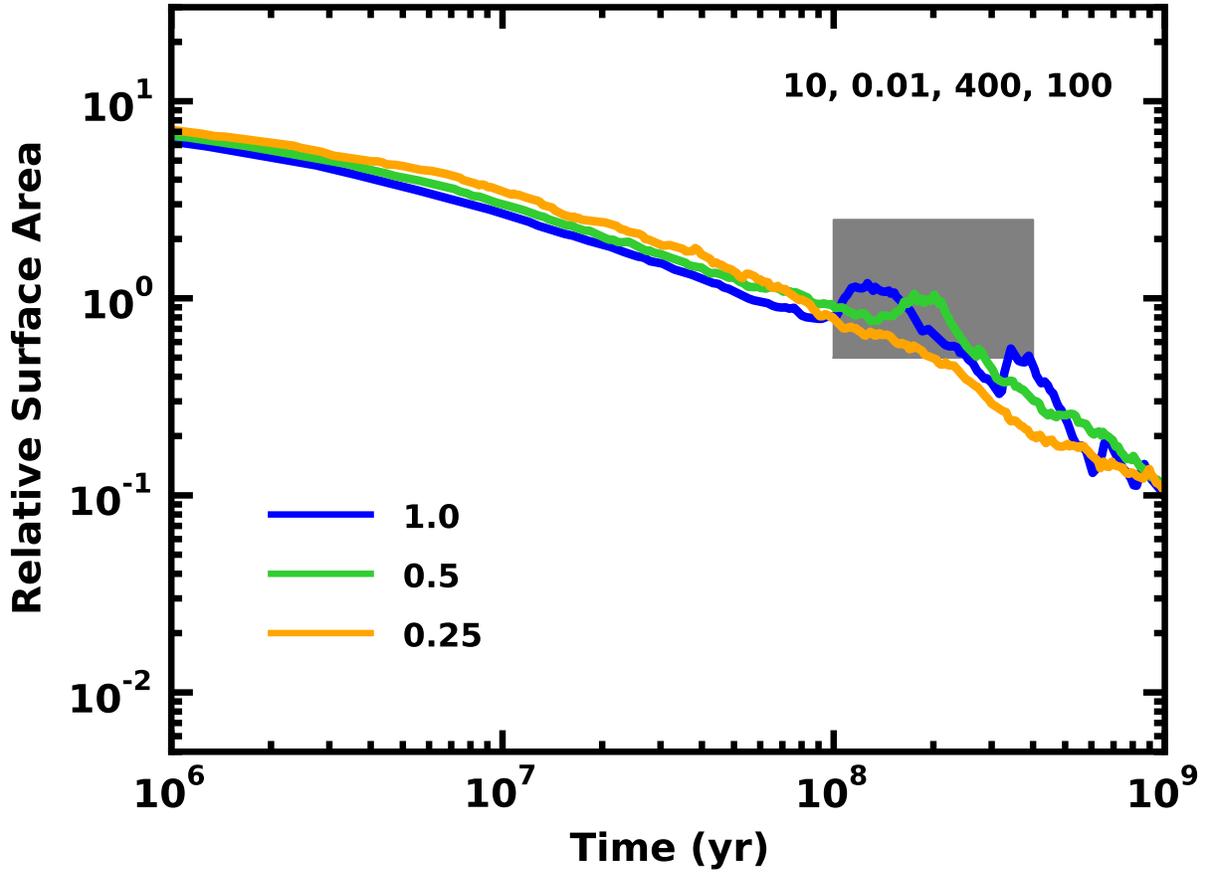}
\vskip 3ex
\caption{
As in Fig.~\ref{fig: area2} for the \mpl, \xcl, \rmax, and
\rmin\ listed in the upper right corner and various \qdstar.
The legend in the lower left corner indicates the value of 
\qdstar\ relative to the nominal value. 
\label{fig: area8}
}
\end{figure}

Fig.~\ref{fig: area8} compares the growth of the surface area for swarms
of satellites with (`sd') and without (`no sd') an initial power-law 
distribution of small particles. When \rmax\ = 100~km, the collision time
is rather short. It takes only a few $\times ~ 10^6$ yr for the surface area
to reach the levels of the baseline model. At 10~Myr, the satellite swarm 
in the mono-disperse model has relatively more small particles than the
swarm with the initial power-law size distribution (Fig.~\ref{fig: sd8}).
Thus, the mono-disperse model has a larger surface area. By $\sim$ 1~Gyr,
collisions have erased the starting conditions; the size distributions 
(and hence the surface areas) are indistinguishable.

When the initial \rmax\ is 400~km, the evolution of satellite swarms with
and without an initial power law size distribution diverge. In these
calculations, the collision time is much longer. Populating the small
end of the size distribution takes 30--40~Myr instead of 3--6~Myr. With
fewer small objects to chip away at the mass of the largest objects, the
largest objects grow more rapidly and contain a larger fraction of the
initial mass (Fig.~\ref{fig: rmax7}). At $\sim$ 100~Myr, these calculations
produce more debris than the baseline model. As the evolution proceeds, 
the larger mass tied up in the largest objects speeds up the decline of 
the population of small particles. After 400--500~Myr, the surface area in 
the small particles is a factor of $\gtrsim$ 2 smaller than in the baseline
model.

\subsection{Growth of Very Large Objects in the Swarm}
\label{sec: calcs-nb}

In some calculations, the largest objects in the swarm grow by factors of 
10--1000 in mass over 0.1--1~Gyr. As these objects grow, their gravity may
stir up smaller objects in the swarm. Small amounts of stirring increase 
typical collision velocities and enhance the impact of the collisional
cascade. Larger amounts can eject material from the swarm. Aside from 
ejections, dynamical interactions between pairs of very large objects 
can place satellites on bound orbits closer to the planet. If the planet
already has a set of closely bound satellites, a `new' satellite might
disrupt the indigenous satellite system.

To explore the impact of these processes within our simulations, we begin
with general principles. When large satellites grow, they try to stir up
smaller satellites to their escape velocity. Satellites with radii 
\rmax\ have escape velocity
\begin{equation}
v_{esc} = 0.9 \left ( \rmax \over 10^3~{\rm km} \right ) ~ \kms\ ~ .
\label{eq: v-esc}
\end{equation}
For a central planet with \mpl\ = 10~\mearth, satellites with \rmax\ $\approx$ 
350--400~km have an escape velocity, 0.3~\kms, comparable to the collision
velocity of satellites in a spherical swarm. Thus, stirring is a crucial issue 
for large satellites with radii of 1000--2000~km. 

Satellites orbiting the planet have Hill spheres where the gravity of the 
satellite dominates the gravity of the planet. For satellites orbiting at 
semimajor axis $a$, the Hill radius is
$R_{H,s} \approx r_s a$ (e.g., eq.~\ref{eq: rhill}) where
\begin{equation}
r_s \approx 0.04 
\left ( { \rmax\ \over {\rm 1000~km} } \right )
\left ( { \mpl \over 10~\mearth\ } \right)^{-1/3} ~ .
\end{equation}
Larger satellites around less massive planets have larger Hill radii.

When the semimajor axes of orbiting satellites differ by 3--4 Hill radii or 
less, they interact dynamically. For simplicity, we define the minimum orbital 
separation for stability as $\Delta a \approx 4 r_s a$. In a system of $N$ 
massive satellites within a spherical shell of width $\delta a = \eta_2 a$, 
the system is dynamically stable when $4 N r_s a \lesssim \eta_2 a$.  Solving 
for N, the maximum number of non-interacting satellites is
\begin{equation}
N_{max} \approx 3 
\left ( { \eta_2\ \over 0.5 } \right )
\left ( { \rmax\ \over {\rm 1000~km} } \right )^{-1}
\left ( { \mpl \over 10~\mearth\ } \right)^{1/3} ~ .
\label{eq: nmaxd}
\end{equation}
When satellites orbit massive planets with \mpl\ $\gtrsim$ 1--300~\mearth, 
large-scale dynamical interactions require a few satellites with 
\rmax\ $\gtrsim$ 1000~km. For fixed \rmax, dynamical interactions are more
common around less massive planets.

In our suite of calculations, stirring of smaller satellites by the largest 
satellites is a relatively minor issue. For objects within the collisional
cascade, the initial collision velocities produce shattering. Larger relative 
velocities produce a little more debris and a somewhat faster decline in the
relative surface area of small particles. Larger objects are already immune
to the cascade; larger relative collision velocities have little impact on
their evolution.

\begin{figure} 
\includegraphics[width=6.5in]{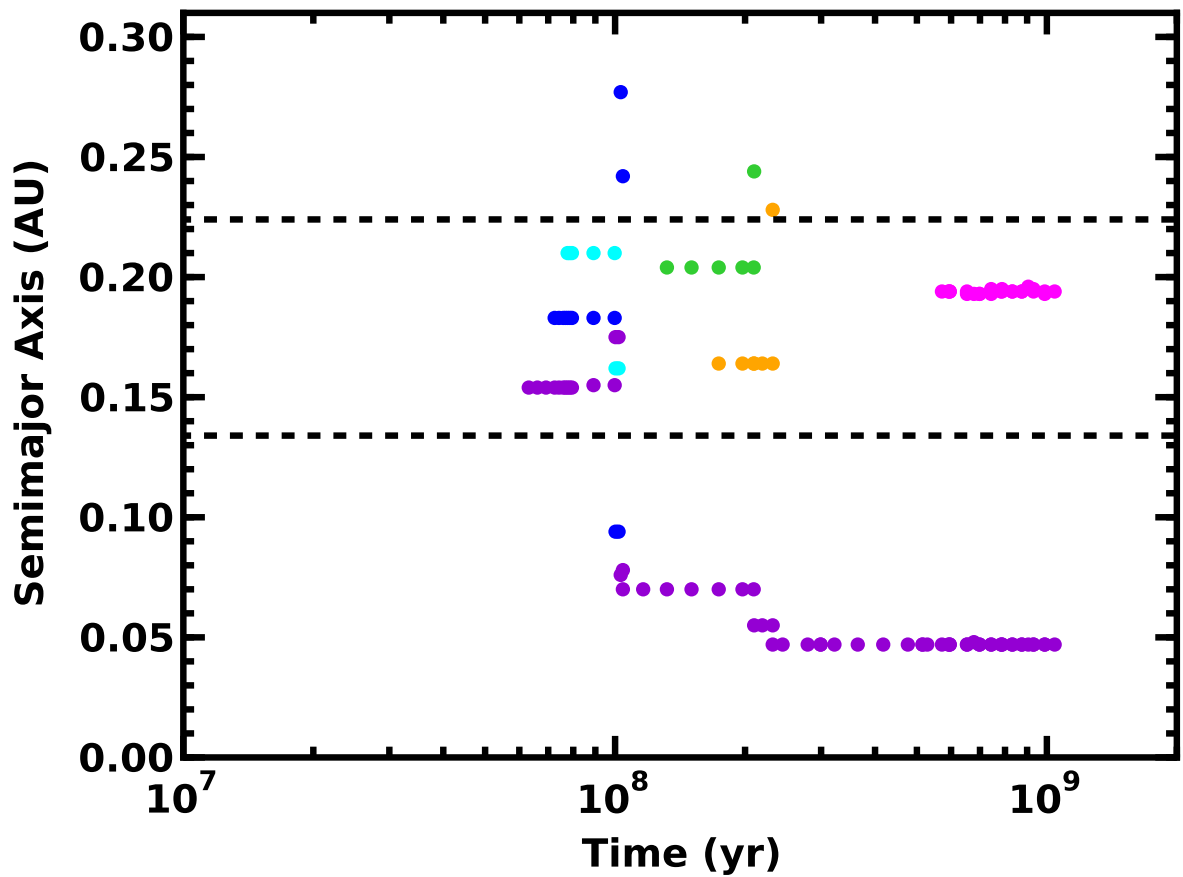}
\vskip 3ex
\caption{
Evolution of semimajor axis for massive ($n$-body) satellites 
orbiting a 1~\mearth\ planet. The dashed lines indicate the 
extent of the satellite swarm within the coagulation code.
Tracks for individual $n$-bodies
are coded by color. At 70--100~Myr, the coagulation code promotes 
three objects into the $n$-body code. After several minor 
encounters, strong interactions lead to a single object on an
$e$ = 0.6 orbit at smaller $a$ and the ejection of two $n$-bodies.
Somewhat later ($\sim$ 200~Myr), interactions between a second 
pair of $n$-bodies leads to a second set of ejections and a 
modest contraction of the orbit of the original $n$-body. At late
times, promotion of a sixth $n$-body leaves the system with a
single relatively low mass object on a fairly circular orbit at 
0.20~AU and a more massive satellite on an eccentric orbit at 0.05~AU.
\label{fig: nb1}
}
\end{figure}

Dynamical interactions among $n$-bodies are more important. As 
one example, Fig.~\ref{fig: nb1} tracks the time evolution of the semimajor
axes for a set of $n$-bodies orbiting a 1~\mearth\ planet. In this calculation,
the initial set of 400~km objects grows throughout the evolution of the satellite
swarm (Fig.~\ref{fig: rmax1}). At $\sim$ 60~Myr, the coagulation code promotes
the first satellite into the $n$-body code with a circular, but highly inclined,
orbit at 0.15~AU. Roughly 10~Myr later, a second $n$-body appears with an orbit 
at 0.18~AU. Within another 5~Myr, a third $n$-body has an orbit at 0.21~AU. 
At 100~Myr, the separations of these satellites are roughly 3.5 mutual Hill radii.
Strong dynamical interactions are inevitable. A scattering event between the
outer two satellites places one on a very close orbit with the inner satellite. 
All develop eccentric orbits. Eventually, the more massive inner satellite ejects
the other two satellites and ends up on an eccentric orbit much closer to the
planet.

In this suite of calculations, the dynamical evolution of the $n$-bodies has
no impact on satellites remaining in the coagulation code. A few larger objects
continue to grow. Promotion of two of these satellites into the $n$-body code
leads to another set of dynamical interactions at 200~Myr, where the two new
(and lower mass) $n$-bodies are ejected and the original massive $n$-body moves
a little closer to the planet. One last satellite promoted into the $n$-body 
code at $\sim$ 500~Myr orbits on a circular, highly inclined orbit well away from
the inner massive satellite. This system remains stable for the rest of the
calculation.

At the end of this calculation, roughly 33\% of the initial mass in solids
remains in orbit around the planet. Nearly all of this material is in the
two large satellites orbiting at 0.05~AU and 0.19~AU. Dynamical interactions 
(44\% $\pm$ 7\%) and radiation pressure (56\% $\pm$ 6\%) eject equal amounts 
of material. 
Despite this rough equality in mass, radiation pressure removes
100~\mum\ particles from the grid. Dynamical interactions place four 
Pluto-mass planets into orbits around the central star.

This evolution of large objects is fairly typical. Roughly half of the
simulations with growing $n$-bodies leave massive satellites orbiting 
the planet. Nearly 15\% of these have satellites orbiting at semimajor 
axes well inside the initial extent of the satellite swarm. Less than 
5\% have satellites outside the initial boundary of the swarm. In the
rest, 1--2 satellites orbit stably within the swarm.

\subsection{Comparing the Analytic and Numerical Models}
\label{sec: calcs-an}

Compared to the analytic model, the numerical simulations yield several clear
differences in the collisional evolution of a satellite swarm. 
The sizes of the largest objects change considerably in 0.1--1~Gyr. When 
catastrophic and cratering collisions dominate, \rmax\ declines by 30\%
or more. The largest objects then have roughly 30\% of the mass of the
largest objects in an analytic model where \rmax\ is constant in time.

In some systems, the largest satellites grow substantially. Left unchecked,
this growth yields massive objects capable of disrupting the satellite swarm
(and perhaps satellite systems closer to the planet). Swarms orbiting lower
mass planets are more prone to this evolution than swarms around more massive
planets. 

The size distribution does not follow a simple power law.
Although the numerical simulations 
roughly follow this power law for satellites with $r \gtrsim$ 0.1~km, all of
these models produce large (factor of 3--5) waves about the power law. At
small sizes, there is a large deficit in 0.1~cm to 10--30~m particles relative
to the analytic power-law. In calculations with an initial power-law size
distribution, smaller \rmin\ and larger $b_L$ reduce the size of the deficit.
Calculations starting with a mono-disperse size distribution also yield smaller
deficits.

The smallest particles with $r \lesssim$ 0.1~cm follow a steep power law with 
$q \approx$ 4--5. In our simulations, the lack of particles with $r < \rmin$
limits mass loss among larger particles. When these particles lose mass less
rapidly, they stay in the grid for longer periods of time, steepening the size
distribution. 

Despite these differences, the time evolution of the surface area in the 
baseline model (Fig.~\ref{fig: area2}) is similar to the time evolution 
of the analytic model (Fig.~\ref{fig: area1}). In the analytic model, 
satellite swarms orbiting 10~\mearth\ planets evolve through the middle
of the target. Swarms around 30~\mearth\ planets have a surface area in
the upper half of the target box. Swarms orbiting smaller (3~\mearth) or
larger (30~\mearth) planets have areas that graze or just miss the target.
In the baseline numerical model, satellites orbiting 10--30~\mearth\ planets
match the observations. Swarms around 3~\mearth\ or 100~\mearth\ planets
fall below or graze the upper edge of the target. Overall, the numerical
simulations require slightly more massive planets to match the observations
than the analytical models. 

\begin{figure} 
\includegraphics[width=6.5in]{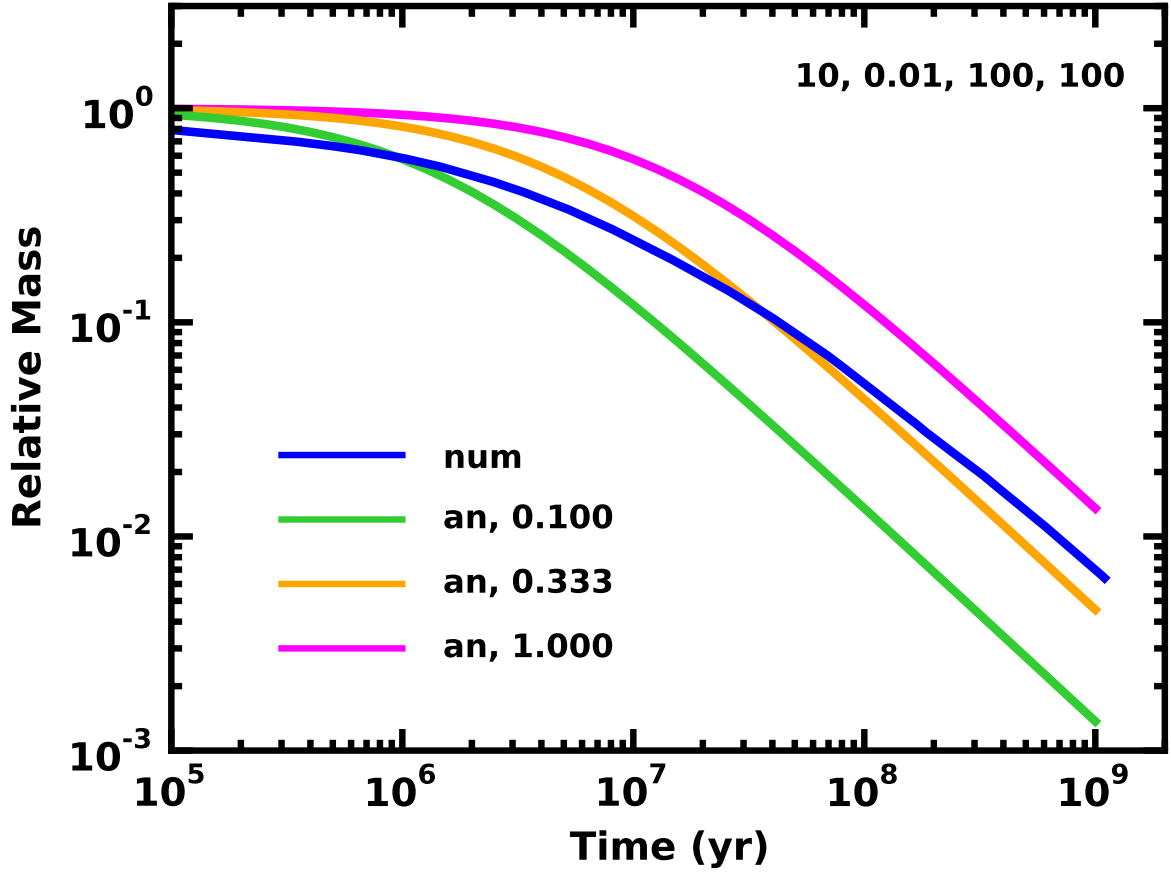}
\vskip 3ex
\caption{
Evolution of the mass of satellite swarms relative to their 
initial mass for the baseline numerical model ('num') and
analytic models ('an') with three values of $\alpha$ as listed
in the legend. As indicated in the upper right corner, the
baseline model has \mpl\ = 10~\mearth, \xcl\ = 0.01, 
\rmax\ = 100~km, and \rmin\ = 100~\mum. The evolution of an
analytic model with $\alpha \approx$ 1 provides a reasonable 
match to the numerical model.
\label{fig: mass1}
}
\end{figure}

Fig.~\ref{fig: mass1} compares the evolution of the mass in a baseline model 
with the predictions of the analytic model using three different values for 
the correction factor $\alpha$. For \mpl\ = 10~\mearth, \xcl\ = 0.01, 
\rmax\ = 100~km, and \rmin\ = 100~\mum, the mass in the numerical model 
begins to decline at 0.1~Myr. It takes 40--50~Myr for the mass to reach 10\% 
of the initial mass and 700--800~Myr to reach 1\% of the initial mass. 
At late times, the mass evolves with time as $M(t) \propto t^{-n}$ and 
$n \approx$ 0.8--0.9. Thus, the mass declines somewhat less rapidly than 
in the analytic model.

In the analytic model, the mass loss rate is initially smaller than in the
numerical model. As time proceeds, analytic models with $\alpha$ = 0.333 (3.333)
decline more rapidly (slowly) than the numerical model. When $\alpha$ = 1, 
the analytic and numerical models match at 20--30~Myr. After this time, the 
$t^{-1}$ decline of the analytic model results in a faster rate of mass loss 
than the numerical model.

For comparisons between the analytic model and the complete suite of numerical 
simulations, the `best' $\alpha$ depends on the input parameters. For all models 
with the baseline set of parameters, $\alpha \approx$ 1. Changing \xcl, \rmin,
\qdstar, and the initial size distribution has little impact on $\alpha$.  When 
$b_L \approx$ 1, the faster removal of material in the grid leads to more rapid
evolution and smaller $\alpha \approx$ 1/2 to 1.

\section{DISCUSSION}
\label{sec: disc}

Our suite of simulations paints an interesting picture for the evolution
of swarms of satellites orbiting 1--300~\mearth\ planets at 120~AU from 
a central 1.9~\msun\ star. Depending on 
\mpl, \xcl, and \rmax, the masses of the largest satellites grow or shrink 
with time. Satellites with large \qdstar\ grow; satellites with small
\qdstar\ shrink.  Clouds with smaller \xcl\ and larger \rmax\ evolve more
slowly.  Outcomes are insensitive to \rmin, the shape of the initial size 
distribution, or the algorithm for distributing debris among the mass bins.

However the largest objects evolve, the cumulative size distribution 
transforms into a standard shape which is fairly independent of the 
input parameters. This standard shape has three distinct pieces:
(i) a steep power law with $n(>r) \propto r^{-q_c}$ and $q_c \approx$ 4 
($q_r \approx$ 2) for small particles ($r \lesssim$ 10--30~cm),
(ii) a flat portion where $q_c \approx 0--1$ ($q_r \approx -2$ to $-1$)
for intermediate size particles ($r \approx$ 30~cm to 0.1~km), and 
(iii) a shallow, wavy power law with $q_c \approx$ 1--2 
($q_r \approx -1$ to 0) for large particles ($r \gtrsim$ 0.1~km). 
The development of this standard shape depends on the collision time:
swarms with longer collision times take longer to establish this size
distribution.

\begin{figure} 
\includegraphics[width=6.5in]{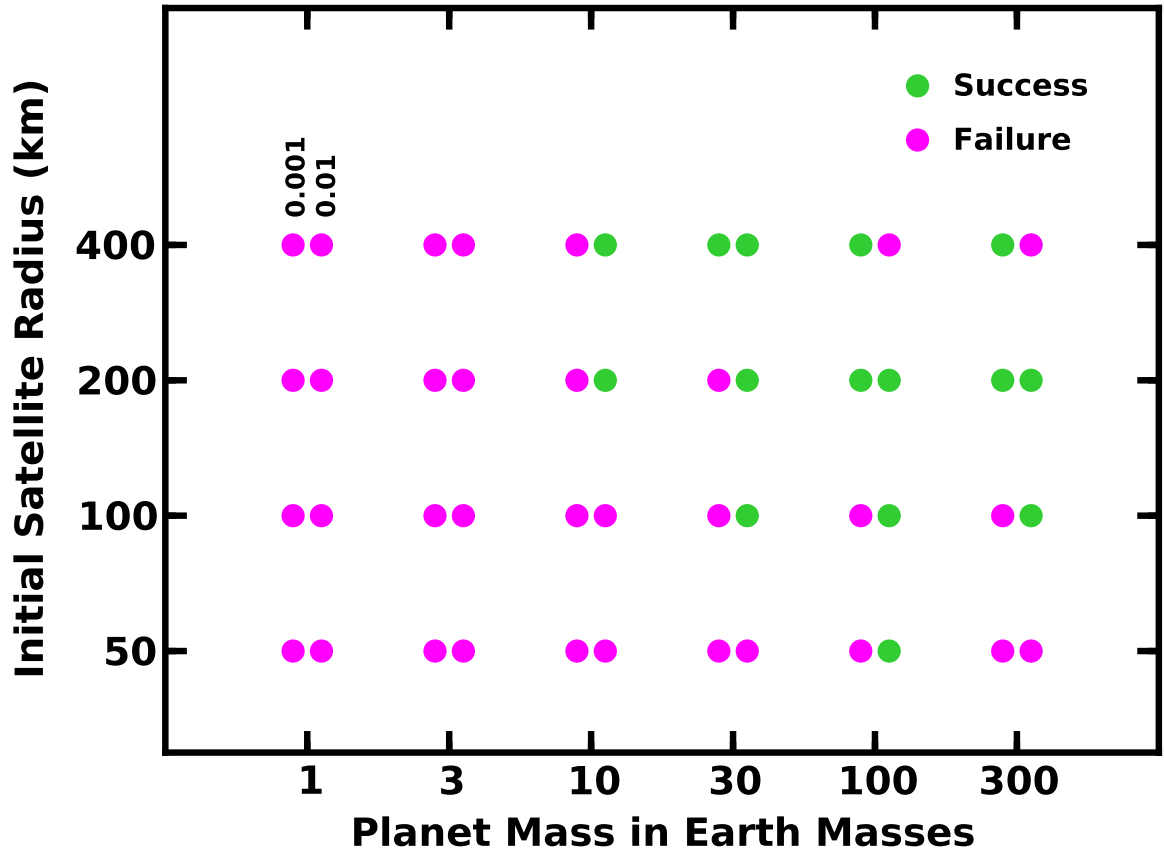}
\vskip 3ex
\caption{
Grid of outcomes for satellites swarms in Fomalhaut b. 
For initial relative cloud mass \xcl\ = 0.001 and 0.01
(as indicated above the first column of points), green 
(red) points indicate models which match (do not match) 
the observed surface area of Fomalhaut b at 100--400~Myr.
\label{fig: succ}
}
\end{figure}

Despite the diverse outcomes, a broad set of satellite swarms produce a
cross-sectional area which matches the observed area in Fomalhaut b.
For models with the nominal \rmin, \qdstar, and $b_L$, Fig.~\ref{fig: succ} 
summarizes these outcomes a function of \mpl, \rmax, and \xcl.  Swarms 
orbiting low mass planets always fail. Satellites around more massive 
planets are often successful. Successful models have a factor of 2--3 
range in the initial mass in satellites relative to a `best' model with
an initial mass of $\sim 10^{27}$~g. 

Aside from the initial mass, the size of the smallest stable particle 
orbiting the planet and the mass distribution of debris from a collision
establish the ability of a model to match the observed cross-sectional
area. If particles with \rmin\ = 10~\mum\ can stably orbit the planet,
smaller initial cloud masses are possible. Larger \rmin\ and $b_L >$ 0 
reduce the grid of successful models.

\subsection{Theoretical Issues}
\label{sec: disc-theory}

In these coagulation calculations, the standard size distribution has
several features in common with results for debris disks orbiting main
sequence stars. There are also a few major differences. In addition,
several uncertain parameters establish whether model satellite swarms 
have cross-sectional areas at 100--400~Myr comparable to the observed 
area in Fomalhaut b. Assigning different values for \rmin, \qdstar, and 
$b_L$ allows swarms with different combinations of \mpl, \xcl, and 
\rmax\ to match the observations.  Here, we discuss features of the
size distribution and consider the flexibility of the theory in setting 
the various input parameters and the likely consequences of our choices.

\subsubsection{Size Distribution}
\label{sec: disc-theory-sd}

\begin{figure} 
\includegraphics[width=6.5in]{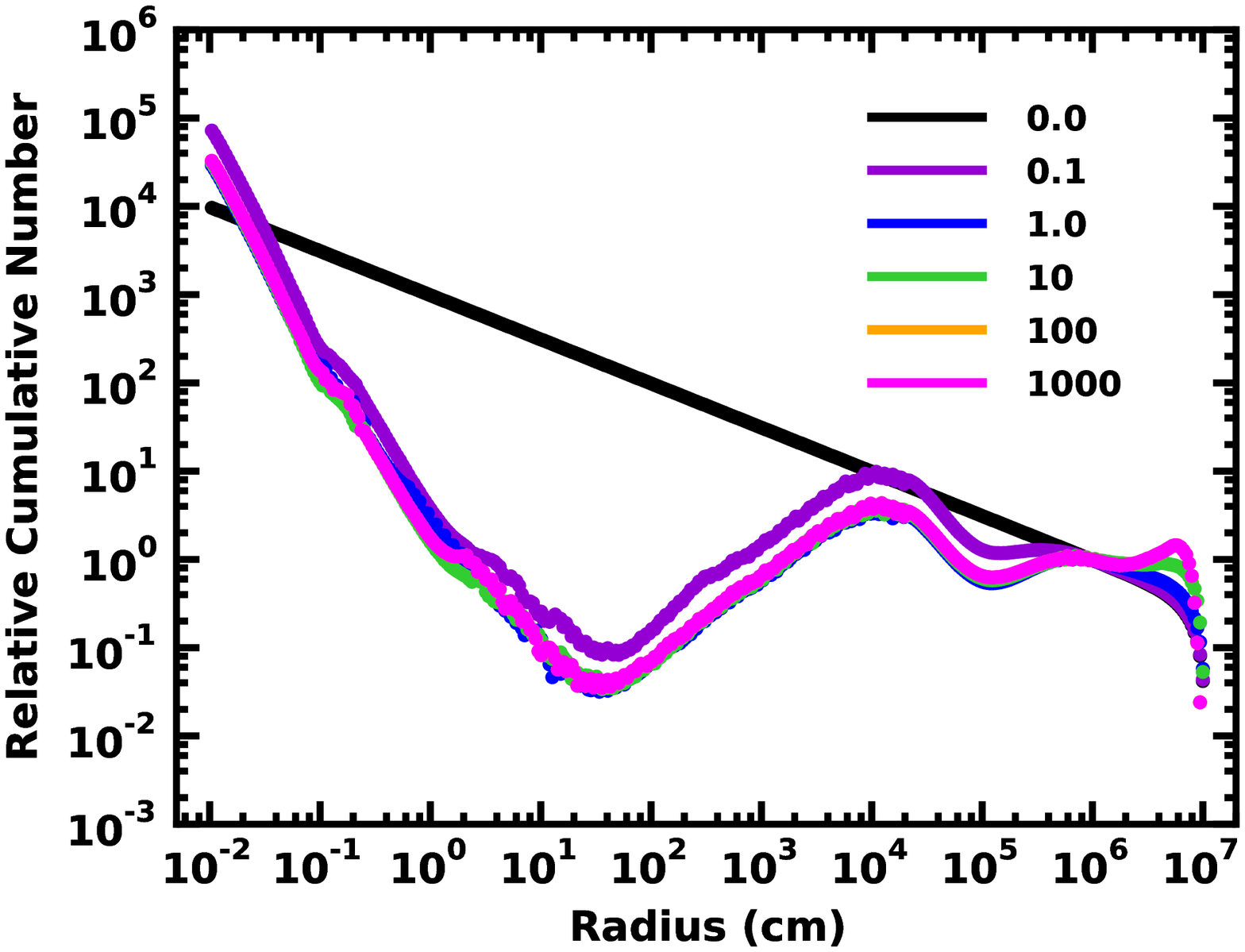}
\vskip 3ex
\caption{
Snapshots evolution of the relative cumulative size distribution for 
calculations with \mp\ = 10~\mearth, \xcl\ = 0.01, \rmax\ = 100~km,
\rmin\ = 100~\mum, \mlz\ = 0.2, 
and \bl\ = 0.0.  The legend in the upper right indicates the evolution
time in Myr. Within roughly 1~Myr, collisions produce several distinct 
features
in the relative size distribution: (i) a steep rise at the largest sizes 
($ r \gtrsim$ 50~km), (ii) a shallower rise which approximately follows
the original power law, (iii) a sharp drop at intermediate sizes 
($r \approx$ 30~cm to 0.1~km), and (iv) a steep rise at the smallest
sizes ($r \approx$ 0.1~mm to 30~cm).
\label{fig: sd1}
}
\end{figure}

\begin{figure} 
\includegraphics[width=6.5in]{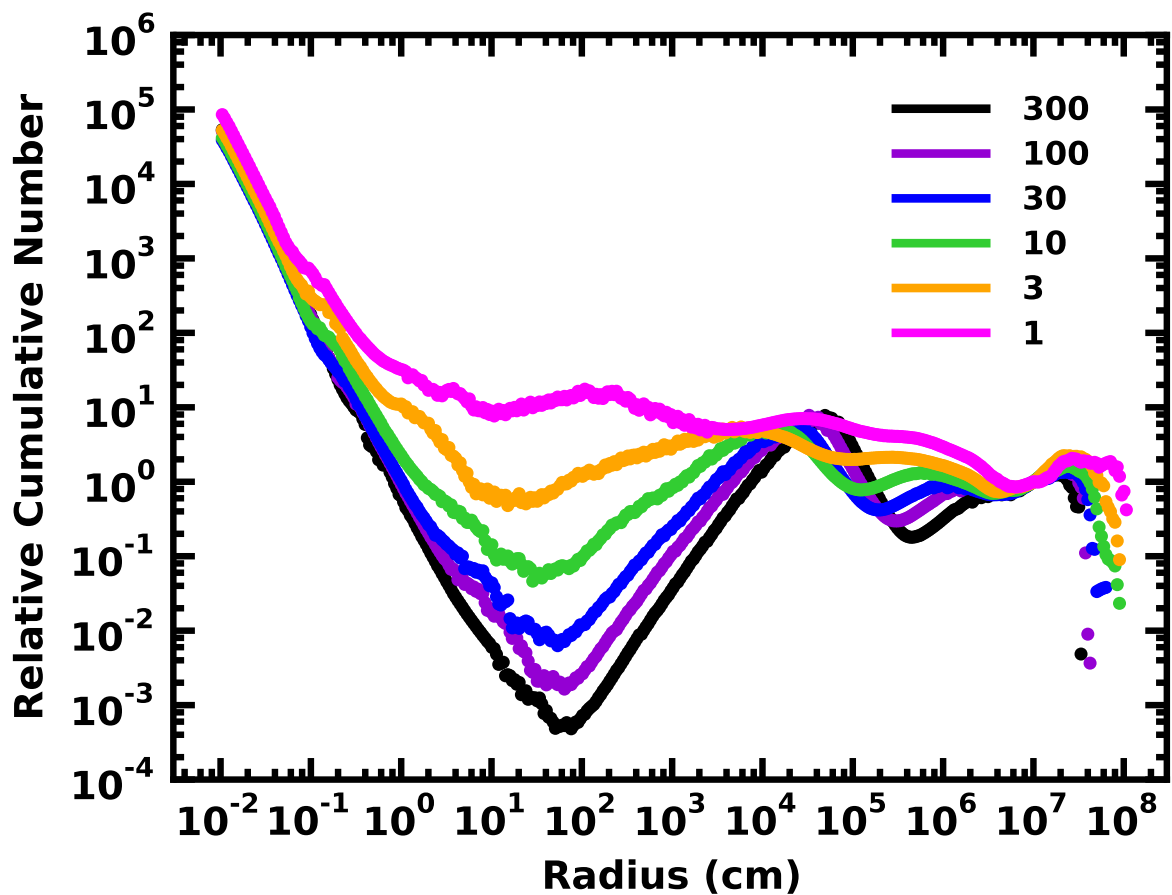}
\vskip 3ex
\caption{
As in Fig.~\ref{fig: sd1} for a range of masses for the central planet
at $t$ = 100~Myr.  The legend in the upper right indicates the mass of
the planet in \mearth. For all \mp, the size distribution is very steep
for particle sizes $r \approx$ 0.1~mm to 1--30~cm. The depth of the 
minimum at 10--30~cm grows with the mass of the planet. For $r \gtrsim$
0.1~km, fluctuations about the original power law grow with the mass
of the planet.
\label{fig: sd2}
}
\end{figure}

\begin{figure} 
\includegraphics[width=6.5in]{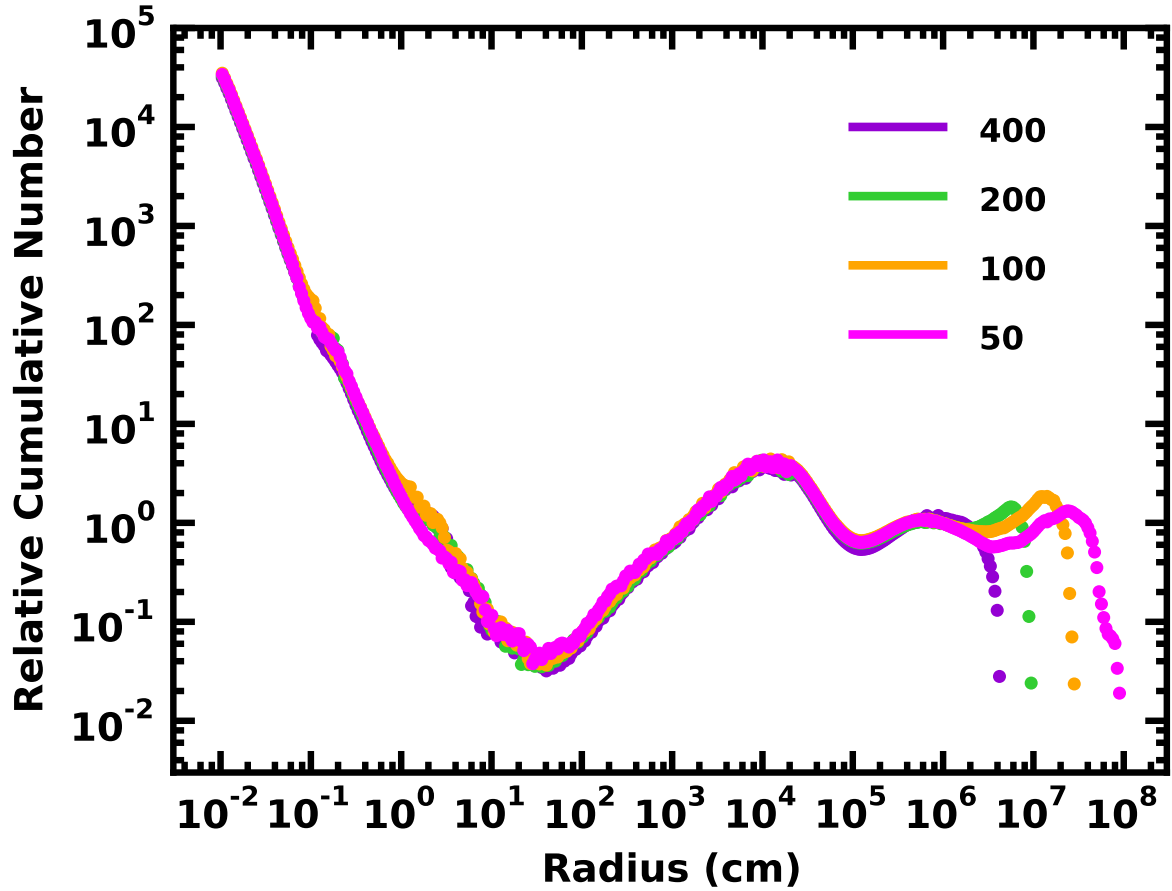}
\vskip 3ex
\caption{
As in Fig.~\ref{fig: sd2} for \mpl\ = 10~\mearth\ and various initial 
\rmax\ as indicated in the legend. Aside from differences at 
$r \approx$ \rmax, the relative size distribution at 100~Myr is
independent of initial \rmax.
\label{fig: sd3}
}
\end{figure}

\begin{figure} 
\includegraphics[width=6.5in]{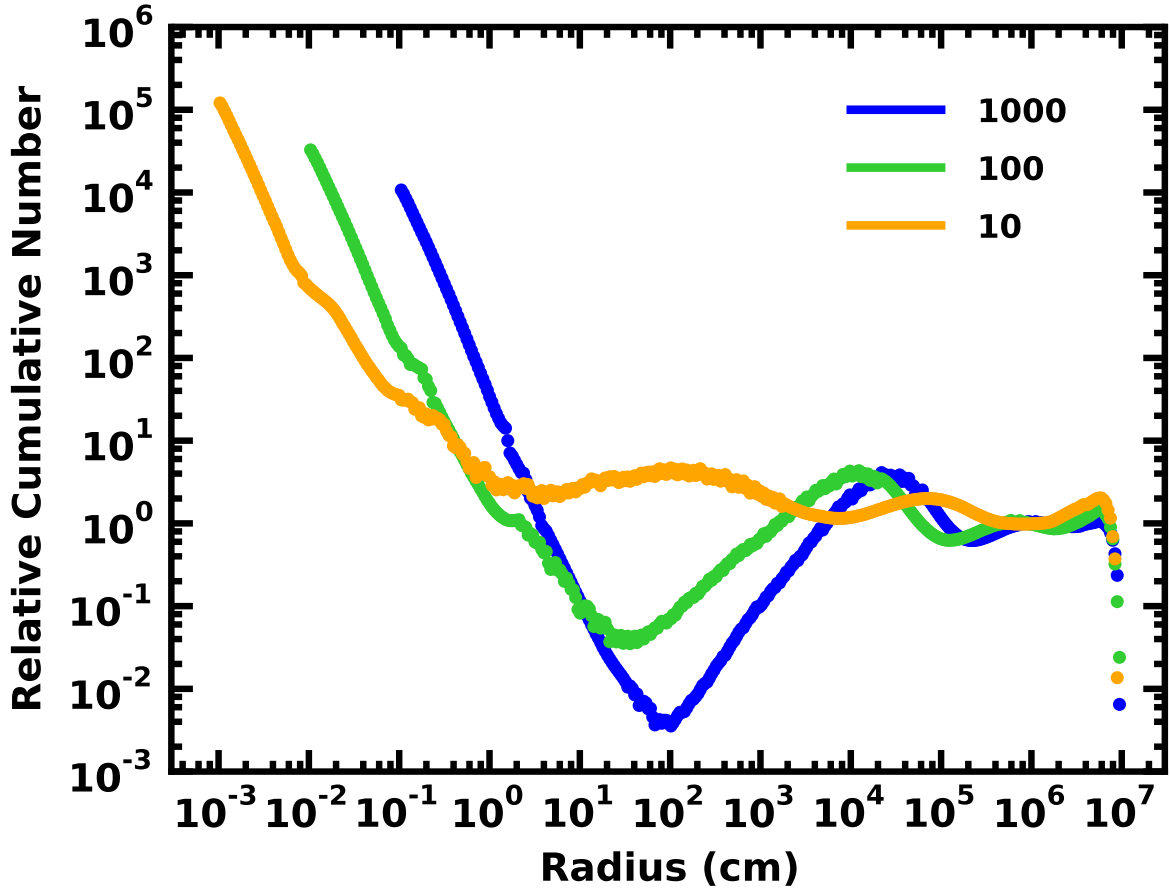}
\vskip 3ex
\caption{
As in Fig.~\ref{fig: sd3} for \rmax\ = 100~km and various
\rmin\ as indicated in the legend. When \rmin\ is smaller, 
the relative size distribution is closer to a single power
law for $r \gtrsim$ 1~cm and has a smaller deficit of
particles at 10--1000~cm.
\label{fig: sd4}
}
\end{figure}

\begin{figure} 
\includegraphics[width=6.5in]{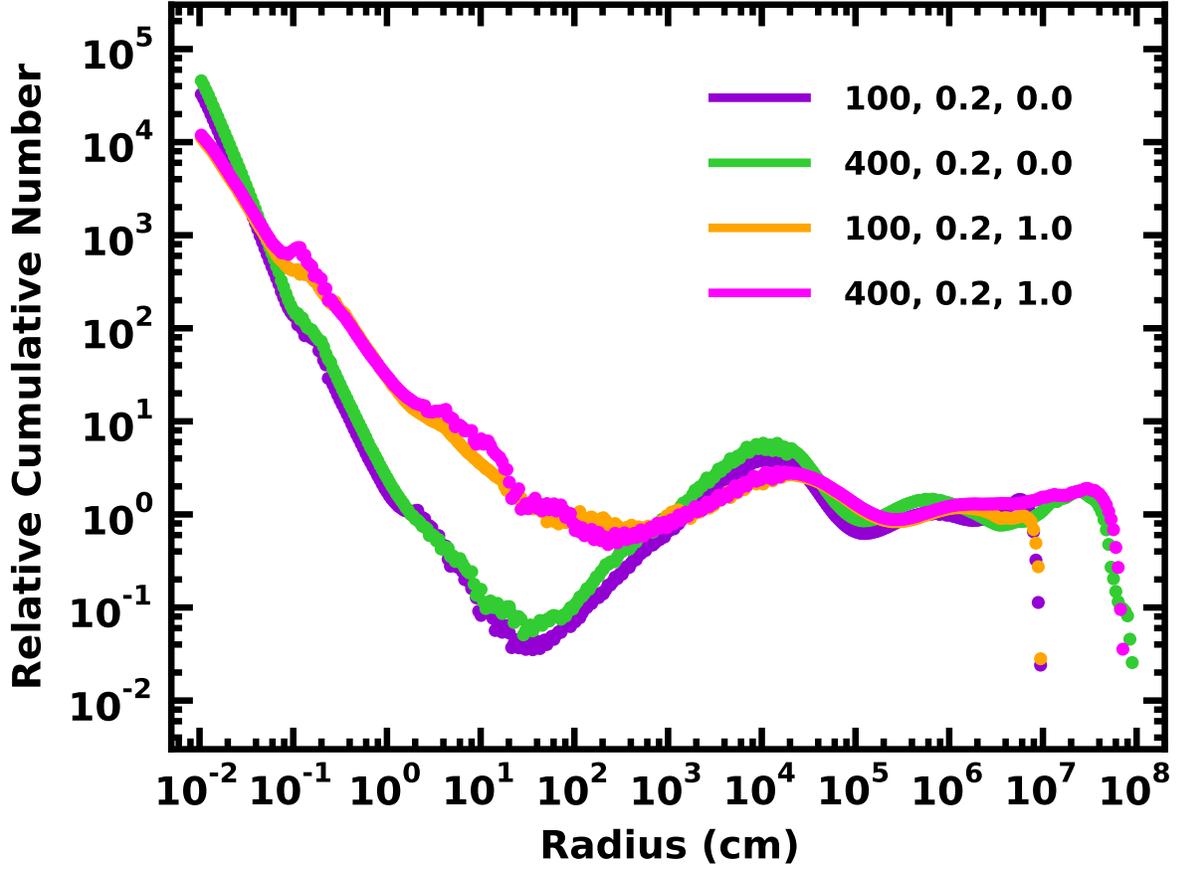}
\vskip 3ex
\caption{
As in Fig.~\ref{fig: sd2} for \mpl\ = 10~\mearth\ and various
combinations of \rmax, $m_{L,0}$, \ and $b_L$ as indicated in the 
legend.  For $r \gtrsim 10^3$~cm, the relative size distribution is
fairly independent of $b_L$. Among smaller particles, calculations
with $b_L$ = 1 yield a shallower size distribution than those 
with $b_L$ = 0.
\label{fig: sd5}
}
\end{figure}

\begin{figure} 
\includegraphics[width=6.5in]{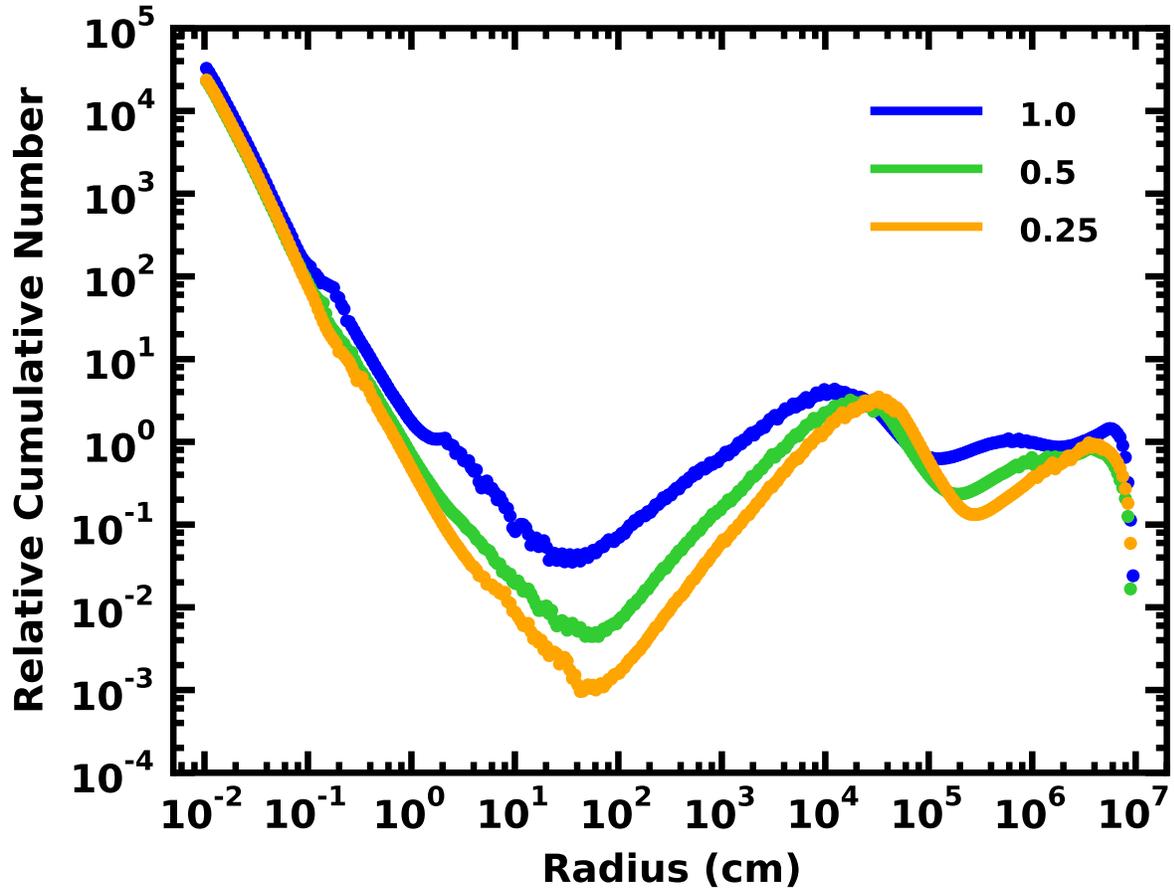}
\vskip 3ex
\caption{
As in Fig.~\ref{fig: sd3} for \rmax\ = 100~km, \rmin\ = 
100~\mum, and different values for \qdstar. The legend indicates 
the value of \qdstar\ relative to the nominal fragmentation 
parameters listed in the main text.
\label{fig: sd6}
}
\end{figure}

\begin{figure} 
\includegraphics[width=6.5in]{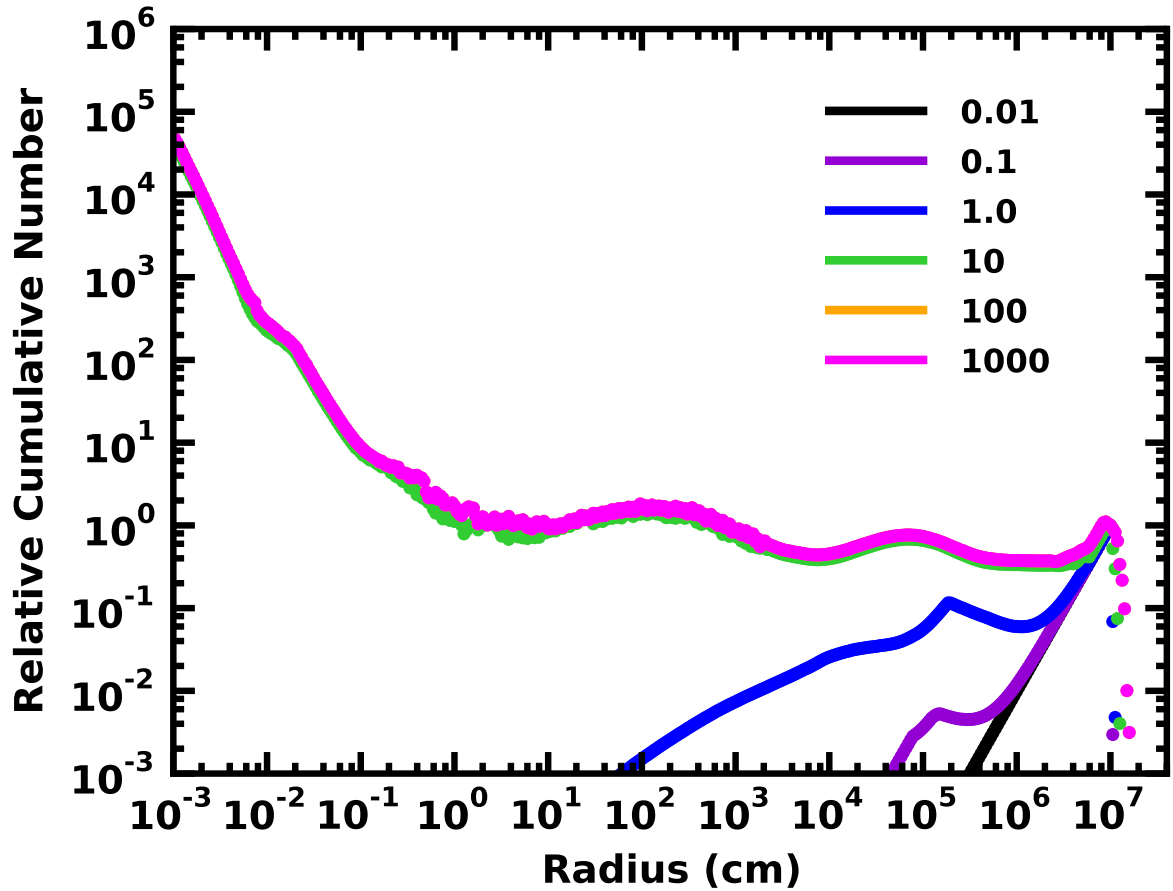}
\vskip 3ex
\caption{
As in Fig.~\ref{fig: sd1} for calculations starting with a 
mono-disperse set of particles. The legend indicates the
evolution time in Myr.  At early times, collisions 
among 100~km objects produce a small amount of debris
populating the small size end of the size distribution. 
After 10~Myr, debris from these collisions and the debris 
from collisions of smaller objects yields a smooth 
equilibrium size distribution from 10~\mum\ to roughly 100~km.
\label{fig: sd7}
}
\end{figure}

In Figs.~\ref{fig: sd1}--\ref{fig: sd8}, the size distributions derived
in our calculations are radically different from the smooth power law,
$n(r) \propto r^{-q}$ with $q \approx$ 3.5--3.7, expected from an
equilibrium collisional cascade \citep{dohn1969,obrien2003,koba2010a}.
The general trend of $n(r)$ is much shallower than this power law. 
Pronounced waves are superposed on this general trend.

Wavy size distributions are a common feature in numerical calculations 
of debris disks orbiting 1--3~\msun\ stars \citep[e.g.,][]{campo1994,
obrien2003,theb2003,kriv2006,lohne2008,gaspar2012b,kral2013}. These waves 
have two typical sources: the low mass cutoff of the size distribution and
the transition between the bulk strength and gravity regimes in analytic 
expressions for \qdstar\ (see eq.~\ref{eq: qd}). Although the amplitudes 
(in $n(r)$) of the waves depend on the radius of the low mass cutoff, 
the ratio $Q_c / \qdstar$, and the parameters in the relation for 
\qdstar, typical values are a factor of 10 or smaller.

By typical debris disk standards, the waves in 
Figs.~\ref{fig: sd1}--\ref{fig: sd8} are somewhat extreme. For our 
calculations, the positions and relative spacing of minima and maxima 
follow general predictions from analytic models \citep[e.g.,][]{obrien2003}. 
Factor of 3--10 amplitudes at 0.1--100~km are also normal. However, the 
amplitudes of waves at small sizes are 10--1000 times larger than those 
reported from numerical simulations of debris disks. 

The long-term collisional evolution of satellite swarms is responsible
for this difference.  Compared to debris disks around stars, satellite
swarms orbiting planets are much more collisionally evolved 
\citep[e.g.,][]{bottke2010}. More collisional evolution enhances the
excess of particles with $r \approx$ 1--10~\rmin\ and the deficit of particles
with $r \approx$ 10--100~\rmin. As a result, the amplitude of the wave simply
grows larger and larger with time. Fig.~\ref{fig: sd2} clearly shows the 
impact of longer collisional evolution: massive systems with shorter 
collision times have much stronger waves than low mass systems with
longer collision times.

\subsubsection{The minimum particle size}
\label{sec: disc-theory-rmin}

In the baseline model, \rmin\ is the blowout radius $r_b$ for a
10~\mearth\ planet.  The slow variation of $r_b$ with \mpl\ justifies this 
assumption for all \mpl\ (eq. \ref{eq: rblow}). However, radiation pressure 
cannot remove small particles when (i) the collision time is comparable
to or shorter than the orbital period around the planet and (ii) the
optical depth of the cloud is one or larger. If satellite swarms meet 
either of these conditions, smaller particles stably orbit the planet.

The optical depth $\tau$ of the cloud is roughly the ratio of $A_d$ to 
the cross-sectional area defined by the physical extent of the swarm
$A_s = \pi (\eta_1 R_H)^2$. Setting $\tau = A_d / A_s$ and adopting
the nominal parameters, 
\begin{equation}
\tau \approx 9 \times 10^{-4} 
\left ( { \eta_1 \over 0.2 } \right )^{-2}
\left ( { \mp \over 10 ~ \mearth } \right )^{-2/3}
\left ( { \mstar \over 2 ~ \msun } \right )^{2/3}
\left ( { a_p \over {\rm 120~AU} } \right )^{-2}
\left ( { A_d \over {\rm 10^{23}~cm^2} } \right ) ~ .
\label{eq: ratio1}
\end{equation}
For planets with \mpl\ $\lesssim$ 30~\mearth, satellite swarms have 
$A_d \lesssim 10^{25}$~cm$^2$ throughout their evolution. These
swarms are never optically thick.  Early in the evolution of swarms 
orbiting more massive planets, $A_d \gtrsim 10^{26}$~cm$^2$; these 
swarms are optically thick.

For a single small particle, it is straightforward to derive the ratio 
$\xi$ of the collision time to the orbital period. The collision
time is roughly $t_c \approx V / v A_d $; the orbital period is
$T = 2 \pi a / v$. The ratio is then:
\begin{equation}
\xi \approx 1.85 \times 10^{2} 
\left ( { \eta_1 \over 0.2 } \right )^2
\left ( { \eta_2 \over 0.5 } \right )
\left ( { \mp \over 10 ~ \mearth } \right )^{2/3}
\left ( { \mstar \over 2 ~ \msun } \right )^{-2/3}
\left ( { a_p \over {\rm 120~AU} } \right )^2 
\left ( { A_d \over {\rm 10^{23}~cm^2} } \right )^{-1} ~ .
\label{eq: ratio2}
\end{equation}
Around low mass planets with \mpl\ $\lesssim$ 10~\mearth, the initial 
cross-sectional area of the swarm is $\lesssim 10^{24}$~cm$^2$. 
Radiation pressure removes small particles faster than collisions.
Among more massive planets, $A_d \gtrsim 10^{26}$~cm$^2$ for 
$t \lesssim$ 1~Myr. High speed collisions destroy small particles 
faster than radiation pressure removes them. The smallest particles
in the swarm are then much smaller than 100~\mum. Although the
cross-sectional area of these swarms is then formally very large, 
the observed area is limited by the optical depth. With 
$\tau \approx$ 1, the maximum area is roughly $10^{26}$~cm$^2$. 

This discussion implies that \rmin\ is rarely smaller than the nominal
blowout size $r_b$. Early in the evolution of satellite swarms around 
massive planets, particles with sizes smaller than 10--100~\mum\ remain
bound. This phase is short-lived, $\sim$ 1~Myr. As these systems evolve,
$A_d$ declines rapidly as \rmin\ returns to its nominal value. The
evolution then continues as outlined in \S\ref{sec: calcs-area}.

\subsubsection{The particle strength}
\label{sec: disc-theory-qd}

The binding energy of solid particles establishes collision outcomes.
When the collision energy \qc\ exceeds the binding energy \qdstar, 
more than half of the mass of the colliding pair of particles ends up 
in debris. In circumstellar disks, collisional damping and gravitational 
interactions between particles often limit the impact of \qdstar\ on
the evolution \citep[e.g.,][]{kb2008,kb2010}. For satellite swarms
within a spherical shell, damping and gravitational interactions are
minimal. With \qc\ solely a function of $\eta_1$ and \mpl, the growth 
of the largest particles is a strong function of \qdstar. 

For icy objects with sizes much larger than 1~cm, astronomical observations, 
laboratory experiments, and numerical simulations paint a disparate 
picture for \qdstar. Recent experiments colliding cm-sized icy solids 
in the lab suggest tensile strengths of roughly 
$10^6$~\ergg\ \citep[e.g.,][] {shimaki2012,yasui2014}, which agrees 
with previous results \citep[e.g.,][]{ryan1999}. Numerical simulations
of high speed collisions between icy objects are the basis for the 
expression in eq.~\ref{eq: qd} \citep{benz1999,lein2009}. Typically,
$Q_b \approx$ $10^5 - 10^8$ erg~cm$^{0.4}$~g$^{-1}$ and $Q_g$ = 
0.1--2~erg~cm$^{1.65}$~g$^{-1}$. For $r \approx$ 1--10~cm,
the simulations suggest a binding energy of roughly $10^6$~\ergg. 
Experiments and simulations agree on the strength for small objects.

Observations of comets yield much smaller binding energies. Models for 
comet D3/1993 F2 (Shoemaker-Levy) and other disrupted comets suggest 
binding energies of 1--$10^3$~\ergg\ \citep[e.g.,][]{asp1996,rich2007,
skorov2012}. Data from Deep Impact imply a strength in the middle of
this range \citep[e.g.,][]{hols2007,ahearn2011}. With $r \approx$
0.1--1~km for the nuclei of these comets, the maximum strength of
$10^3$~\ergg\ is a factor of 10 or more smaller than expected from
eq.~\ref{eq: qd} and the results from laboratory and numerical experiments.

Our adopted values for the parameters in \qdstar\ lie intermediate between 
observations and numerical simulations. Values for \qdstar\ similar to
results from the studies of comets preclude the growth of large 
satellites around planets with \mpl\ = 1--30~\mearth\ (see 
Fig.~\ref{fig: rmax1}). More material then participates in the 
collisional cascade. Although evolution times are somewhat shorter, 
the evolution of the cross-sectional area is unchanged. Thus, 
significantly smaller \qdstar\ does not change our conclusions.

Larger values for \qdstar\ enhance the growth of large satellites 
around all planets. For sufficiently large \qdstar\ as in \citet{benz1999},
the collisional cascade is limited. Few satellite swarms have sufficient
surface area to match observations of Fomalhaut b. In these systems, 
the largest satellites probably grow large enough to disrupt the 
satellite swarm completely (see \S\ref{sec: calcs-nb}). Then, all
models fail: none match observations of Fomalhaut b.

\subsubsection{The size of the largest particle in the debris}

Theory currently provides limited guidance on \mmaxd, the mass 
of the largest particles in a cloud of debris ejected during a high 
velocity collision.  For cratering collisions, $m_{esc}$ is a simple 
function of the collision energy and the gravity of the planet 
\citep[e.g.,][]{housen2011,svet2011}.  However, there are no direct 
calculations of \mmaxd.  \citet{weth1993} examined laboratory 
data and adopted the simple relation $\mmaxd = 0.2 m_{esc}$ used 
in our baseline model. Recent experiments confirm this choice
\citep[e.g.,][and references therein]{burchell2005,poelchau2014}.

For catastrophic impacts, numerical simulations provide somewhat 
conflicting advice for \mmaxd.  In \citet{lein2012}, collisions 
of 1--10~km icy objects yield a broad range,
\mmaxd/\mesc\ $\approx$ 0.001--1. Power-law fits to the distribution 
of debris particles require assigning either a lower bound to the 
size of a debris particle or a slope to the size distribution. 
Choosing the slope leads to a fixed value \mmaxd/\mesc\ $\approx$ 
0.026 independent of \qdstar. 

\citet{durda2004,durda2007} describe numerical simulations of collisions
for 10--100~km rocky objects. In these simulations, the mass of the 
largest object within the debris is 1--2 orders of magnitude smaller 
and very sensitive to the ratio of \qc\ to 
\qdstar\ \citep{morby2009b,bottke2010}. These calculations predict
much steeper size distributions than those of \citet{lein2012}.
However, including these results in coagulation codes requires 
adopting a shallow slope for small sizes to conserve mass which
introduces additional input parameters.

For high velocity collisions of small objects, laboratory experiments 
suggest the mass in the largest debris particle scales roughly inversely 
with the ratio of the collision energy to the binding energy 
\citep[e.g.,][]{arakawa1999,arakawa2002,shimaki2012}. Available data 
imply larger particle sizes for collisions between more porous and
stronger targets.

Our approach to placing debris in mass bins roughly follows the spirit 
of \citet{lein2012}. In our baseline model, debris from cratering 
collisions agrees with experimental results; the mass of the largest
object in catastrophic collisions is roughly 10 \citep{lein2012} to 
100 \citep{bottke2010} times larger than inferred from numerical 
simulations. Compared to the predictions of numerical collision codes, 
these calculations probably underestimate the rate of decline for the 
cross-sectional area around massive planets. In models with $b_L = 1$, 
catastrophic collisions yield results more similar to \citet{lein2012}.
Although this treatment of cratering collisions leaves too much mass in 
large objects, most collisions are catastrophic. Thus, the cratering
algorithm has little impact on our results.

Adopting the \citet{bottke2010} treatment of \mmaxd\ speeds up the 
collisional cascade. When the debris evolves more rapidly, swarms
orbiting 10--30~\mearth\ planets cannot match the observed surface 
area of Fomalhaut b for ages of 100--400~Myr. However, more rapid
evolution for swarms around 100--300~\mearth\ planets allows these
systems to match the observations.

\subsection{Predictions for Fomalhaut b and Other Exoplanetary Systems}
\label{sec: disc-fb}

Two aspects of our calculations allow tests from existing observations
of Fomalhaut b or new observations of other debris disks. All collisional 
cascade models predict a mass loss rate from the production of particles
with sizes less than the size of the smallest stably orbiting particle
\citep[e.g.,][]{koba2010a}. Most of these particles should lie close to
the orbit of the planet around the central star \citep[e.g.,][]{kcb2014}.
Numerical results for this mass loss rate at 100--400~Myr yield an expected 
surface brightness along the path of Fomalhaut b. For younger systems, the
mass loss rate and the cross-sectional area of the swarm are much larger.
For sufficiently large $A_d$, satellite swarms are detectable around stars
with ages of 1--10~Myr.

To quantify our first prediction, we consider baseline models of satellite
swarms orbiting 10--100~\mearth\ planets. At 100--200~Myr, the mass loss 
rate in small particles is $0.6-3 \times 10^{18}$~\gyr. For particle sizes 
of 100~\mum, mass loss leaves behind a trail with a cross-sectional area 
of roughly $10^{20}$~cm$^2$ every year. Fomalhaut b has an orbital period 
of roughly 1000~yr. Every orbit, mass loss produces a ring of material with 
a cross-sectional area comparable to the observed $A_d$ of Fomalhaut b.

The total surface area of this ring depends on the long-term evolution of 
small particles. If the particles have a velocity dispersion similar to 
their escape velocity from the planet, they have orbits with eccentricity
$e \approx$ 0.1 around Fomalhaut.  Interactions with Fomalhaut b are 
probably rare.  With inclinations $i \approx e/2$, the collision time for
a single 100~\mum\ particle is roughly 1~Gyr for $A_d \approx 10^{23}$ and 
1~Myr for $A_d \approx 10^{26}$~cm$^2$. With roughly 1~Myr required to eject 
particles with $A_d \approx 10^{26}$~cm$^2$, we envision an approximate 
steady-state where ejections of 100~\mum\ particles from Fomalhaut b roughly
balance particles lost from destructive collisions.

To estimate the surface brightness of this ring, we consider bound orbits 
along a ring with semimajor axis $a \approx$ 120~AU, width $\delta a \approx$
0.1 $a$, and $e \approx$ 0.8 \citep[e.g.,][]{beust2014}. Along this ring,
there are roughly $10^4$ resolution elements on HST images \citep{kcb2014}.
With $A_d \approx 10^{26}$~cm$^2$, each resolution element has a 
cross-sectional area of $10^{22}$~cm$^2$ in 100~\mum\ particles. 

If the orbit of Fomalhaut b is stable on Myr time scales, tracing a ring of 
dust along this orbit is challenging \citep[e.g.,][]{currie2012,galich2013,
kalas2013}. However, comparing the average surface brightness of coadded 
pixels along the orbit with similarly coadded pixels 20--30~AU away should 
yield a clear measure of the surface brightness along the ring and a strong 
test of the model.

To make predictions for the brightness of satellite swarms orbiting 
any star, we consider $f_o$ the observed flux of the swarm relative 
to $f_{\star}$ the observed flux from the central star \citep[see also 
\S2.3 of][]{kcb2014}.  For a swarm with optical depth $\tau$ and 
scattering efficiency $Q_s$, $f_o/f_\star = Q_s A_d / 4 \pi a_p^2$. 
We set $A_d = \tau A_s = \tau \pi (\eta_1 r_H a_p)^2$ with $\eta_1$ = 
0.2.  Defining $\Delta m = -2.5~{\rm log}~(f_o/f_\star)$, the predicted 
contrast between the satellite swarm and the central star is
\begin{equation}
\Delta m \approx 15.83 - 2.5~{\rm log}~\tau - 
5~{\rm log}~\left ( { \eta_1 \over 0.2 } \right ) -
1.67 ~ {\rm log}~\left ( { \mp \over 10~\mearth } \right ) +
1.67 ~ {\rm log}~\left ( { \mstar \over 1~\msun } \right ) ~ .
\label{eq: contrast}
\end{equation}
Although the contrast depends on $\tau$, $\eta_1$, \mp, and \mstar, 
it is formally independent of the semimajor axis of the planet and 
the distance to the star. However, swarms with $\tau \approx$ 0.1--1
have longer lifetimes at larger $a$; thus, observations are more
likely to detect bright swarms at 100~AU than at 10~AU.  

Among nearby associations of young stars, the $\beta$ Pic moving 
group provides the best testing ground for this prediction. With
$\sim$ 30 members having V $\approx$ 4--9, $d \approx$ 20~pc, 
and ages $\sim$ 20~Myr \citep{zuck2001,mama2014}, this association 
is closer than the somewhat younger TW Hya
\citep[60~pc, 10~Myr;][]{schneider2012} and 
the somewhat older Tuc-Hor \citep[40~pc, 30~Myr;][]{zuck2011} 
associations. Roughly 30\% of the members have luminous debris 
disks \citep[e.g.,][]{rebull2008,nilsson2009,riv2014}; 
$\beta$~Pic contains at least one gas giant \citep{lagrange2010}. 
From current planet detection statistics, $\gtrsim$ 50\%--60\% of 
stars with \mstar\ $\lesssim$ 1--2 \msun\ have at least one planet 
with $m_p \gtrsim$ 5--10~\mearth\ and $a \lesssim$ 10--20~AU
\citep[e.g.,][and references therein]{najita2014}. If satellite 
swarms around massive planets are relatively common, there is a
a reasonably high probability of finding at least one satellite 
swarm within the $\beta$ Pic moving group.

Although only one planetary mass companion has been detected orbiting
members of the $\beta$ Pic moving group, current detection limits are 
encouraging. \citet{kasper2007} derive $\Delta m \approx$ 9--10~mag
at 0\secpoint5 in the broadband L filter; \citet{biller2013} report 
$\Delta m \approx$ 14--15 at 1--2\arcsec\ in the broadband H and 
narrow band CH$_4$ filters. As the sample sizes grow and the data
acquisition/reduction techniques improve, it should be possible to
constrain the frequency of luminous satellite swarms around the 
nearest young stars.

\subsection{Predictions for Irregular Satellites in the Solar System}
\label{sec: disc-ss}

Satellite swarm models for Fomalhaut b are based on the ensemble of 
irregular satellites orbiting the four gas giants in the solar system
\citep[e.g.,][]{kw2011a}. The $\sim$ 160 known satellites have radii
$r \lesssim$ 100--200~km and lie on eccentric, high inclination orbits
with semimajor axes of 20\% to 50\% of the Hill radius
\citep[e.g.,][]{jewitt2007a,brozovic2011,alexa2012}. Among the largest
objects with $r \approx$ 10--100~km, the cumulative size distribution is 
shallow and reasonably close to a power law with $n(>r) \propto r^{-q_c}$
and $q_c \approx$ 1 \citep{jewitt2007a,bottke2010}. For smaller objects, 
the size distribution may steepen.

Using a set of coagulation calculations, \citet{bottke2010} show that 
several Gyr of collisional evolution naturally produces satellite 
swarms with shallow size distributions. For model satellites with 
$r \approx$ 0.05--100~km orbiting Jupiter, the slope ranges from 
$q_c \approx$ 1.5 for $r \lesssim$ 5~km to $q_c \lesssim$ 1 at $r$ = 
5--100~km.  Swarms with longer collision times orbiting Saturn and 
Uranus have somewhat steeper power laws.

Our calculations confirm and extend these results. For 0.1--10~km 
satellites orbiting 100--300~\mearth\ planets, size distributions 
at 1~Gyr are wavy power laws with $q_c \approx$ 1--2; 10--100~km 
objects have steeper power laws $q_c \sim$ 2--3. We extended several 
of these calculations to 5 Gyr; large objects then have 
$q_c \sim$ 1--2. 

Satellites with $r \lesssim$ 0.1~km have wavy cumulative size
distributions with a broad range of power law slopes 
(Figs.~\ref{fig: sd1}--\ref{fig: sd8}). Independent of various
input parameters, very small particles with $r \lesssim$ 1~cm
have steep power laws with $q_c \approx$ 4. Intermediate size
particles with $r \approx$ 1~cm to 100~m have flatter size 
distributions, $q_c \approx$ 0--2. These power laws are sensitive 
to the size of the smallest particles 
(Figs.~\ref{fig: sd4}--\ref{fig: sd5}).  Calculations with smaller 
particles have steeper power laws than calculations with larger 
particles. 

Although recent surveys detect several irregular satellites with
$r \approx$ 1~km around Jupiter \citep[e.g.,][]{brozovic2011,
alexa2012,jacob2012,gomes2015}, testing our predictions is 
challenging.  With expected optical magnitudes $\gtrsim$ 28, 
irregular satellites with $r \lesssim$ 0.1~km are too faint 
for any current and planned ground-based telescope. However, 
many of our calculations predict changes in the slope of the 
size distribution at 0.1--1~km. Extending the discovery space 
to this size range is feasible and would place interesting 
constraints on the coagulation models.

\section{CONCLUSIONS}
\label{sec: conc}

We describe results from a large suite of coagulation calculations for
irregular satellite swarms orbiting 1--300~\mearth\ planets at $a$ = 
120~AU from a 1.9~\msun\ central star. The calculations follow the 
evolution of the size distribution for 10~\mum\ to 3000~km particles 
for 1~Gyr as a function of the initial mass of the swarm, the size of
the smallest particle in the swarm, the initial size distribution of 
particles, the binding energy of the particles, and the method for
distributing debris from a collision into smaller mass bins. 

Throughout the evolution, the largest satellites may grow or shrink.
Growing satellites may scatter other satellites out of the planet's 
Hill sphere or into tighter orbits around the planet. Among smaller
satellites, the size distribution develops a characteristic shape 
with a steep power law at small sizes, a flat portion at intermediate
sizes, and a shallow power law at larger sizes. The growth (shrinkage)
of satellites and the time for the size distribution to develop a
standard shape depend on the initial cloud mass, the initial size
distribution, the initial \rmin\ and \rmax, and the binding energy
of satellites.

In our baseline models, swarms orbiting 10--100~\mearth\ planets have
cross-sectional areas at 100--400~Myr comparable to the observed 
cross-sectional area of Fomalhaut b. In these models, \xcl\ = 0.01 and
\rmin\ = 100~\mum. Smaller \xcl\ and larger \rmin\ allow swarms 
orbiting somewhat more massive planets to match observations of 
Fomalhaut b.  Calculations with smaller \rmin\ require swarms around
somewhat less massive planets. Changing the initial size distribution
of satellites has little impact on these conclusions. Modifying the 
binding energy and the algorithm for distributing debris in smaller 
mass bins generally lowers the cross-sectional area, requiring swarms
around more massive planets to match Fomalhaut b.

Aside from discussing the impact of these calculations on our
understanding of planet formation theory (\S\ref{sec: disc-theory}), 
we derive 
predictions for (i) irregular satellites in the solar system and 
(ii) Fomalhaut b and satellites swarms in other exoplanetary systems. 
Identifying 0.1--1~km irregular satellites orbiting Jupiter would set
interesting constraints on coagulation models.  In Fomalhaut b, we 
predict a detectable trail of small particles within a few AU of 
the nominal orbit of the planet candidate.  For exoplanetary systems 
with ages of 1--10~Myr, detectable satellite swarms orbiting 
30--300~\mearth\ planets provide a way to estimate the frequency 
of sub-Jupiter mass planets at 50--150~AU around 1--2~\msun\ stars.

\acknowledgements

We acknowledge generous allotments of computer time on the NASA 
`discover' cluster. Comments and suggestions from M. Geller, 
G. Kennedy, and an anonymous referee improved our discussion.  
Portions of this project were supported by the {\it NASA } 
{\it Astrophysics Theory} and {\it Origins of Solar Systems} 
programs through grant NNX10AF35G and the {\it NASA Outer 
Planets Program} through grant NNX11AM37G.

\appendix

\section{Appendix}

To test the algorithms used in \orch, we compare numerical results 
with analytic solutions to the coagulation equation 
\citep{kl1998,kb2015} and published results from other investigators 
\citep{kb2001,bk2006,kb2008,bk2011a,kb2015}.  Here, we examine how 
\orch\ performs for collisional cascades in spherical swarms of 
satellites orbiting a massive planet.

The accuracy of all coagulation calculations depends on the mass 
spacing parameter between adjacent mass bins, 
$\delta_k = m_{k+1} / m_k$ \citep[e.g.,][]{weth1990,kl1998,kb2015}. 
At the start of our calculations, we fix the typical mass $m_k$ and 
the boundaries $m_{k-1/2}$ and $m_{k+1/2}$ of each mass bin.  The 
initial average mass within each bin is $\bar{m}_k = M_k / N_k$; 
typically $\bar{m}_k \approx m_k$. As each calculation proceeds, 
collisions add and remove mass from all bins; the average mass 
$\bar{m}_k$ and the average physical radius of particles 
$\bar{r}_k = (3 \bar{m}_k / 4 \pi \rho_p)^{1/3}$ then change with time. 

To illustrate how the evolution of satellite swarms changes with $\delta$, 
we consider the baseline model described in the main text. An ensemble of
satellites with \xcl\ = 0.01, \rmax\ = 100~km, and \rmin\ = 100~\mum\ orbit
a planet with \mpl\ = 10~\mearth. The initial size distribution is a power
law with $n(r) \propto r^{-q}$ and $q$ = 3.5 between \rmin\ and \rmax.

\begin{figure} 
\includegraphics[width=6.5in]{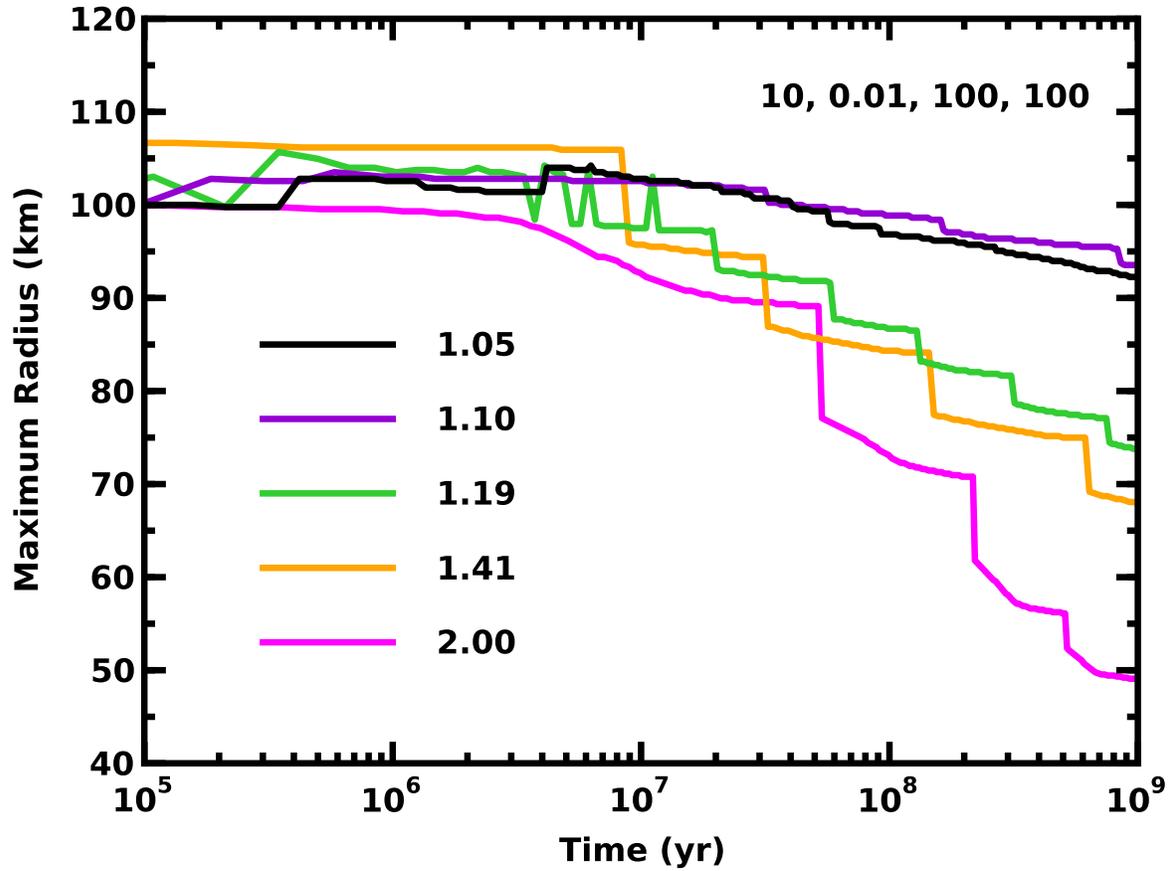}
\vskip 3ex
\caption{
Time evolution of the radius of the largest object for different 
mass spacing factors $\delta$ as listed in the legend. The satellite
swarm -- \xcl\ = 0.01, \rmax\ = 100~km, \rmin\ = 100~\mum\ -- 
orbits a planet with \mpl\ = 10~\mearth. 
\label{fig: rmax0}
}
\end{figure}

Fig.~\ref{fig: rmax0} shows the time-variation of \rmax\ for five different 
values of $\delta$.  In the figure, all curves have the same general shape: 
a brief, $\sim 10^5 - 10^6$~yr period where \rmax\ is roughly constant, 
followed by a gradual decrease in \rmax\ with time.  Tracks with larger 
$\delta$ decline faster. 

Along each track, the decline consists of a gradual reduction in 
\rmax\ interspersed with occasional small jumps to larger \rmax\ and 
large jumps to smaller \rmax.  In this example, collisions between equal 
mass objects with $r \approx \rmax$ increase the mass of the merged pair.
These collisions produce jumps to larger \rmax.  Cratering collisions -- 
where somewhat smaller objects gradually chip away at the mass of larger 
objects -- produce continuous mass loss from the largest objects. Thus, 
the average mass in the largest mass bin falls with time.  Eventually, 
this mass falls below the mass boundary between adjacent bins 
(e.g., $\bar{m}_k < m_{k-1/2}$). Objects in bin $k$ are then placed into 
bin $k-1$.  Averaging the mass of the `old' objects in bin $k-1$ with 
the `new' objects from bin $k$ yields a new average mass $\bar{m}_{k-1}$ 
which is smaller than the average mass of bin $k$. Thus, the size of the 
largest object jumps downward. Because the spacing of mass bins scales 
with $\delta$, calculations with larger $\delta$ have larger jumps than 
those with smaller $\delta$.

Although the mass loss rate from the grid is fairly insensitive to $\delta$,
the mass of the largest object clearly declines faster in calculations with 
larger $\delta$.  Cratering collisions are responsible for this difference. 
For all $\delta$, these collisions are rare. Thus, only a few of the largest 
objects suffer substantial mass loss from cratering collisions every time step.  
When $\delta$ is small (1.05--1.10), these objects are placed into the next 
smallest mass bin; the average mass of the remaining objects in the mass bin 
is unchanged. When $\delta$ is large (1.41--2.00), the amount of mass loss is 
not sufficient to place objects into the next smallest mass bin; the average 
mass of all objects in the bin then decreases. As a result, the average mass 
of the largest objects declines faster when $\delta$ = 2 than when $\delta$ 
= 1.05.

\begin{figure} 
\includegraphics[width=6.5in]{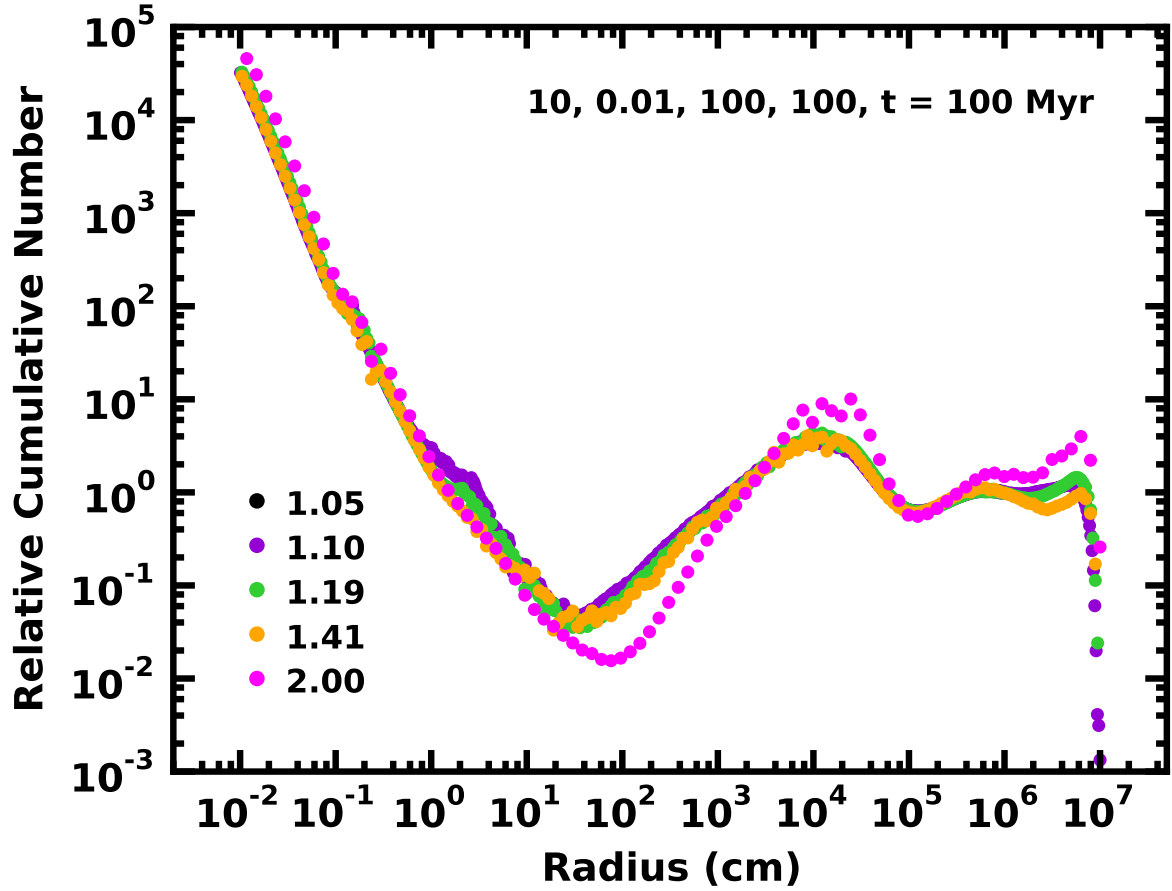}
\vskip 3ex
\caption{
Snapshot of the size distribution for the baseline models in 
Fig.~\ref{fig: rmax0} at 100~Myr. Solutions for $\delta$ = 2
have more waviness than those with smaller $\delta$. Solutions
for $\delta$ = 1.05--1.4 are nearly identical. 
\label{fig: sd0}
}
\end{figure}

Despite this difference, other aspects of the evolution are fairly insensitive 
to $\delta$.  Fig.~\ref{fig: sd0} shows snapshots of the relative size distributions 
at 100~Myr. Each curve follows a standard pattern, with a steep power law at
0.1~mm to 10~cm, a minimum at $\sim$ 30~cm, a rise from 1~m to 100~m, a
wavy pattern from 100~m to 50~km, and then an abrupt decline at the largest
sizes. When $\delta$ = 2, the fluctuations about a reference model with 
$\delta$ = 1.05 are large. For other $\delta$, deviations from the reference
model are minimal.

\begin{figure} 
\includegraphics[width=6.5in]{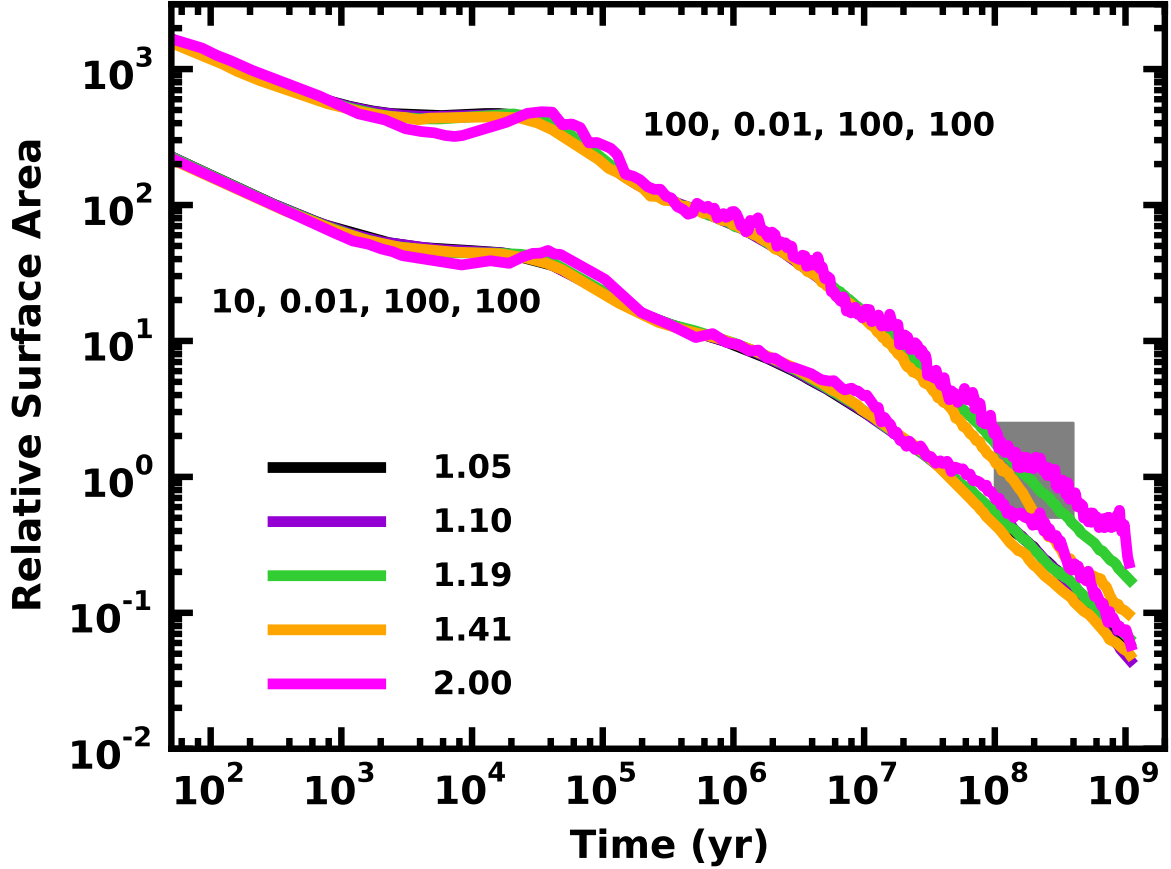}
\vskip 3ex
\caption{
Time evolution of the cross-sectional area for two baseline models
as a function of $\delta$ (as listed in the legend). For the upper
(lower) set of curves, \mpl\ = 100~\mearth\ (10~\mearth); in both,
\xcl\ = 0.01, \rmax\ = 100~km, and \rmin\ = 100~\mum.
The surface area is not a strong function of $\delta$.
\label{fig: area0}
}
\end{figure}

Fig.~\ref{fig: area0} illustrates the evolution of the relative surface area
for two different baseline models as a function of $\delta$. When \mpl\ = 
10~\mearth, the relative area declines from roughly $10^2$ at 100--1000~yr 
to roughly 0.1 at 1~Gyr. When $\delta$ = 1.05, the decline is smooth, with 
a minor change in slope at roughly $10^4$~yr. Adjustments from the initial
power law to the non-power law equilibrium size distribution (e.g., 
Fig.~\ref{fig: sd0}) cause this change in slope.  Evolution with $\delta$ = 2
is more ragged, with modest fluctuations relative to the reference model with
$\delta$ = 1.05. As $\delta$ declines, the evolution of the relative area
follows the reference model more closely. 

When \mpl\ = 100~\mearth, the initial relative area for models with \xcl\ = 0.01
is a factor of ten larger. The long term evolution is similar: a slow decline
with an inflection point around $10^4$~yr. Once again, the evolution of the
relative area is somewhat more ragged in calculations with $\delta$ = 2 than 
in calculations with smaller $\delta$.

Although there are clear differences in the evolution as a function of $\delta$,
the ability of an initial set of model parameters to match the observations rarely
depends on $\delta$. For these two examples, all of the 100~\mpl\ calculations
pass through the target box for Fomalhaut b. All of the 10~\mpl\ calculations 
graze the lower edge of the target box. 

For swarms of satellites in a spherical shell around a massive planet, 
calculations with $\delta \lesssim$ 1.2 yield a better understanding of the 
long term evolution of \rmax\ and the size distribution. Evolution of the 
total mass and relative surface area are fairly insensitive to $\delta$.
Single annulus calculations with $\delta$ = 1.2 run quickly, with execution 
times of 25 cpu hours for 1~Gyr evolution times using 10~\mum\ to 1000~km 
particles. Thus, we perform most calculations with $\delta$ = 1.2 and use
occasional calculations with smaller $\delta$ to verify interesting features
of the evolution.

\bibliography{ms.bbl}

\clearpage

%
%

\clearpage

\clearpage

\clearpage

\clearpage

\clearpage

\clearpage

\clearpage

\clearpage

\clearpage

\clearpage

\clearpage

\clearpage

\clearpage

\clearpage

\clearpage

\clearpage

\clearpage

\clearpage

\clearpage

\clearpage

\clearpage

\clearpage

\clearpage

\begin{figure} 
\includegraphics[width=6.5in]{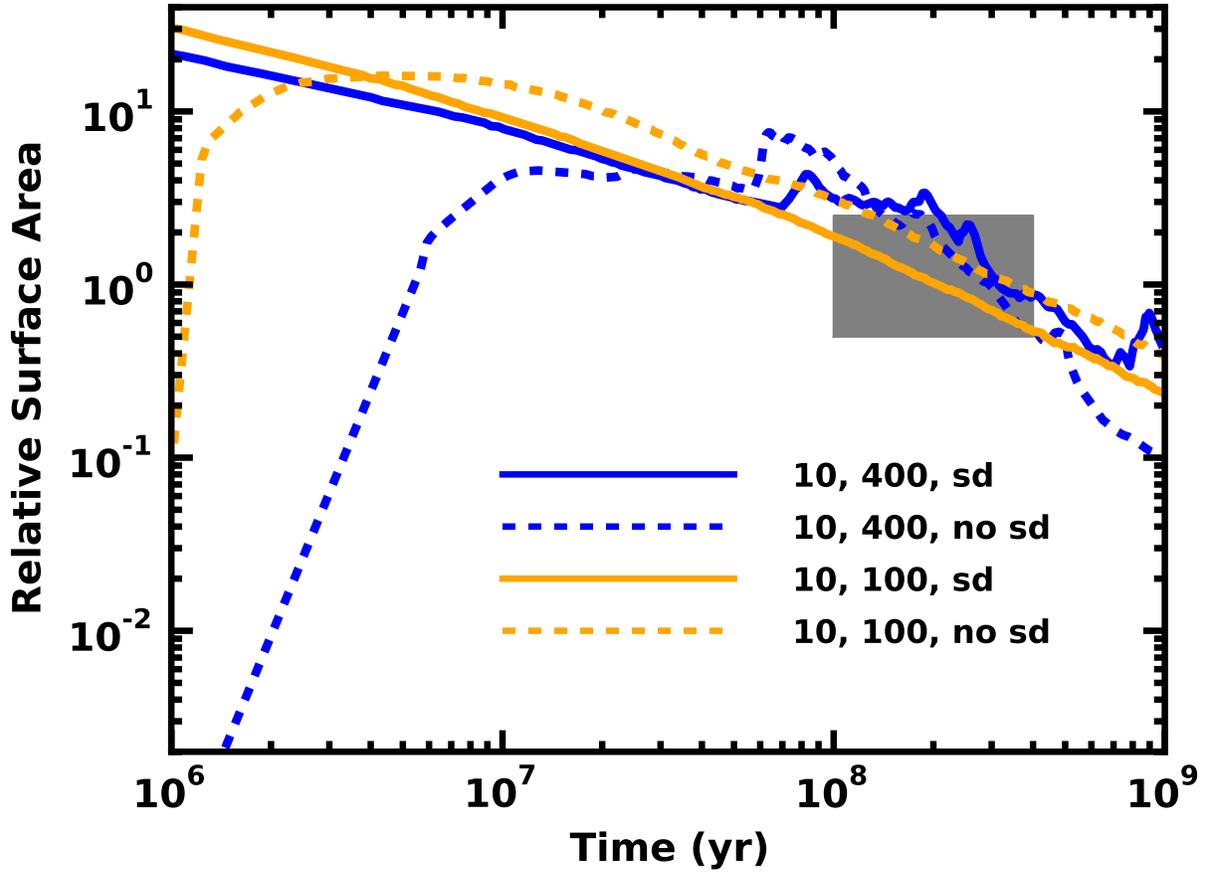}
\vskip 3ex
\caption{
As in Fig.~\ref{fig: area2} for calculations with (sd) and
without (no sd) an initial power law size distribution of
solid particles. The legend indicates \mpl, \rmax, and the
initial size distribution for each model curve.
\label{fig: area9}
}
\end{figure}
\clearpage

\clearpage

\clearpage

\clearpage

\clearpage

\clearpage

\clearpage

\end{document}